\documentclass{ametsocV6.1}

\usepackage{subfigure}

\newcommand{\OHC}{\mathrm{OHC}}
\newcommand{\W}{\mathcal{W}}

\newcommand{\s}{\bm{s}}
\renewcommand{\u}{\bm{u}}
\newcommand{\iid}{\overset{\text{i.i.d.}}{\sim}}
\newcommand{\Var}{\text{Var}}

\title{Locally stationary Argo ocean heat content estimates: Modeling, validation and uncertainty quantification}

\authors{Thea Sukianto,\aff{a}\correspondingauthor{Mikael Kuusela, mkuusela@andrew.cmu.edu}
Mikael Kuusela,\aff{a}$^\dagger$\thanks{$^\dagger$These authors share co-second authorship.}
Donata Giglio,\aff{b}$^\dagger$
Anirban Mondal,\aff{c}
Pulong Ma,\aff{d}
and Douglas W. Nychka\aff{e}
}

\affiliation{\aff{a}{Department of Statistics and Data Science, Carnegie Mellon University}\\
\aff{b}{Department of Atmospheric and Oceanic Sciences, University of Colorado Boulder}\\
\aff{c}{Department of Mathematics, Applied Mathematics, and Statistics, Case Western Reserve University}\\
\aff{d}{Department of Statistics, Iowa State University}\\
\aff{e}{Department of Applied Mathematics and Statistics,
Colorado School of Mines}
}

\abstract{Argo profiling floats measure seawater temperature and salinity in the upper 2000 meters of the ocean. These floats are uniquely capable of measuring the global Ocean Heat Content (OHC), a quantity that is of central importance for understanding Earth Energy Imbalance. Yet, producing Argo-based OHC estimates with reliable uncertainties is statistically challenging due to the complex structure and large size of the Argo dataset. Here we present an end-to-end mapping and uncertainty quantification framework for Argo-based OHC estimation using state-of-the-art methods from spatio-temporal statistics. The framework is based on modeling vertically integrated Argo temperature profiles as a locally stationary Gaussian process defined over space and time. This enables us to produce computationally tractable OHC anomaly maps based on data-driven decorrelation scales estimated from the Argo observations. Our modeling choices are validated using statistical cross-validation, which demonstrates the importance of including a climatological time trend in the mean field and accounting for time in the covariance function. We quantify the uncertainty of these maps using local conditional simulation ensembles, a novel approach that leads to principled spatially and temporally correlated uncertainty quantification. A new paired cross-validation technique is presented to validate these uncertainties. The mapping framework is implemented in an open-source codebase that is designed to be modular, reproducible and extensible. To demonstrate the mapping and uncertainty quantification capabilities of this approach, we present new Argo OHC maps with uncertainties for 2004--2022 and report on various downstream climatological estimates and their uncertainties. \vfill \vspace{5cm} {\small
This Work has not yet been peer-reviewed and is provided by the contributing Author(s) as a means to ensure timely dissemination of scholarly and technical Work on a noncommercial basis. Copyright and all rights therein are maintained by the Author(s) or by other copyright owners. It is understood that all persons copying this information will adhere to the terms and constraints invoked by each Author's copyright. This Work may not be reposted without explicit permission of the copyright owner.}
}

\begin{document}

\maketitle

\section{Introduction}

Monitoring Ocean Heat Content (OHC) is crucial for understanding the state and evolution of the climate system and how changes in the Earth's climate can impact ecosystems and human societies. Since more than 90\% of the excess heat in the climate system is absorbed by the ocean, monitoring changes in global OHC is to date the most accurate way to estimate the absolute magnitude of the Earth's energy imbalance (EEI) (\cite{IPCC6}, \cite{Meyssignac2019}, \cite{Hakuba2024}) driving the warming of the planet. It is therefore a key observable to quantify the effectiveness of mitigation efforts to reduce EEI. Monitoring OHC changes is also key to our understanding of global and regional sea level changes, as rising ocean temperatures raise sea level through thermal expansion and by accelerating the melt of polar ice sheets (\cite{IPCC6-ch9}). Furthermore, an increasing OHC (e.g., \cite{Johnson2025}, \cite{Trenberth2025}) has been associated with increasing intensity of tropical cyclones (e.g., \cite{Knutson2010}, \cite{Gilford2024}, \cite{Van_Oldenborgh2017}, \cite{Trenberth2018-jw}) and has already been linked to ecosystem impacts as, e.g., alterations in global biogeochemical cycles, shifts in fish populations, and coral bleaching (\cite{Hughes2017}).

Estimating OHC requires observing subsurface temperature. Hence, the Argo array of profiling floats (\cite{Roemmich2009_ArgoProgram}, \cite{Wong2020}) is key for global estimates of OHC, as it provides unprecedented (near-)global measurements of temperature, salinity and pressure in the upper 2,000~m of the ocean, with no seasonal bias. The program has been collecting data since the early 2000s and has played an essential role in our understanding of OHC and its changes in the past two decades (e.g., \cite{Roemmich2015}, \cite{Riser2016}).

Estimating OHC using Argo is not trivial: Most floats collect data every 10 days and the array was designed to achieve a nominal sampling resolution of $3^\circ \times 3^\circ$ in longitude and latitude.
To produce an estimate of global or regional OHC, one therefore needs to statistically interpolate these in-situ observations onto a regular grid in space and time. This yields mapped fields which can then be integrated over space to estimate OHC. Producing these gridded maps involves making modeling assumptions about the statistical distribution of the oceanographic fields. Different available products are based on different modeling assumptions, which contributes to the substantial observed spread in OHC estimates (\cite{Von_Schuckmann2016-rf}, \cite{von_Schuckmann2023}, \cite{Savita2022}, \cite{Hakuba2024}). While some work has been done to analyze the sensitivity to modeling choices (e.g., \cite{boyer_etal2016}, \cite{Allison2019}, \cite{Savita2022}), a comprehensive assessment of the ecosystem of mapping methods is ongoing (\cite{ME4OH}). Most importantly, open-source code for reproducing the different products is often not available and statistical uncertainties are usually not provided.

In this work, we introduce a new statistical mapping framework for estimating global and regional OHC from Argo observations. Our framework is based on the locally stationary Gaussian process approach introduced for Argo mapping in \cite{Kuusela2018}. Here we adapt and extend that methodology to inferring OHC: a key contribution of this work is a novel method for rigorously quantifying the statistical uncertainty of the mapped fields. The uncertainty quantification method is based on producing a conditional simulation ensemble of the locally stationary mapped fields. The ensemble members reflect the variability and spatio-temporal dependence of the maps conditional on the Argo observations. The ensemble can therefore be used to obtain uncertainties on downstream scientific quantities, such as the spatial integral for OHC, by simply processing each ensemble member and looking at the distribution of the resulting values. Crucially, this approach accounts for the spatio-temporal correlation in the conditional distribution of the mapped fields and fully propagates this dependence into the uncertainty of the OHC estimates and their dependence across time. The spatio-temporal dependence encoded by the conditional simulation ensemble is validated using a new paired cross-validation approach.

In addition to providing statistically rigorous mapping uncertainties, our framework carries over the benefits of the local mapping approach of \cite{Kuusela2018} into the OHC estimates. Namely, we use flexible, data-driven local models for both the mean field and the anomalies. Specifically, the anomalies are mapped using an anisotropic local covariance function that includes both space and time and whose variance and decorrelation scale parameters are estimated from the data. When compared with a model that does not estimate covariance parameters directly from the input data and does not include time in the covariance function  \citep[e.g.][]{Roemmich2009}, the model used here was shown to yield more accurate estimates of Argo temperature fields in \cite{Kuusela2018}. In the following, we demonstrate a similar improvement for OHC estimates.

Furthermore, we investigate the sensitivity of our estimates to the inclusion of a climatological time trend in the mean field used to produce the maps. We find that the estimated OHC trend can nearly double depending on this modeling choice (an increase that is consistent with previous studies, e.g., \cite{Cheng2015}). Using statistical cross-validation, we use Argo data itself to show that the model without the time trend is misspecified which leads to downward biased estimates of the OHC trend. On the other hand, we find that cross-validation does not indicate any obvious misspecification in the model where a local time trend is included in the mean field.

Alongside this paper, we release the full source code of the mapping framework at \url{https://github.com/ttsukianto/LocalGP_OHC} under a permissive open-source license. The codebase enables reproducing every step of our pipeline, starting from the near-real-time Argo observations available at the Argo Global Data Assembly Centres (GDACs). To the best of our knowledge, this is the first fully reproducible Argo-based end-to-end mapping framework for OHC. The codebase is also designed to be modular and easily extensible to a variety of oceanographic mapping tasks beyond OHC. Our hope is that this codebase will serve as a starting point for community efforts to further study, compare and improve Argo mapping methods toward increasingly accurate estimates of OHC and other oceanographic quantities of interest.

We demonstrate our mapping framework by creating OHC estimates and associated conditional simulation ensembles for years 2004--2022 on a $1^\circ \times 1^\circ \times 1 \text{ month}$ grid. We employ the new OHC estimates and, in particular, the uncertainty quantification capability enabled by the conditional simulations in a number of climatological case studies, each of which demonstrates an uncertainty quantification task that would not have been possible to achieve using existing OHC data products.

A LocalGP data product based on the methodology described here is underway as part of an ongoing effort and some of the estimates from the project \citep{Giglio_etal2026_Zenodo} have been used in a number of climate assessments (\cite{WMO2023, WMO2022, WMO2021}, \cite{Johnson2024, Johnson2025}), oceanographic studies (\cite{von_Schuckmann2023}, \cite{Hakuba2024}, \cite{Meyssignac2024}, \cite{Sala2025}) and intercomparison efforts \citep{ME4OH}. This paper serves as a formal description of the mapping and uncertainty quantification methodology underlying the LocalGP product and estimates.

The rest of the manuscript is structured as follows: In Section~\ref{sec:background}, we provide a brief description of Argo data, define the OHC mapping problem and briefly review the related literature. Section~\ref{sec:method} describes in detail each step of our OHC mapping and uncertainty quantification pipeline. We validate our modeling and methodological choices in Section~\ref{sec:validation}. Section~\ref{sec:OHC_product} illustrates the resulting OHC estimates and uncertainties in a number of climatological case studies and Section~\ref{sec:discussion} concludes with a discussion.

\section{Data description, problem definition and related work} \label{sec:background}

\subsection{Argo float data} \label{sec:Argo_data}

Argo autonomous profiling floats (\cite{Roemmich2009_ArgoProgram}, \cite{Riser2016}, \cite{Wong2020}) have been deployed since 1999, with the array reaching sparse (near-)global coverage of temperature, salinity, and pressure observations in the upper 2000~m of the ocean in the mid-2000s. Many mapped Argo data products therefore start from 2004 or 2005. The $3^\circ \times 3^\circ$ global target coverage was achieved around 2007, although the array has kept expanding since then. For the past decade or so, roughly 4000 floats have been consistently available, surpassing the original target of 3000 active floats.

Of particular relevance to this work is the fact that there is a certain degree of heterogeneity in the vertical sampling of the Argo array (\cite{Wong2020}). First, the vertical sampling of the early floats is much sparser than that of the more recent floats. This is because the early floats used the Argos satellite system for data transmission which limited the profiles to 50--100 vertical observations, while the more recent floats use the Iridium communication system enabling the profiles to have up to $\sim$1000 observations. Second, different profiles span the depth range in the upper 2000 dbar to a different extent. For most floats, the shallowest observation is a few meters below the surface, limiting the observations available for the top 10--15 dbar of the vertical column. Similarly, not all profiles actually reach 2000 dbar. As an example, in the early years of Argo, some floats in the tropics did not sample below 1000 dbar due to limitations on buoyancy adjustment. This resulted in, e.g., the Central Atlantic to be poorly sampled below 1000 dbar during those years. Finally, even when the ocean is deep enough and the float is operating as planned, the deepest reported observation is typically tens of meters short of 2000 dbar. Our OHC mapping method needs to take these complexities in Argo's vertical sampling into account, as discussed below in Section~\ref{sec:background}\ref{sec:probDef}.

For the estimates presented in this paper, we use as the input data profiles collected during 2004--2022 and distributed in the January 2023 snapshot of the Argo GDAC (\cite{ArgoDataJan2023}). The data need to be quality-controlled before being used in the mapping pipeline. In this work, we use the quality control procedure described in \cite{Kuusela2018} and implemented in the \texttt{processDACdata.m} script in the codebase accompanying this paper. The procedure is based on the built-in Argo quality control flags and a number of additional filtering steps. We use delayed-mode (D) or adjusted real-time-mode (A) data whenever available but accept unadjusted real-time-mode (R) profiles if D or A mode data are not available. A total of 1,731,330 profiles for 2004--2022 passed the quality control steps. In addition, we exclude profiles at the same spatial location if their timestamp difference is $\leq$ 15 minutes, which removed an additional 220 profiles.

\subsection{Problem definition} \label{sec:probDef}

The ocean heat content (OHC) at time $t$ between depths $z_u$ and $z_d$ within spatial domain $D$ is the time series defined by the integral
\begin{equation}
\OHC_{z_u}^{z_d}(t) = \underset{{(x,y) \in D}}{\iint} \int_{z=z_u}^{z=z_d} \rho_0 \, c_{p,0} \, T(x,y,z,t)\,\mathrm{d}z \,\mathrm{d}S(x,y). \label{eq:OHC_def}
\end{equation}
Here $T(x,y,z,t)$ is the potential temperature (computed from temperature, salinity, and pressure in-situ measurements using the TEOS-10 toolbox; \cite{TEOS-10}) at longitude $x$, latitude $y$, pressure $z$ (a pressure change of 1~dbar corresponds to roughly 1~m) and time $t$; density $\rho_0$ is 1030 kg/m$^3$; and specific heat capacity $c_{p,0}$ is $3989.244 \ \mathrm{J\cdot kg^{-1} \cdot K^{-1}}$.

Thanks to Argo's fine vertical sampling, the vertical integral in \eqref{eq:OHC_def} is straightforward to compute accurately (see Section~\ref{sec:method}\ref{sec:verticalInt}). Denoting the vertical integral by $\widetilde{\OHC}_{z_u}^{z_d}(x,y,t)$, we have
\begin{equation}
\OHC_{z_u}^{z_d}(t) = \underset{{(x,y) \in D}}{\iint} \widetilde{\OHC}_{z_u}^{z_d}(x,y,t) \,\mathrm{d}S(x,y) \approx \sum_{i} \widetilde{\OHC}_{z_u}^{z_d}(x_{i},y_{i},t) \cdot S_{i}, \label{eq:OHC_def_xy}
\end{equation}
where $(x_{i},y_{i})$ are points on a $1^\circ \times 1^\circ$ grid over $D$ and $S_{i}$ is the area of a $1^\circ \times 1^\circ$ grid element whose center point is $(x_{i},y_{i})$. With a slight abuse of terminology, we will call  $\widetilde{\OHC}_{z_u}^{z_d}(x,y,t)$ regional OHC (it is strictly speaking heat content per unit area with units Jm$^{-2}$).

In the following, our primary focus is to estimate the time series $\OHC_{z_u}^{z_d}(t)$ at monthly time intervals. The main challenge in estimating $\OHC_{z_u}^{z_d}(t)$ from Argo is that $\widetilde{\OHC}_{z_u}^{z_d}(x,y,t)$ is only known at those spatio-temporal locations where a float collected a vertical profile. So in order to compute the spatial integral over $D$ in Eq.~\eqref{eq:OHC_def_xy}, we need to interpolate the observed $\widetilde{\OHC}_{z_u}^{z_d}$ values onto a regular $1^\circ \times 1^\circ \times 1 \text{ month}$ grid over space and time. Hence, the key task here is to produce accurate interpolations of $\widetilde{\OHC}_{z_u}^{z_d}$ and quantify the associated mapping uncertainties so as to obtain accurate estimates of the integral defined in \eqref{eq:OHC_def_xy} and its uncertainty.

In this work, the domain $D$ is either the Argo-sampled part of the global ocean or selected subregions thereof. We use the ocean mask $D_{RG}$ provided with the Roemmich--Gilson 2019 data product (\cite{RG2019}) to identify locations that are as deep or deeper than $z_d$. The domain includes the majority of all oceanic basins (with the exception of the Arctic ocean), but excludes polar regions, continental shelves, and other shallow areas. Additionally, we intersect $D_{RG}$ with a data availability mask $D_{\text{data}}$, defined to include only those grid points whose local neighborhood contains enough data to reliably estimate the seasonal cycle. This is needed since the lack of wintertime observations in certain parts of the ocean with seasonal ice cover caused instabilities in estimating the seasonal cycle in those regions. The final domain of integration is then $D = D_s \cap D_{RG} \cap D_{\text{data}}$, where $D_s$ is either the entire ocean or a specific subregion of interest.

Since we can only compute the vertical integral $\widetilde{\OHC}_{z_u}^{z_d}(x,y,t)$ for profiles ranging at least from $z_u$ to $z_d$, data availability in the vertical direction (see Section~\ref{sec:background}\ref{sec:Argo_data}) affects our choice of depth ranges to consider. In this work, we run the mapping pipeline to estimate OHC time series over two disjoint vertical sections. The upper section ranges from 15 dbar to 975 dbar and the lower from 975 dbar to 1850 dbar with the corresponding time series denoted by $\OHC_{15}^{975}(t)$ and $\OHC_{975}^{1850}(t)$, respectively. The choice of 15 dbar and 1850 dbar as the endpoints is motivated by the low availability of profiles with observations shallower than 15 dbar or deeper than 1850 dbar. Additionally, as described in Section~\ref{sec:background}\ref{sec:Argo_data}, many profiles are cut short near 1000 dbar. This led us to divide the ocean into two vertical sections at 975 dbar in order to include the maximal number of profiles in the analysis. Altogether, 1,540,593 profiles were available for the upper section and 1,128,932 for the lower section. More profiles could be included by further subdividing the sections (e.g., \cite{Giglio_etal2026_Zenodo}), but these two sections suffice for our purposes of introducing the framework here.

\subsection{Related work}

The work presented here builds on a number of efforts that, over time, have advanced mapping methods and highlighted the implications of different assumptions in the statistical model used to fill in observational gaps as well as the need for improved uncertainty estimates in gridded ocean heat content products (e.g., \cite{boyer_etal2016}, \cite{Allison2019}, \cite{Savita2022}, \cite{von_Schuckmann2023}). Thanks to these efforts, several Argo-based gridded oceanic fields are currently available to the oceanographic community (e.g., \cite{argo-products}). Methods underlying these estimates include objective analysis (\cite{Gaillard2016}, \cite{Cheng2017}, \cite{Szekely2025}, \cite{Cabanes2013}, \cite{Gouretski2018}, \cite{UdayaBhaskar2007}, \cite{Hosoda2008}, \cite{Good2013}, \cite{Garcia2024}, \cite{Schmidtko2013}, \cite{Roemmich2009}, \cite{Zhang2022}, \cite{Hermanson_etal2023}), the Barnes successive correction method (\cite{Lu2020}, \cite{Barnes1964}, \cite{Barnes1973}; using weighing functions), a variational analysis approach (\cite{IPRC}), hierarchical Bayesian modeling (\cite{Baugh_and_McKinnon2022}), functional data analysis (\cite{Yarger2022}), and random forest regression (\cite{Lyman2023}; merging information from in-situ and satellite data). As in many products currently available, the mapping method presented here uses objective mapping (also known as kriging or optimal interpolation) to fill in observational gaps. Yet, we incorporate the advances from \cite{Kuusela2018} who showed that estimating the variance and spatial/temporal decorrelation scale parameters from the input data in a moving-window fashion allows for an improved representation of the spatial and temporal variability of interest. This enables producing improved maps and is key to rigorous uncertainty quantification, leading to the local conditional simulation algorithm developed in this work.

\section{Mapping and uncertainty quantification methodology} \label{sec:method}

\subsection{Outline}
Our OHC mapping framework is based on the following statistical model:
\begin{equation}
\widetilde{\OHC}_{z_u}^{z_d}(x,y,t) = \mu (x,y,t) + a(x,y,t) + \varepsilon(x,y,t), \label{eq:model_def}
\end{equation}
where $\mu$ is a climatological mean field, $a$ is a zero-mean anomaly field that reflects transient deviations from the mean field and $\varepsilon$ is a zero-mean spatially and temporally uncorrelated nugget effect that describes unresolved fine-scale variation (e.g., from mesoscale processes). This section describes the modeling and mapping process, divided into four steps:
\begin{enumerate}
    \item Compute the vertical integral $\widetilde{\OHC}_{z_u}^{z_d}(x,y,t)$ at the observed profile locations using PCHIP interpolation for the vertical sections defined in Section~\ref{sec:background}\ref{sec:probDef}.
    \item Estimate the mean field $\mu (x,y,t)$, which includes seasonal harmonics and a climatological time trend, using local polynomial regression.
    \item Map the anomalies $a(x,y,t)$ using locally stationary Gaussian process regression with data-driven decorrelation scales.
    \item Quantify the uncertainty of the $\widetilde{\OHC}_{z_u}^{z_d}(x,y,t)$ field using an ensemble of spatially and temporally correlated local conditional simulations.
\end{enumerate}

\subsection{Vertical integration: PCHIP interpolation} \label{sec:verticalInt}
For each observed Argo profile, we first compute the vertical integrals $\widetilde{\OHC}_{15}^{975}(x_p,y_p,t_p)$ and $\widetilde{\OHC}_{975}^{1850}(x_p,y_p,t_p)$ for the upper and lower depth sections defined in Section~\ref{sec:background}\ref{sec:probDef}. Here $(x_p,y_p)$ is the location and $t_p$ the timestamp of the $p$th profile. For each profile and vertical section, we perform vertical PCHIP interpolation (\cite{Fritsch1980}) as implemented in the Matlab \texttt{interp1} function over a fine grid of $5(z_d - z_u)$ evenly spaced depth points. Then, we numerically integrate the interpolated profile from $z_u$ to $z_d$ using the trapezoidal method to obtain the vertical integral $\widetilde{\OHC}_{z_u}^{z_d}(x_p,y_p,t_p)$ for each $p$.

PCHIP is well-suited for interpolating Argo profiles due to its shape-preserving property which helps it avoid spurious bumps in the interpolations (\cite{Barker2020, Li2022}). In practice, the choice of the vertical integration method matters most for the sparser Argos profiles (see Section~\ref{sec:background}\ref{sec:Argo_data}); the more recent Iridium profiles are sampled densely enough that there are only small differences between reasonable integration methods. In the early phases of this work, we performed a study where we subsampled Iridium profiles to look like Argos profiles and then compared various integration methods against the Iridium ``ground truth''. The methods compared were a) trapezoidal integration without interpolation, b) Simpson's method without interpolation, c) trapezoidal integration with spline interpolation and d) trapezoidal integration with PCHIP interpolation. Out of the four, we found method d) to be the most accurate in terms of the mean, standard deviation, maximum and minimum of the signed integration error and hence chose to use it throughout the rest of the pipeline.

\subsection{Mean field: local polynomial regression}
\label{sec:mean-field}

Note that the vertical integral $\widetilde{\OHC}_{z_u}^{z_d}$ is only observed at the profile locations, so in order to produce an estimate of $\OHC_{z_u}^{z_d}(t)$ and other related quantities, we will need to interpolate the above vertically integrated profiles onto a regular grid over the Argo-sampled part of the global ocean. Following the model definition in Eq.~\eqref{eq:model_def}, we first estimate the mean field $\mu (x,y,t)$.

Let us define a square spatial window 
\begin{equation}
\mathcal{W}(x^*,y^*)=[x^*-\lambda, x^*+\lambda] \times [y^*-\lambda, y^*+\lambda]
\end{equation}
centered on location $(x^*,y^*)$ on a $1^\circ \times 1^\circ$ grid over the domain $D$ defined in Section \ref{sec:background}\ref{sec:probDef}, where $\lambda$ is a window size parameter in degrees.

Following \cite{Ridgway2002,Roemmich2009}, we estimate a climatological mean for each $(x^*,y^*)$  using a local polynomial regression function within $\mathcal{W}(x^*,y^*)$:
{\allowdisplaybreaks
\begin{equation}
\begin{split}
m_{x^*,y^*}(x,y,t) =&\beta_0+\beta_1(x-x^*)+\beta_2(y-y^*)\\
&+\beta_3(x-x^*)(y-y^*)+\beta_4(x-x^*)^2+\beta_5(y-y^*)^2\\
&+\sum_{k=1}^6\gamma_k\sin\left(2\pi k \frac{\tau(t)}{365}\right)+\sum_{k=1}^6\delta_k\cos\left(2\pi k \frac{\tau(t)}{365}\right)\\
&+\nu_1(t-t_D)+\nu_2(t-t_D)^2, \quad (x,y) \in \mathcal{W}(x^*,y^*),
\end{split}
\label{eq:mean_field}
\end{equation}}where $t$ is the time in Julian days, $\tau(t) \in [0,365]$ is the time in yeardays and $t_D$ is the midpoint of the time range of interest in Julian days. In this work, we add the linear and quadratic climatological time trend terms $\nu_1(t-t_D)+\nu_2(t-t_D)^2$ to the regression function definition in \cite{Ridgway2002,Roemmich2009}. We investigate the effect of introducing these terms on the OHC estimates in Section~\ref{sec:validation}\ref{sec:val_mean_field}. For each grid point $(x^*,y^*)$, the regression parameters are estimated using least-squares estimates $(\bm{\hat\beta},\bm{\hat\gamma},\bm{\hat\delta},\bm{\hat\nu})$ over all profile locations $(x_p,y_p,t_p)$ within the window $\mathcal{W}(x^*,y^*)$. The time-dependent mean field estimate at $(x^*,y^*)$ is then given by
\begin{equation}
\begin{split}    
    \hat{\mu}(x^*,y^*,t) & = \hat{m}_{x^*,y^*}(x^*,y^*,t) \\ & = \hat{\beta}_0+\sum_{k=1}^6\hat{\gamma}_k\sin\left(2\pi k \frac{\tau(t)}{365}\right)+\sum_{k=1}^6\hat{\delta}_k\cos\left(2\pi k \frac{\tau(t)}{365}\right)
+\hat{\nu}_1(t-t_D)+\hat{\nu}_2(t-t_D)^2.
\end{split}
\end{equation}

\subsection{Anomalies: locally stationary Gaussian process regression}
\label{sec:anom}
We next turn our attention to modeling the stochastic terms $a(x,y,t)$ and $\varepsilon(x,y,t)$ in Eq.~\eqref{eq:model_def}. Note that
\begin{equation}
    r(x,y,t) := \widetilde{\OHC}_{z_u}^{z_d}(x,y,t) - \mu (x,y,t) = a(x,y,t) + \varepsilon(x,y,t),
\end{equation}
so we can focus on modeling the zero-mean residual field $r(x,y,t)$ which we approximate by $\hat{r}(x,y,t) = \widetilde{\OHC}_{z_u}^{z_d}(x,y,t) - \hat{\mu} (x,y,t)$. To obtain the observed values of $r$, we compute $\hat{r}$ at each observed location $(x_p,y_p,t_p)$ by mean-centering each vertically integrated profile, i.e., by subtracting $\hat{\mu}(x^*,y^*,t_p)$ from the profile, where $(x^*,y^*)$ is the spatial grid point closest to $(x_p,y_p)$.

We now interpolate $r$ onto a regular grid over the Argo-sampled global ocean using the locally stationary Gaussian process (GP) regression model from \cite{Kuusela2018}. Fitting a global nonstationary model would be computationally challenging due to the large number of Argo profiles. Furthermore, it would be challenging to define a sufficiently flexible global covariance function that would accurately describe regional variability. The locally stationary moving-window approach introduced in \cite{Kuusela2018} yields smoothly varying model parameter estimates and predictions over the global ocean while reducing the computational burden. 

Let us first define a spatio-temporal window
\begin{equation}
\widetilde{W}(x^*,y^*,t^*)=[x^*-\lambda_G,x^*+\lambda_G] \times [y^*-\lambda_G,y^*+\lambda_G] \times [t^*-\lambda_t,t^*+\lambda_t]
\end{equation}
centered on points $(x^*,y^*,t^*)$ on a $1^\circ \times 1^\circ \times 1$ month grid over $D \times D_t$, where $D_t$ is the time domain of interest within the Argo sampling period. Now, let us consider the collection of these windows over all years in $D_t$,
\begin{equation}
W(x^*,y^*,\tau^*)= \{ \widetilde{W}(x^*,y^*,t^*) \::\: t^* \in D_t, \: t^* \bmod 365 = \tau^* \}, \end{equation}
where $\tau^* \in [0,365]$ is a yearday on the same monthly grid as $t^*$.
Then, within a given $W$, let us assume that the $\widetilde{\OHC}$ anomaly process for year $i$ is a zero-mean Gaussian process with the following definition:
\begin{equation}
a_i \iid \text{GP} (0,k(\s_1,t_1,\s_2,t_2;\boldsymbol{\theta}_{W})),
\label{eq:anomaly_def}
\end{equation}
where $(\s,t)=(x,y,t)$ is an $i$th year spatio-temporal location within $W$, $\boldsymbol{\theta}_{W}=(\phi_{W},\theta_{W,\text{lon}},\theta_{W,\text{lat}},\theta_{W,t})$ is a vector of covariance parameters for $W$, and the independent and identically distributed (i.i.d.) assumption is across the years in $W$. More specifically,  $k=\phi_{W}\exp(-d(\s_1,t_1,\s_2,t_2))$ is an anisotropic exponential space-time covariance function with a distance function defined by
\begin{equation}
    d(\s_1,t_1,\s_2,t_2) = \sqrt{\left(\frac{x_1-x_2}{\theta_{W,\text{lon}}}\right)^2+\left(\frac{y_1-y_2}{\theta_{W,\text{lat}}}\right)^2+\left(\frac{t_1-t_2}{\theta_{W,t}}\right)^2}, 
    \label{eq:dist_def}
\end{equation}
where $\theta_{W,\text{lon}}$, $\theta_{W,\text{lat}}$ and $\theta_{W,t}$ are the decorrelation scales within $W$.

Following the model definition in \eqref{eq:model_def}, for a given $W$, we additionally define an additive Gaussian nugget effect $\varepsilon \iid N(0,\sigma^2_W)$, where $\sigma^2_W$ is the nugget variance for $W$ and the i.i.d.~assumption is over both space and time in $W$. We then numerically find the maximum likelihood estimates of the model parameters for each $W$ using those mean-centered residuals $\hat{r}$ that fall within $W$. For each model parameter $\theta^k =\phi,\theta_{\text{lon}},\theta_{\text{lat}},\theta_{t},\sigma^2$, this produces spatially and seasonally varying parameter estimates $\hat{\theta}^k_{W(x^*,y^*,\tau^*)}$ as a function of $(x^*,y^*,\tau^*)$. We found empirically that the effect of the seasonal variation in the $\hat{\theta}^k$'s is small. To improve robustness, we therefore opt to use the temporal medians of the parameter estimates as the final non-seasonally varying plug-in estimates. That is, for any $\widetilde{W}$ centered at $(x^*,y^*) \in D$, the final parameter estimates are given by $\hat{\theta}^k_{x^*,y^*} = \underset{\tau^*}{\text{med}}(\hat{\theta}^k_{W(x^*,y^*,\tau^*)})$, where the median is taken over the 1-month grid of $\tau^*$'s.

Since any finite subset of a Gaussian process follows a multivariate Gaussian distribution, the point prediction of $r(x^*,y^*,t^*)$ at a grid point $(x^*,y^*,t^*)$ is the mean of a Gaussian distribution conditioned on the data within $\widetilde{W}(x^*,y^*,t^*)$ and with the $\hat{\theta}^k_{x^*,y^*}$'s plugged in as the model parameters.  The mathematical details are given, for example, in \citet{Kuusela2018}. It then remains to add the estimated mean field value $\hat\mu(x^*,y^*,t^*)$ to obtain the mapped regional OHC estimate $\widehat{\widetilde{\OHC}}_{z_u}^{z_d}(x^*,y^*,t^*)$.

\subsection{Uncertainty quantification: local conditional simulations}
\label{sec:uq-condsim}

To quantify the uncertainty of our OHC estimate for month $t$ defined by Eq.~\eqref{eq:OHC_def_xy}, we are ultimately interested in the predictive variance:
\begin{equation}
\begin{split}
\Var(\OHC_{z_u}^{z_d}(t)|\text{data}) \approx &\sum_{i} \Var(\widetilde{\OHC}_{z_u}^{z_d}(x_{i},y_{i},t)|\text{data}) \cdot S_{i}^2\, +\\ 
&\sum_{i,j} \text{Cov}(\widetilde{\OHC}_{z_u}^{z_d}(x_{i},y_{i},t),\widetilde{\OHC}_{z_u}^{z_d}(x_{j},y_{j},t)|\text{data}) \cdot S_{i}S_{j}.
\end{split}
\end{equation}
The predictive variance $\Var(\widetilde{\OHC}_{z_u}^{z_d}(x_{i},y_{i},t)|\text{data})$ at a grid point $i$ is readily available using the local GP described above. However, the predictive covariance $\text{Cov}(\widetilde{\OHC}_{z_u}^{z_d}(x_{i},y_{i},t),\widetilde{\OHC}_{z_u}^{z_d}(x_{j},y_{j},t)|\text{data})$ does not immediately follow from the above model definition and would in any case be infeasible to compute, store and work with for all pairs of grid points $(i,j)$ over the Argo-sampled global ocean. Instead of attempting to compute the predictive covariance, we propose to simulate an ensemble of realizations from the conditional distribution $p(\{\widetilde{\OHC}_{z_u}^{z_d}(x_{i},y_{i},t_{i})\}_i|\text{data})$. If we assume that the true regional OHC field is a Gaussian process, then the field conditional on the observed data is also a GP. Our goal is to simulate from this GP such that the simulations respect the spatio-temporal dependence structure implied by the local GPs within each window $\widetilde{\W}(x_i,y_i,t_i)$.

Our local conditional simulation method, which has similarities with the unconditional simulation approach in \cite{Nychka2018}, is based on the following argument: A broad class of non-stationary zero-mean spatio-temporal Gaussian processes can be expressed as convolutions of Gaussian white noise \citep{Higdon1998,Higdon1999,Higdon2002}
\begin{equation}
    f(\s,t) = \int_{D \times D_t} h_{\s,t}(\u,v) w(\u,v) \,\mathrm{d}\u \,\mathrm{d} v, \quad (\s,t) \in D \times D_t, \label{eq:kernelConv}
\end{equation}
where $w(\u,v)$ is a Gaussian white noise process over the spatio-temporal domain $D \times D_t$ and the kernel $h_{\s,t}(\u,v)$ is localized so that it has most of its mass near $(\s,t)$. The process $f(\s,t)$ has covariance function
\begin{equation}
    k(\s_1,t_1,\s_2,t_2) = \int_{D \times D_t} h_{\s_1,t_1}(\u,v) h_{\s_2,t_2}(\u,v) \,\mathrm{d}\u \,\mathrm{d} v. \label{eq:convCov}
\end{equation}
Let us now focus on a specific window $\widetilde{\W} (\s^*,t^*)$. Within this window, the predictive covariance of the local Gaussian process model provides a reasonable approximation of $k(\s_1,t_1,\s_2,t_2)$. We can therefore use Eq.~\eqref{eq:convCov} restricted to this window to find a kernel $h_{\s^*,t^*}(\u,v)$ for the midpoint of the window based on the local predictive covariance function within the window. By repeating this for all windows within $D \times D_t$, we can find kernels $h_{\s,t}(\u,v)$ for all $(\s,t) \in D \times D_t$. Kernels defined this way have bounded support within $\widetilde{\W} (\s,t)$, so Eq.~\eqref{eq:kernelConv} becomes
\begin{equation}
    f(\s,t) = \int_{\widetilde{\W} (\s,t)} h_{\s,t}(\u,v) w(\u,v) \,\mathrm{d}\u \,\mathrm{d} v,  \quad (\s,t) \in D \times D_t.
\end{equation}
This shows that to simulate $f(\s,t)$ under this approach, it suffices to convolve white noise within $\widetilde{\W} (\s,t)$ with the kernel $h_{\s,t}(\u,v)$.

\begin{figure}[t] 
  \centering
  \noindent\includegraphics[width=15cm,trim = 0 0 0 4cm, clip=true]{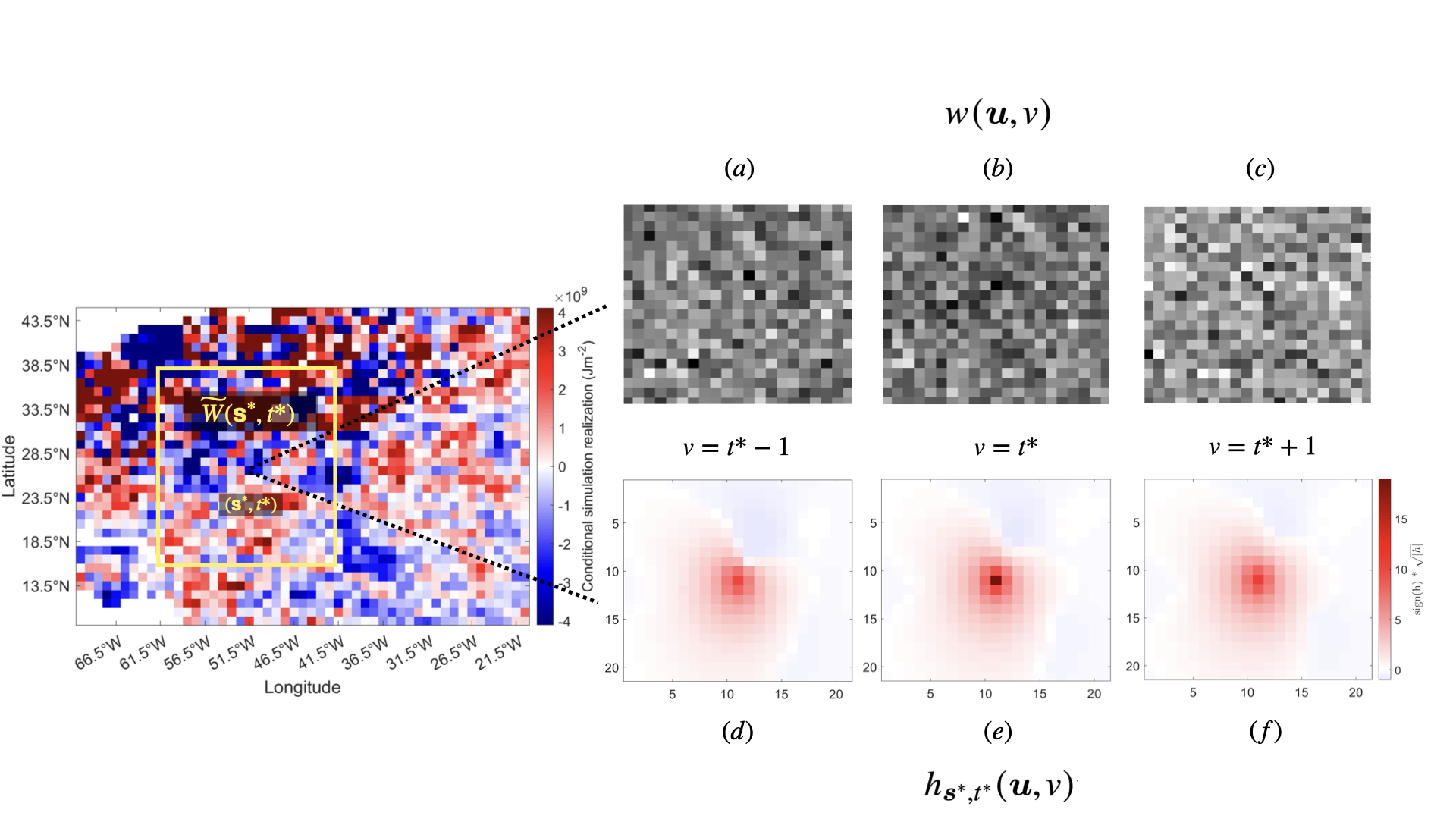}\\
  \caption{Illustration of the local conditional simulation algorithm for a region in the North Atlantic (left panel) for $t^* = $ February 2005. The simulated value at the center point $(\s^*,t^*)$ of the window $\widetilde{W}(\s^*,t^*)$ (yellow box, left panel) is obtained by convolving the white noise $w$ (panels (a)--(c)) with the spatio-temporal kernel $h$ (panels (d)--(f), showing $\text{sign}(h)\cdot\sqrt{|h|}$) within the window. To better show the simulated spatial dependence, this figure does not show the conditional mean or the simulated nugget effect; see Appendix~\ref{sec:condSimExamples} for example conditional simulations with these effects included.}\label{fig:condsim-ex}
\end{figure}

When discretized to a spatio-temporal grid and applied to the regional OHC fields $\widetilde{\OHC}_{z_u}^{z_d}(x,y,t)$, this leads to the following local conditional simulation algorithm:
\begin{enumerate}
    \item Simulate Gaussian white noise on the $1^\circ \times 1^\circ \times 1$ month grid over $D \times D_t$ and keep fixed.
    \item Compute the local predictive covariance matrix $\bm{\Sigma}_i$ using the local GP model for the window $\widetilde{\W}(x_i,y_i,t_i)$ centered on grid point $i$ with the model parameters estimated in Section~\ref{sec:method}\ref{sec:anom}.
    \item Compute the symmetric matrix square root $\bm{\Sigma}_i^{1/2} = \bm{U} \bm{D}^{1/2} \bm{U}^T$, where $\bm{\Sigma}_i = \bm{U} \bm{D} \bm{U}^T$ is the eigendecomposition of $\bm{\Sigma}_i$.
    \item Multiply $\bm{\Sigma}_i^{1/2}$ by the white noise from Step 1 within the window $\widetilde{\W}(x_i,y_i,t_i)$. Retain the value $f_i$ at the center point of the window. The conditional simulation ensemble member for grid point $i$ is then $\widetilde{\OHC}_{z_u,\text{sim}}^{z_d}(x_{i},y_{i},t_i) = \widehat{\widetilde{\OHC}}_{z_u}^{z_d}(x_i,y_i,t_i) + f_i$, where $\widehat{\widetilde{\OHC}}_{z_u}^{z_d}(x_i,y_i,t_i)$ is the mapped value from Section~\ref{sec:method}\ref{sec:anom}.
    \item Repeat steps 2--4 for all grid points in the domain  $D \times D_t$.
    \item Repeat all previous steps for the desired number of conditional simulation ensemble members.
\end{enumerate}
Figure~\ref{fig:condsim-ex} illustrates the approach for a region in the North Atlantic; see also Appendix~\ref{sec:condSimExamples} for example conditional simulations. These conditional simulation ensemble members reflect the predictive spatio-temporal dependence implied by the local GP model. In addition, the use of moving windows makes this approach computationally efficient. The most expensive operation in this algorithm is the eigendecomposition in Step 3, which needs to be computed only once for each window, irrespective of the number of conditional simulation realizations. The algorithm therefore scales well in the size of the conditional simulation ensemble. In Section \ref{sec:validation}\ref{sec:val_UQ}, we perform a cross-validation study to validate the simulated dependence against the observed dependence in the Argo-sampled ocean in 2005.

Now, to obtain an estimate of $\Var(\OHC_{z_u}^{z_d}(t)|\text{data})$, we simply substitute $\{\widetilde{\OHC}_{z_u,\text{sim}}^{z_d}(x_{i},y_{i},t)\}_i$ into Eq.~\eqref{eq:OHC_def_xy} for every ensemble member and compute the sample variance. We can similarly obtain uncertainties for other linear transformations of $\widetilde{\OHC}_{z_u}^{z_d}(x,y,t)$, such as derivatives, fitted trends and temporal averages. We simply transform each ensemble member one at a time and compute the sample variance of the transformed quantities. (Several concrete examples are provided in Section~\ref{sec:OHC_product}.) We can also use the ensemble to obtain predictions and potentially asymmetric uncertainties for nonlinear transformations of $\widetilde{\OHC}_{z_u}^{z_d}(x,y,t)$. This is described in more detail in Section~\ref{sec:OHC_product}\ref{sec:OHC_ENSO_ccf}.

We have so far discussed OHC uncertainty quantification for a single vertical section. Recall from Section~\ref{sec:background}\ref{sec:probDef} that we estimate $\OHC(t)$ over two vertical sections due to data availability. In this case, the predictive variance of the total OHC, $\OHC(t) = \OHC_{z_{u}}^{z}(t) + \OHC_{z}^{z_{d}}(t)$, would be expressed as $\Var(\OHC(t)|\text{data})=\Var(\OHC_{z_u}^{z}(t)|\text{data})+\Var(\OHC_{z}^{z_d}(t)|\text{data})+2\,\text{Cov}(\OHC_{z_u}^{z}(t),\OHC_{z}^{z_d}(t)|\text{data})$, where $z$ is a cutoff depth between $z_u$ and $z_d$. Our present model does not incorporate the dependence between vertical sections, so we use a conservative upper bound, only requiring the predictive variance estimates for each individual vertical section, instead:
\begin{equation}
\Var(\OHC(t)|\text{data})\leq \left(\sqrt{\Var(\OHC_{z_u}^{z}(t)|\text{data})}+\sqrt{\Var(\OHC_{z}^{z_d}(t)|\text{data})}\right)^2. \label{eq:UQ_UB}
\end{equation}
Modeling the vertical dependence to improve the mapped anomalies and uncertainty quantification is the subject of an upcoming extension (a brief discussion is provided in Section \ref{sec:discussion}).

\subsection{Modular software pipeline}

\begin{figure}[!h] 
  \centering
  \noindent\includegraphics[width=15cm,angle=0]{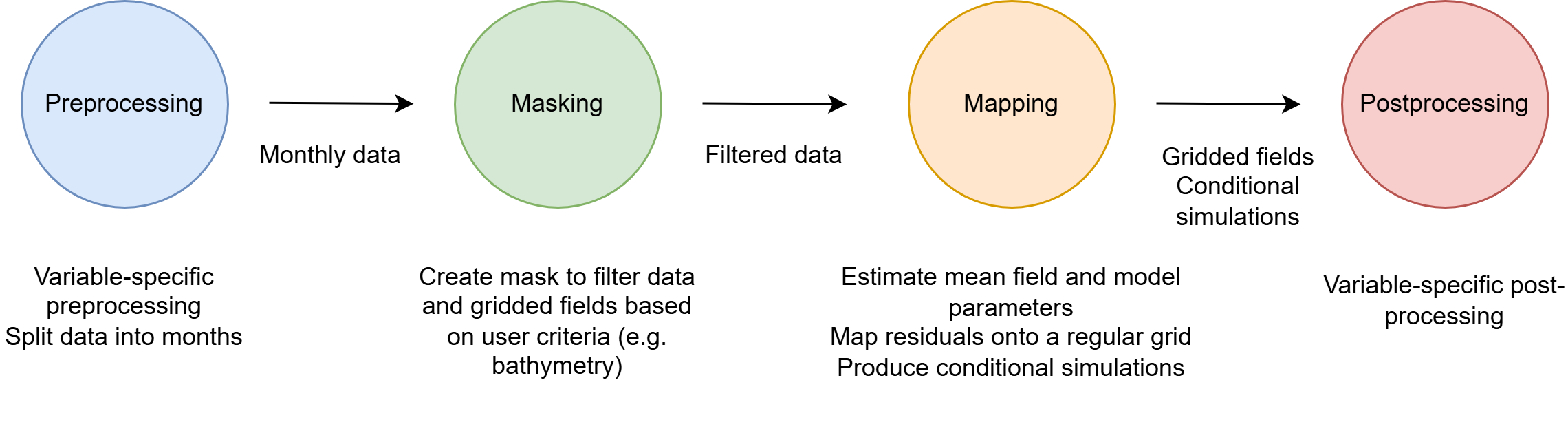}\\
  \caption{4-step software pipeline for mapping Argo data.}\label{fig:uml-diagram}
\end{figure}

We implement the mapping approach in a modular and extensible open-source software pipeline that enables the user to fully reproduce the above methodology from processing the Argo GDAC data to the final uncertainty quantification. We implement the pipeline in MATLAB and the codebase is freely available at \url{https://github.com/ttsukianto/LocalGP_OHC/}. The preprocessing and postprocessing modules are currently tailored for OHC mapping using Argo temperature profile data, but the scripts are easily extensible to processing other variables of interest (e.g., salinity, steric height, mixed layer depth, etc.; an effort ongoing as part of a different project). The masking and mapping modules are already applicable to other variables of interest. Figure \ref{fig:uml-diagram} shows the flow of data between each of the four main modules in the software pipeline. A full UML diagram for the mapping pipeline is provided in the above GitHub repository.

The preprocessing module (blue; \texttt{preProcessing.m}) contains scripts that prepare the input data for mapping. \texttt{processDACdata.m} implements the quality control method described in \cite{Kuusela2018} and Section~\ref{sec:background}\ref{sec:Argo_data} of this paper on profiles from the Argo GDAC. Next, we vertically integrate these processed profiles (\texttt{selectionAndVerticalIntegration.m}) as described in Section~\ref{sec:method}\ref{sec:verticalInt}. The output of this module is then the vertically integrated temperature profiles grouped into monthly files.
                                   
The masking module (green; \texttt{masking.m}) filters the preprocessed data in these monthly files according to the domain $D$. In this work, this involves creating the data availability mask $D_{\text{data}}$ (\texttt{createDataMask.m}) as described in Section~\ref{sec:background}\ref{sec:probDef} and intersecting it with the mask $D_{\text{RG}}$ from the Roemmich--Gilson 2019 data product (\cite{RG2019}) (\texttt{createCombinedMask.m}), but it is possible to extend the latter script to intersect with other custom masks. Since the mapping method relies on moving windows, we additionally create a basin mask (\texttt{createBasinMask.m}) to ensure that the subsequent mapping steps for grid points in the Pacific Ocean do not use observations from the Atlantic Ocean and vice versa near Central America (other regions could be included in a similar manner). Finally, \texttt{filterUsingMasks.m} filters out the data outside $D_{\text{data}} \cap D_{\text{RG}}$.

The mapping module (orange; \texttt{mapping.m}) produces estimates and conditional simulation ensembles for gridded fields according to the methodology described in Sections \ref{sec:method}\ref{sec:mean-field}--\ref{sec:uq-condsim}. After estimating the mean field parameters (\texttt{estimateMeanField.m}) using the filtered data and subtracting the mean to obtain the set of observed residuals (\texttt{subtractMean.m}), we estimate the monthly decorrelation scales and variances (\texttt{localMLE.m}). We then use the annual median of the monthly maximum likelihood estimates as the final parameter estimates to map the residuals (\texttt{mapResiduals.m}) and produce the conditional simulation ensembles (\texttt{localCondSim.m}). To convert from the residuals to the full fields (i.e., $\widetilde{\OHC}_{z_u}^{z_d}(x,y,t) = \mu (x,y,t) + r(x,y,t)$), we use \texttt{computeFullFields.m}.

Finally, the postprocessing module (red; \texttt{postProcessing.m}) performs post-mapping transformations for the variable of interest. In this work, this involves substituting the mapped regional OHC values and conditional simulation ensemble members for each grid point into Eq.~\eqref{eq:OHC_def_xy} (i.e., integrating horizontally) to obtain the monthly estimates of OHC. Here, we implement postprocessing for OHC, but it is our expectation that the framework in this codebase will facilitate mapping and uncertainty quantification for other gridded fields as well. The codebase is designed so that this can simply be done by replacing the pre- and postprocessing modules with versions specific to the particular variable of interest.

\section{Sensitivity to modeling choices and method validation} \label{sec:validation}

In Sections 4 and 5, we illustrate the OHC mapping pipeline using Argo temperature profiles from 2004--2022. As mentioned in Section 2b, we consider two vertical sections: 15--975 dbar (``Upper ocean'') and 975--1850 dbar (``Midocean''). Unless otherwise stated, in obtaining the following results, we set the mean field window size parameter to $\lambda=5$ degrees, the Gaussian process spatial window size parameter to $\lambda_G=10$ degrees, time window size parameter to $\lambda_{t}=1.5$ months, and the number of conditional simulation ensemble members to 500.

Section 4 studies the sensitivity of the OHC estimates to the inclusion of a time trend in the mean field (Section~4a) and temporal dependence in the model definition (Section~4b). In addition, Section~\ref{sec:validation}\ref{sec:val_UQ} validates the dependence structure in the conditional simulation ensembles. As in \citet{Kuusela2018}, we compare the prediction errors for each modeling choice in two cross-validation scenarios: leave-one-observation-out (LOOO) and leave-one-float-out (LOFO) cross-validation. The latter is further extended to paired cross-validation in Section~\ref{sec:validation}\ref{sec:val_UQ}. We might imagine predicting OHC at a densely observed location to be the least difficult and predicting in an ocean region with few Argo floats to be the most difficult (if the decorrelation scales and the overall complexity of the ocean dynamics are similar in the two regions). These two cross-validation modes therefore represent approximate lower and upper bounds, respectively, on the difficulty of predicting the OHC at an unobserved location. 

\subsection{Effect of mean field time trend} \label{sec:val_mean_field}

\begin{figure}[!h]
    \centering
    \subfigure[Upper ocean]{
        \includegraphics[width=8cm,trim = 0.5cm 0cm 1cm 0cm,clip = true]{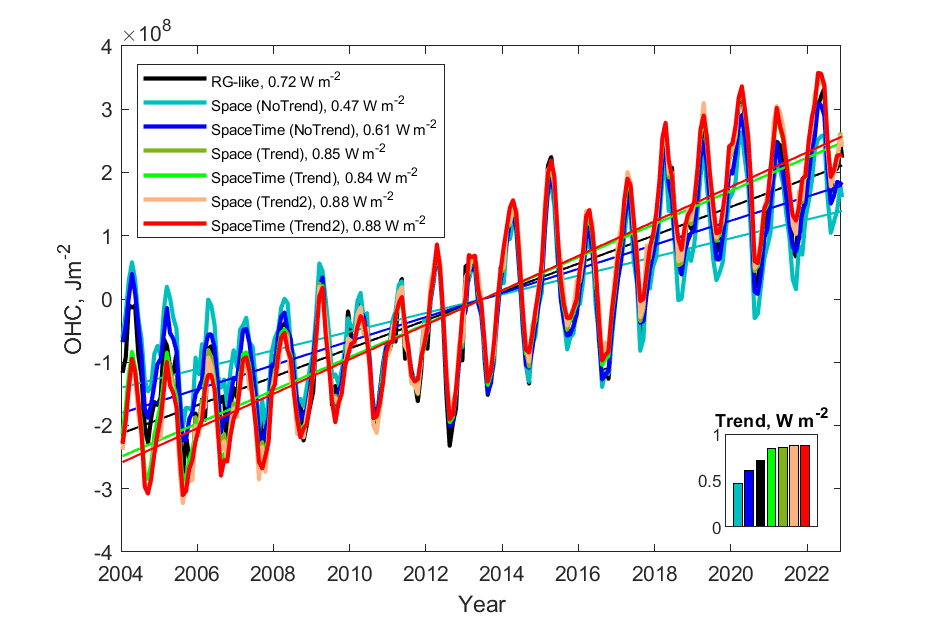}}
    \subfigure[Midocean]{
        \includegraphics[width=8cm,trim = 0.5cm 0cm 1cm 0cm,clip = true]{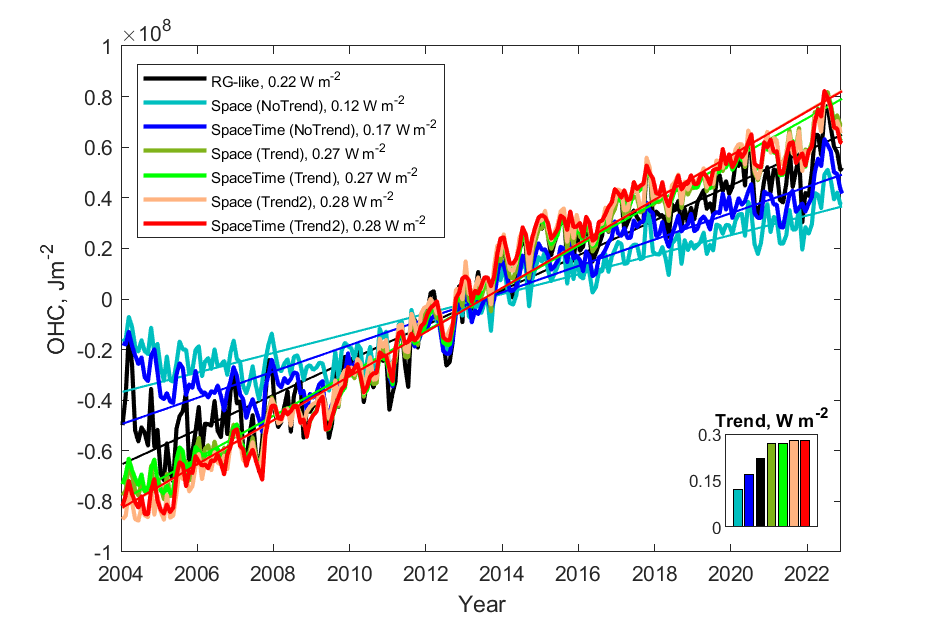}}
    \caption{Ocean heat content time series and trends for different modeling choices.}
    \label{fig:OHC_val_trend}
\end{figure}

We first explore the effect of including a time trend in the climatological mean field definition \eqref{eq:mean_field} on the global OHC trend estimate. To this end, we produce variants of our OHC maps with $\nu_1=\nu_2=0$ (``NoTrend''), $\nu_1 \neq 0, \nu_2=0$ (``Trend'') and $\nu_1 \neq 0, \nu_2 \neq 0$ (``Trend2''). In addition, we produce variants with a space-only covariance ($\theta_{W,t} = \infty$ in Eq.~\eqref{eq:dist_def}) as well as a reproduction of the \cite{Roemmich2009} product (``RG-like'') with $\nu_1=\nu_2=0$ and the covariance as described for the RG-like model in \citet{Kuusela2018}. In these latter two cases, we set $\lambda_{t}=0.5$ months for consistency with the original Roemmich--Gilson maps. (\cite{Kuusela2018} investigated space-only maps for both $\lambda_{t}=0.5$ months and $\lambda_{t}=1.5$ months and found that the two cases have comparable predictive performance in the absence of the temporal decorrelation scale.)

The results (normalized to the area of the Argo-sampled global ocean) are shown in Figure~\ref{fig:OHC_val_trend}, where we obtain each trend line by fitting a linear regression model with monthly factors for the seasonal cycle to each OHC time series. Consistent with \cite{Cheng2015}, the RG-like (black) and models without the mean field time trend (cyan and blue) produce smaller OHC trends. The increase in the estimated upper ocean OHC trend is as large as 87\% (cyan vs. pink or red) between the no trend and quadratic trend models. This difference is also apparent in the midocean, where the percent increase is as large as 133\% (cyan vs. pink or red). The results also suggest that including a quadratic time trend in the mean field has minimal effect on the OHC trend estimate for this time period when compared to a linear trend. However, it may still be prudent to eventually include a quadratic time trend in the mean field as the time span of the Argo observations grows in order to capture a potential accelerating or decelerating warming trend.

\begin{figure}[!h]
    \centering
    \subfigure[Space-only covariance, LOOO]{
        \includegraphics[width=6cm,trim = 0.5cm 0cm 1cm 0cm,clip = true]{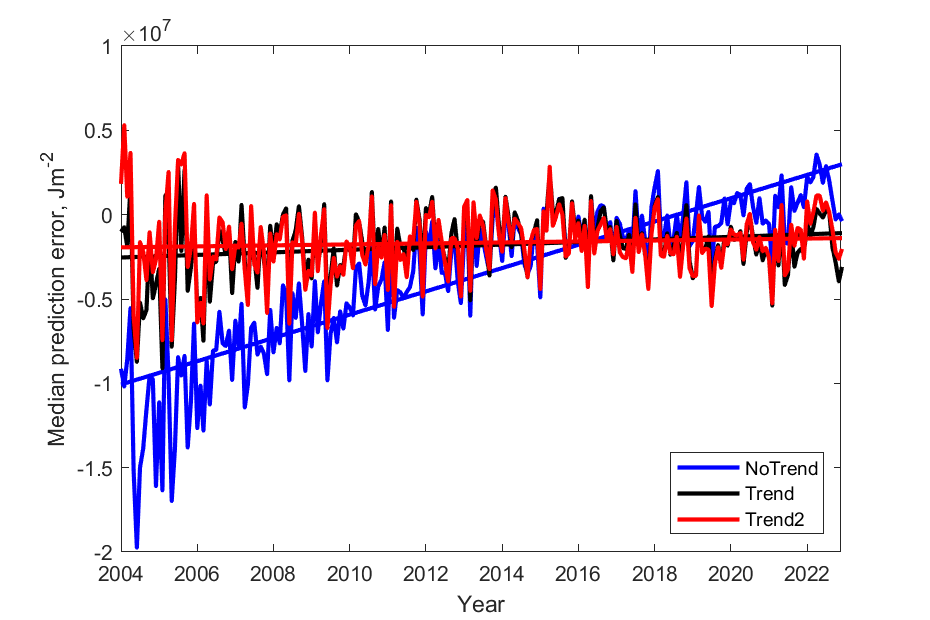}}
    \subfigure[Space-time covariance, LOOO]{
        \includegraphics[width=6cm,trim = 0.5cm 0cm 1cm 0cm,clip = true]{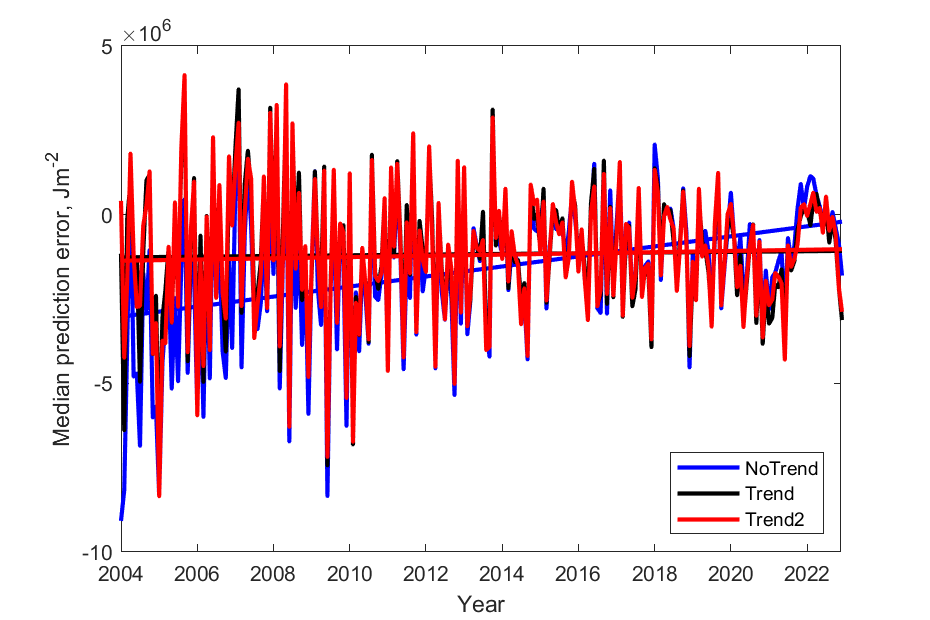}}\\
    \subfigure[Space-only covariance, LOFO]{
        \includegraphics[width=6cm,trim = 0.5cm 0cm 1cm 0cm,clip = true]{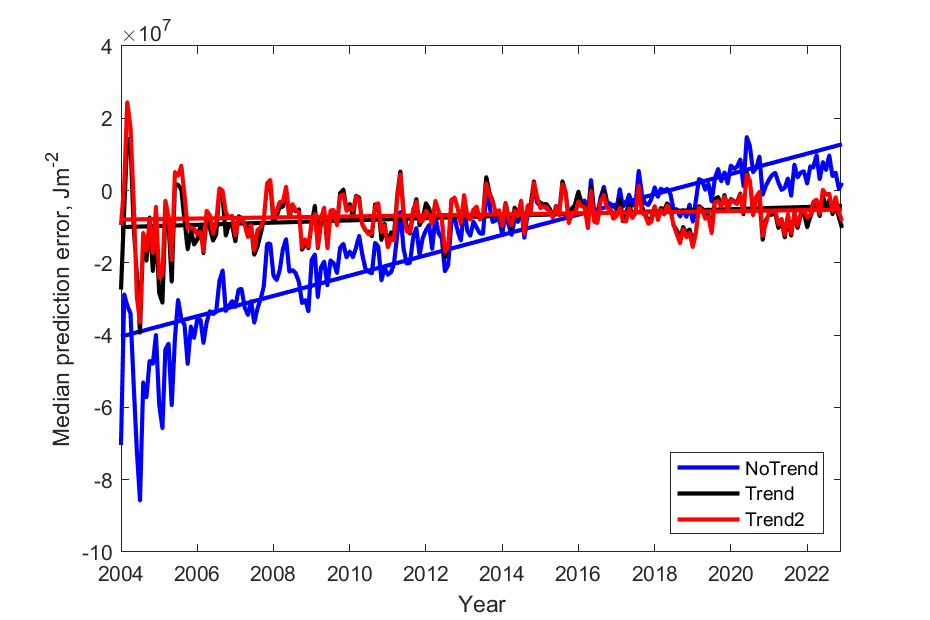}}
    \subfigure[Space-time covariance, LOFO]{
        \includegraphics[width=6cm,trim = 0.5cm 0cm 1cm 0cm,clip = true]{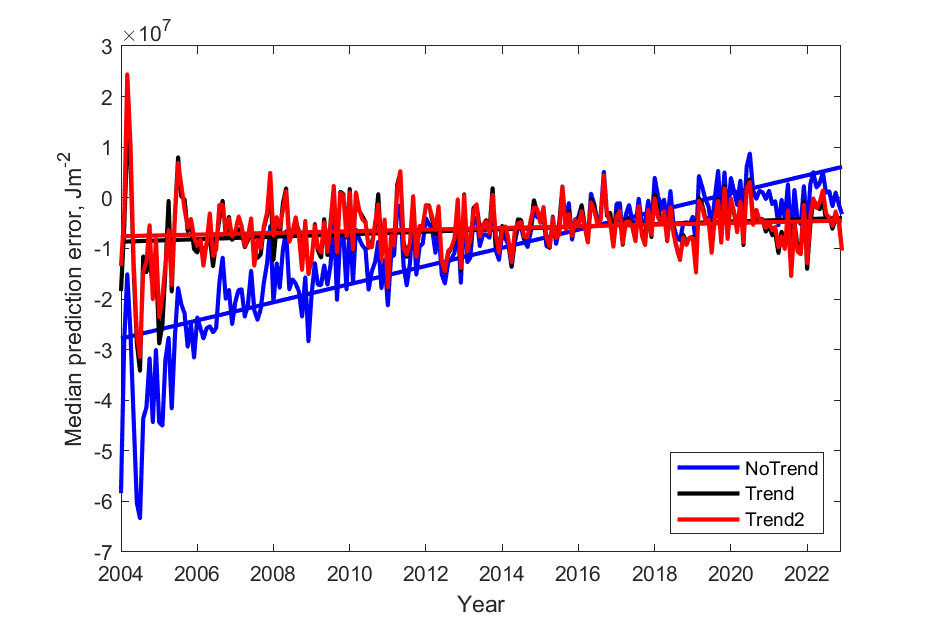}}\\
    \caption{Cross-validated monthly median prediction errors (midocean) for different modeling choices.}
    \label{fig:OHC_val_trend_CV_midocean}
\end{figure}

The large difference in the estimated OHC trends depending on the mean field modeling choices raises the question of which of these estimates is best supported by the data. We can answer this question by performing a cross-validation study comparing the different models. Figure~\ref{fig:OHC_val_trend_CV_midocean} shows time series of monthly median cross-validation prediction errors. For brevity, we only show the midocean results; the upper ocean results are shown in Figure~\ref{fig:OHC_val_trend_CV_upper_ocean} in the appendix and yield similar conclusions as the midocean results. The upper two panels show the LOOO median prediction errors, while the lower two panels show the LOFO median prediction errors. If the model was specified correctly, we would expect the errors to be centered on zero with no obvious patterns. However, we see that when we do not include a trend in the mean field (blue), there is a positive trend in the prediction errors regardless of whether we include time in the covariance function (left vs.~right panels) or perform LOOO or LOFO cross-validation (upper vs.~lower panels). From the sign of the prediction errors, we can conclude that the no-trend model is consistently predicting too large OHC in the early years and too small OHC in the later years, which indicates that it is underestimating the overall trend. Statistically, the no-trend model is misspecified so that the i.i.d. assumption in Eq.~\eqref{eq:anomaly_def} does not hold and this misspecification causes the early predictions to be warm-biased and the later predictions to be cold-biased. In contrast, the models with linear or quadratic trend (black and red) do not show a trend in the prediction errors, which suggests that any trend in the observations is more adequately captured. We conclude that the models with trend included in the mean field yield more reliable estimates of the final OHC trend.

\subsection{Effect of anomaly mapping method}
\label{sec:val_covariance}

\begin{figure}[!h]
    \centering
    \subfigure[Upper ocean]{
    \includegraphics[width=8cm, trim = 0.5cm 0cm 1cm 0cm, clip = true]{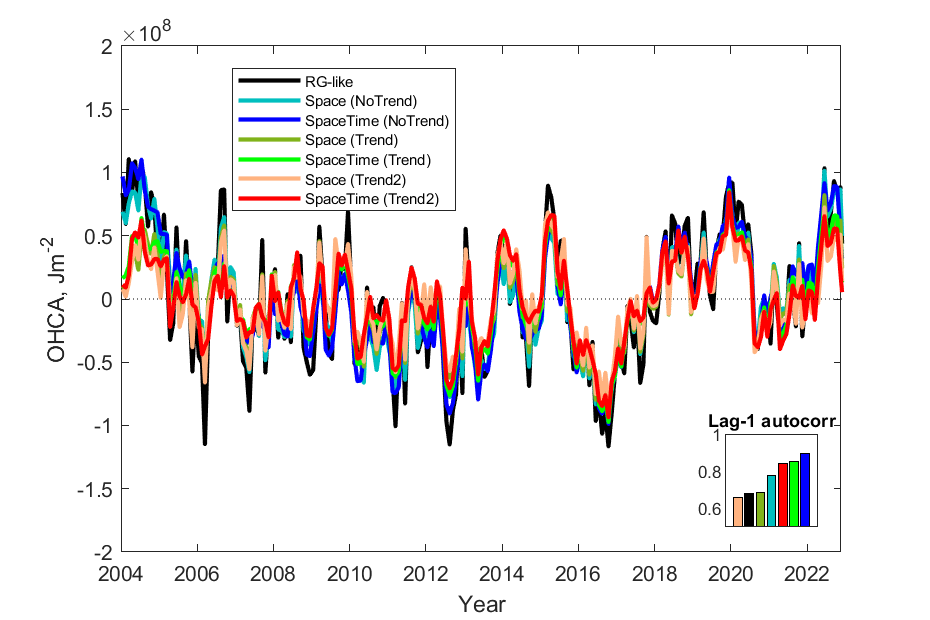}}
    \subfigure[Midocean]{
    \includegraphics[width=8cm, trim = 0.5cm 0cm 1cm 0cm, clip = true]{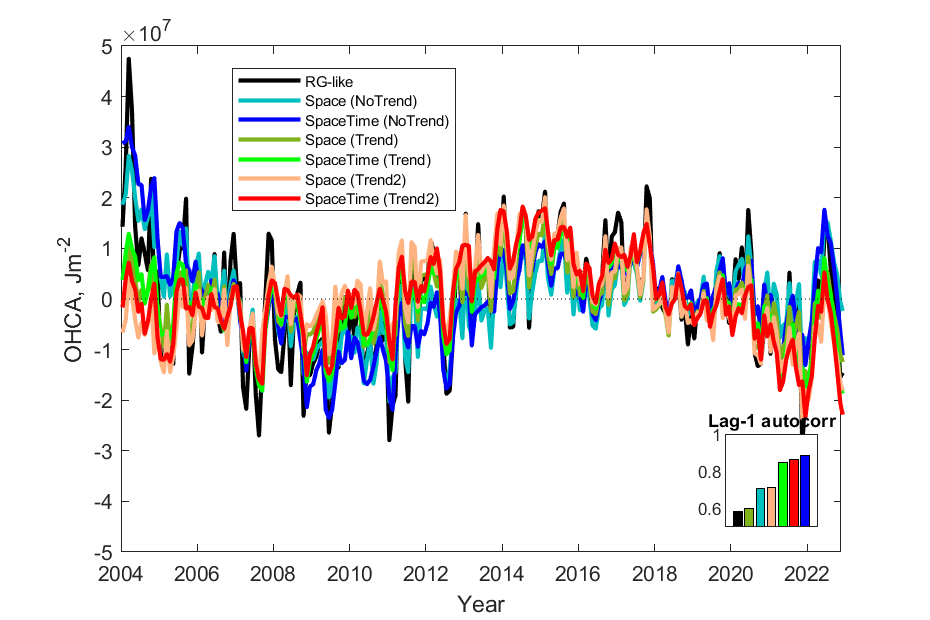}}
    \caption{OHC anomalies for different modeling choices. The insets show the estimated lag-1 autocorrelations.}
    \label{fig:OHC_val_cov_upper_ocean}
\end{figure}

In this section, we explore the effect of including time in the covariance function on the mapped OHC anomalies using the same model variants as in Section~\ref{sec:validation}\ref{sec:val_mean_field}. In that section, we saw that the effect of including time is negligible on the estimated long-term global OHC trend, but we will now see that it makes a big difference on the mapped monthly anomalies. Figure~\ref{fig:OHC_val_cov_upper_ocean} shows the mapped global OHC anomalies (with trend and seasonal cycle removed; normalized to the area of the Argo-sampled global ocean) for the different mean field trend (no trend, linear, or quadratic) and covariance function (space-time or space only) modeling choices. We see that all the models produce anomaly estimates with similar overall patterns and magnitudes, but the space-time models (blue, green, red) seem to produce smoother estimates across time with fewer spurious fluctuations compared to their space-only counterparts (cyan, yellow-green, pink). The lag-1 autocorrelations (bottom right insets) are also noticeably higher in the space-time models, indicating that the dependence between months may be better resolved. The RG-like model, which is a space-only model, also has lag-1 autocorrelation comparable to our locally stationary space-only models.

\begin{figure}[t]
    \centering
    \subfigure[Upper ocean, LOOO]{
        \includegraphics[width=8cm,trim = 0.5cm 0cm 1cm 0cm,clip = true]{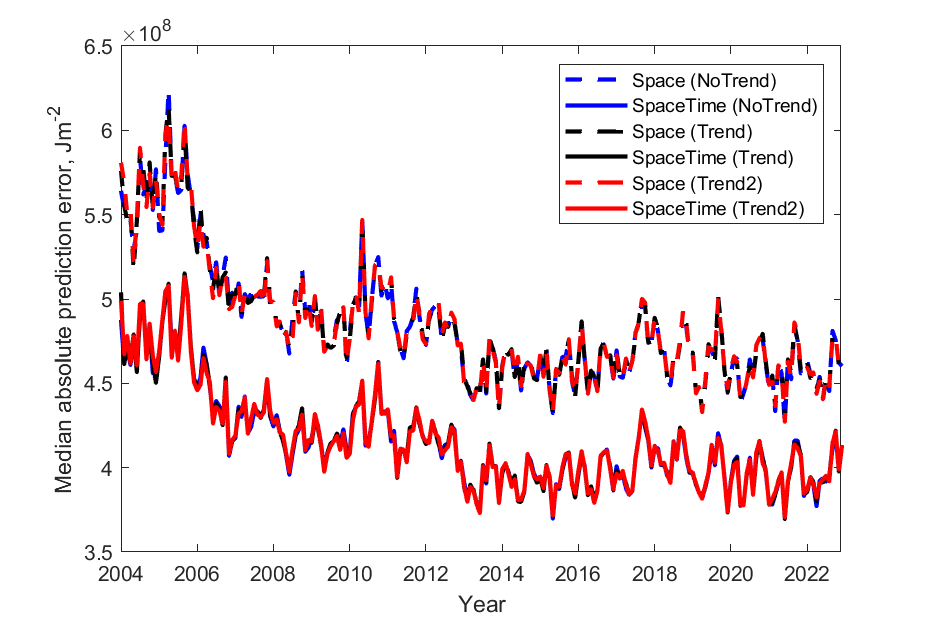}}
    \subfigure[Upper ocean, LOFO]{
        \includegraphics[width=8cm,trim = 0.5cm 0cm 1cm 0cm,clip = true]{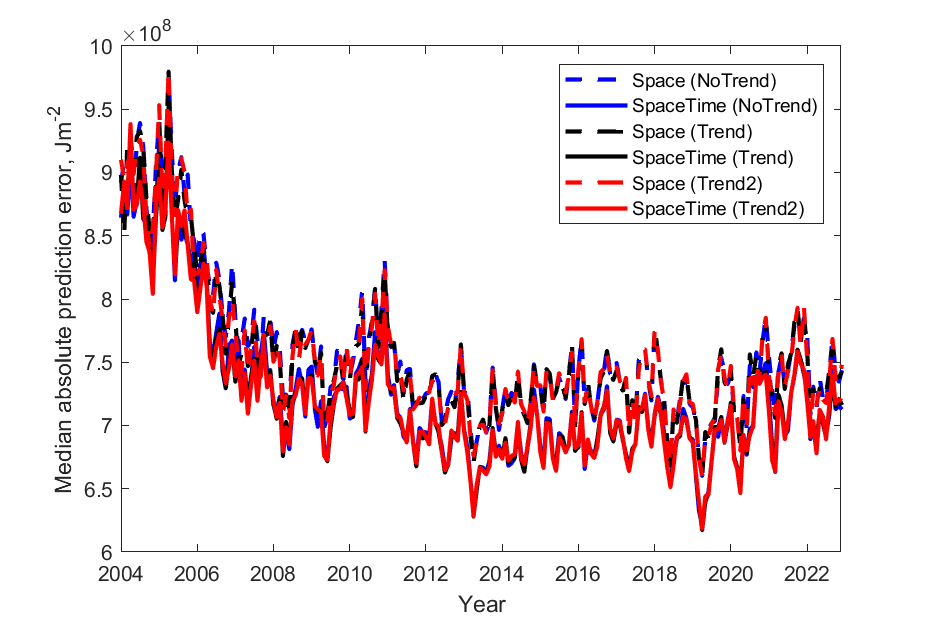}}
    \caption{Cross-validated monthly median absolute prediction errors (upper ocean) for different modeling choices.}
    \label{fig:OHC_val_cov_CV_upper_ocean}
\end{figure}

As before, we can use cross-validation to study whether the space-time models actually resolve the OHC anomaly signal better than the space-only models. The upper ocean cross-validation results are shown in Figure~\ref{fig:OHC_val_cov_CV_upper_ocean}, while the midocean results are in Figure~\ref{fig:OHC_val_cov_CV_midocean} in the appendix. When we examine the LOOO cross-validation results, we see that the space-time models (solid lines) consistently produce smaller median absolute prediction errors than their space-only counterparts (dashed lines) in both the upper ocean and midocean. We observe a similar but less pronounced pattern in the LOFO cross-validation study. These results confirm that the space-time models are indeed able to resolve the OHC anomaly signal better. Given the interpretation for the two cross-validation schemes discussed at the beginning of this section, the difference between the LOOO and LOFO results indicates that the improvement from the space-time model is especially pronounced at smaller spatio-temporal scales.

Overall, the above OHC trend and anomaly estimates together with the cross-validation results suggest that the best models out of those considered are the models with a trend in the mean field and time in the covariance function. Since we do not observe large differences between the linear and quadratic trend models, we will focus on the linear trend model for simplicity and will use the space-time variant of that model to produce the remaining results discussed in this work. For completeness, the covariance parameter estimates for this model are shown in Appendix~\ref{sec:covParamPlots}.

\subsection{Validation of local conditional simulations}
\label{sec:val_UQ}

Finally, we perform a cross-validation study to compare the modeled and observed conditional dependencies at different spatial and temporal lags. Since we aim to validate the modeled spatio-temporal dependence, we will perform the cross-validation study by leaving out \emph{pairs} of floats. Our goal is then to show that across a population of left-out pairs, stratified by spatial and temporal separation, the conditionally simulated dependence matches the observed dependence.

Let us first define a temporal lag $\Delta t$ in months and a spatial distance bin $\mathcal{B} = [\Delta d -0.5,\Delta d+0.5)$, where $\Delta d$ is a spatial lag in degrees. For a given year $i$ and for each $\Delta t$ and $\mathcal{B}$, we consider up to 120,000 (constraint imposed due to computational limitations) vertically integrated temperature profile pairs $(\widetilde{\OHC}(x_1,y_1,t_1),\widetilde{\OHC}(x_2,y_2,t_2))$ with temporal lag $|t^*_2-t^*_1|=\Delta t$, where $t^*_i$ is the closest temporal grid point to $t_i$, and spatial lag $\sqrt{(x_2-x_1)^2+(y_2-y_1)^2}$ in $\mathcal{B}$. For each pair of observations, let $(f_1,f_2)$ be the corresponding float IDs and $(x^*_1,y^*_1)$ and $(x^*_2,y^*_2)$ the closest spatial grid points. We then exclude all profiles with float ID $f_1$ or $f_2$, produce a spatio-temporal conditional simulation realization and extract from it the simulated pair $(\widetilde{\OHC}_{\text{sim}{,-(f_1,f_2)}}(x^*_1,y^*_1,t^*_1),\widetilde{\OHC}_{\text{sim}{,-(f_1,f_2)}}(x^*_2,y^*_2,t^*_2))$. In addition, we store the corresponding Gaussian process predictions $\widehat{\widetilde{\OHC}}_{-(f_1,f_2)}(x^*_1,y^*_1,t^*_1)$ and $\widehat{\widetilde{\OHC}}_{-(f_1,f_2)}(x^*_2,y^*_2,t^*_2)$ and predictive variances $\sigma^2_{-(f_1,f_2)}(x^*_1,y^*_1,t^*_1)$ and $\sigma^2_{-(f_1,f_2)}(x^*_2,y^*_2,t^*_2)$. We use these to standardize the marginal distributions, which vary from pair to pair. Then, if the spatio-temporal model is correctly specified, the conditionally simulated standardized pair
\begin{small}
\begin{equation}
    \left(\frac{\widetilde{\OHC}_{\text{sim}{,-(f_1,f_2)}}(x^*_1,y^*_1,t^*_1)-\widehat{\widetilde{\OHC}}_{-(f_1,f_2)}(x^*_1,y^*_1,t^*_1)}{\sigma_{-(f_1,f_2)}(x^*_1,y^*_1,t^*_1)},\frac{\widetilde{\OHC}_{\text{sim}{,-(f_1,f_2)}}(x^*_2,y^*_2,t^*_2)-\widehat{\widetilde{\OHC}}_{-(f_1,f_2)}(x^*_2,y^*_2,t^*_2)}{\sigma_{-(f_1,f_2)}(x^*_2,y^*_2,t^*_2)}\right)
\end{equation}
\end{small}
and the observed standardized pair
\begin{small}
\begin{equation}
    \left(\frac{\widetilde{\OHC}(x^*_1,y^*_1,t^*_1)-\widehat{\widetilde{\OHC}}_{-(f_1,f_2)}(x^*_1,y^*_1,t^*_1)}{\sigma_{-(f_1,f_2)}(x^*_1,y^*_1,t^*_1)},\frac{\widetilde{\OHC}(x^*_2,y^*_2,t^*_2)-\widehat{\widetilde{\OHC}}_{-(f_1,f_2)}(x^*_2,y^*_2,t^*_2)}{\sigma_{-(f_1,f_2)}(x^*_2,y^*_2,t^*_2)}\right)
\end{equation}
\end{small}should both follow the bivariate Gaussian distribution with standard Gaussian margins and common correlation $\rho_{-(f_1,f_2)}(x^*_1,y^*_1,t^*_1,x^*_2,y^*_2,t^*_2)$. For all pairs
in $\mathcal{B}$ with temporal lag $\Delta t$, we then compute the empirical Pearson correlation coefficient across the standardized conditionally simulated pairs and the standardized observed pairs to validate that the two correlations match in aggregate.

\begin{figure}[!h]
    \centering
    \subfigure[Upper ocean]{
        \includegraphics[width=8cm,trim = 0.5cm 0cm 1cm 0.5cm,clip = true]{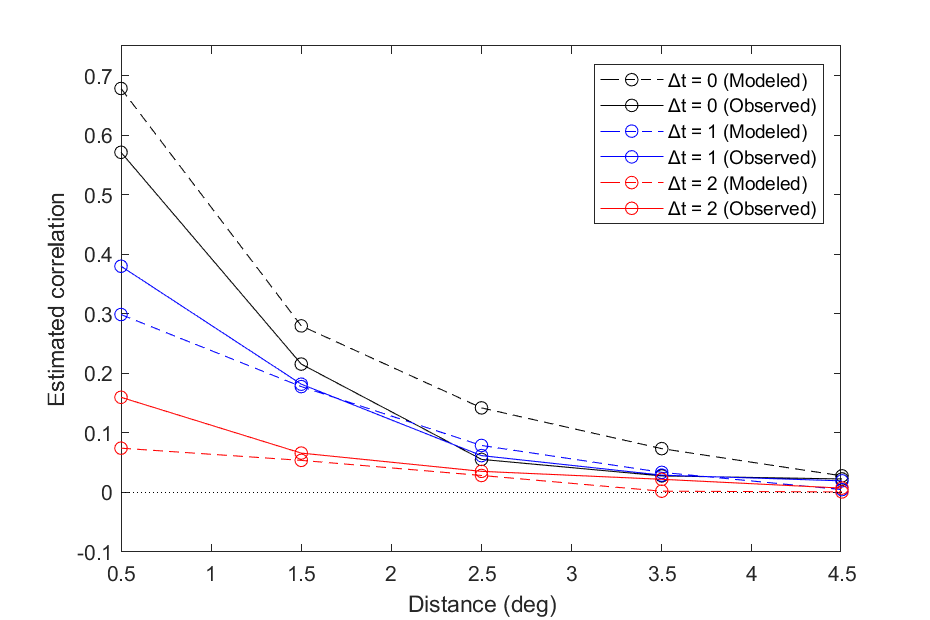}}
    \subfigure[Midocean]{
        \includegraphics[width=8cm,trim = 0.5cm 0cm 1cm 0.5cm,clip = true]{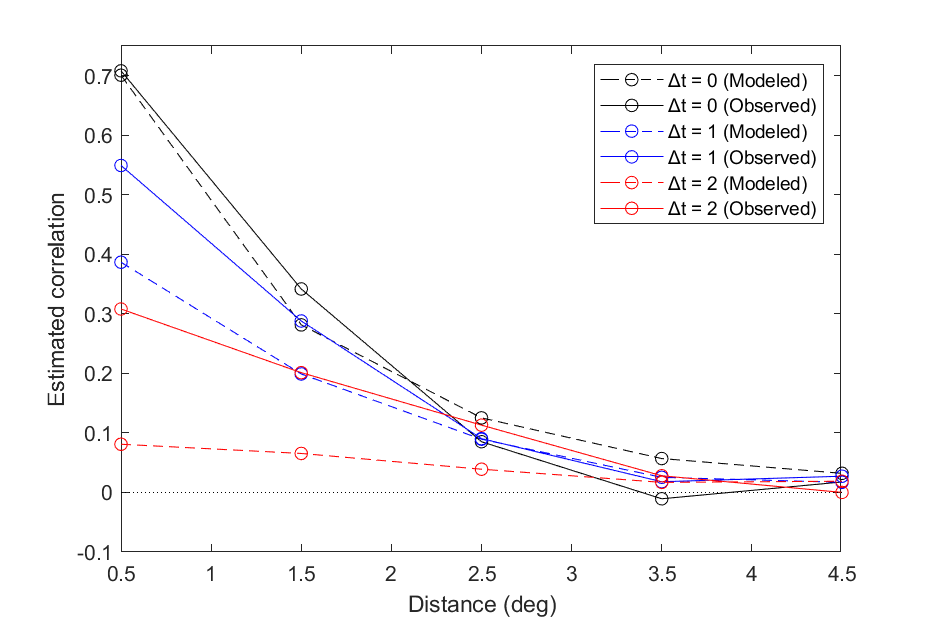}}
    \caption{Cross-validation for spatio-temporal conditional dependence. The solid lines show the observed correlation while the dashed lines show the conditionally simulated modeled correlation.}
    \label{fig:OHC_val_CV_cond_sim}
\end{figure}

Figure~\ref{fig:OHC_val_CV_cond_sim} shows the cross-validation results for year 2005, where $\Delta t=0,1,2$ months and $\Delta d$ ranges from 0.5 to 4.5 degrees in 1-degree increments. We observe that the modeled correlation is in broad agreement with the observed correlation. Pairs that are close in space and time are more strongly correlated than more distant pairs with remarkably similar magnitude and shape for the modeled and observed correlation curves, especially for the upper ocean section. In a more careful inspection, we observe that the model seems to slightly underestimate the conditional dependence, especially for the midocean and longer time lags (blue and red curves in the right panel). This shows that even though the model seems to broadly capture the conditional dependence structure well, there is room for future improvement in the spatio-temporal structure of the model, perhaps in terms of how space and time interact in the covariance function.

For completeness, we also provide in Figure~\ref{fig:OHC_val_cond_sim} in the appendix the non-cross-validated empirical correlations for the conditional simulation ensembles computed using a similar binned spatio-temporal aggregation approach, but with the bins defined by the native spatio-temporal grid of the simulations.

\section{Example applications for ocean heat content estimates with uncertainties}
\label{sec:OHC_product}
In this section, we use the mapping pipeline described in Section~\ref{sec:method} to demonstrate novel or improved mapping uncertainty quantification capabilities enabled by conditional simulation. That is, we use the conditional simulation ensembles of the OHC fields as a starting point to obtain uncertainties for other derived quantities, e.g., OHC trend, ocean heat uptake and correlation of the OHC anomalies with ENSO. The common feature of these examples is that rigorous mapping uncertainty quantification for the final quantity of interest requires accounting for the spatial and temporal dependence in the OHC mapping uncertainties, a unique capability enabled by our conditional simulation approach.

\subsection{Global ocean heat content trend}
\begin{figure}[!h]
    \centering
    \noindent\includegraphics[width=10cm,angle=0]{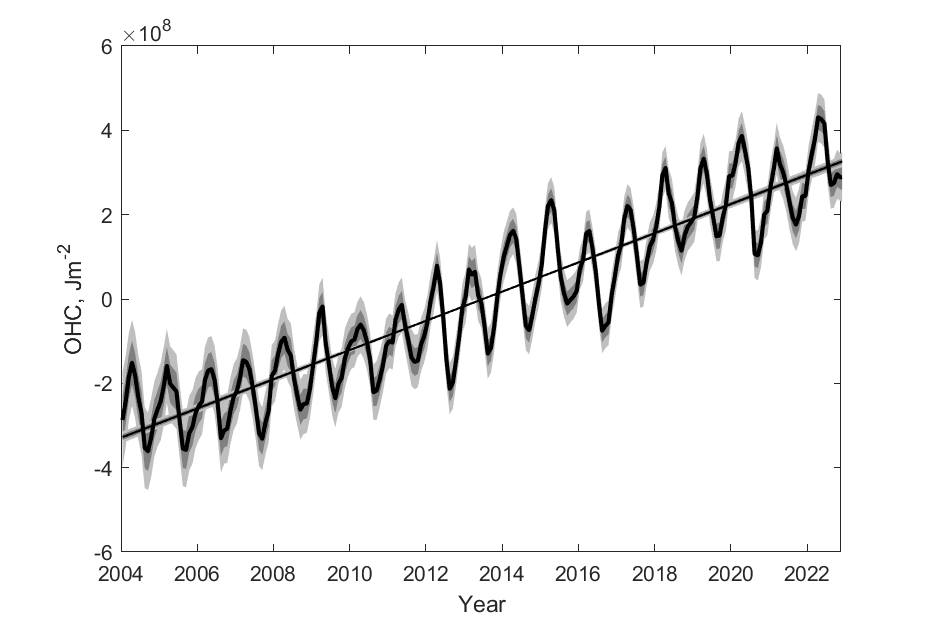}\\
    \caption{Total global OHC (15--1850 dbar) time series with fitted trend. The shaded regions show 68\% (dark gray) and 95\% (light gray) uncertainties obtained using the conditional simulation ensembles. The fitted trend and its 68\% uncertainty are $1.114 \pm 0.024$ W/m$^2$. }\label{fig:OHC_trend_sum}
\end{figure}

Given the importance of estimating long-term changes in global OHC for the quantification of the EEI absolute magnitude, the estimate of the trend in global OHC has been the subject of a number of intercomparison studies (e.g., \cite{Hakuba2024}, \cite{von_Schuckmann2023}). Here, we estimate that, during 2004--2022, OHC (i.e., $\OHC_{15}^{975}(t) + \OHC_{975}^{1850}(t)$) increased at a rate of $1.114$ W/m$^2$ with 95\% uncertainty interval $[1.066,1.162]$ W/m$^2$ (Figure~\ref{fig:OHC_trend_sum}; expressed as a flux relative to the area of the Argo-sampled global ocean; with Figures~\ref{fig:OHC_trend_upper_ocean} and \ref{fig:OHC_trend_midocean} in the appendix showing the corresponding time series and trends for the individual vertical sections). To obtain the uncertainty on the trend, we first fit a linear regression with monthly factors for the seasonal cycle to each conditional simulation ensemble member in each vertical section. Similar to the conservative estimate in Eq.~\eqref{eq:UQ_UB}, the conservatively estimated standard error of the total trend is the sum, across the vertical sections, of the sample standard errors of the trend ensemble members. We note that this uncertainty estimate for the global OHC trend would not have been possible to obtain without accounting for the time dependence in the conditional simulation ensembles.

\subsection{Global ocean heat content anomalies}
\begin{figure}[!h]
    \centering
    \subfigure[Monthly]{
        \includegraphics[width=6cm,trim = 0.5cm 0cm 1cm 0.2cm,clip = true]{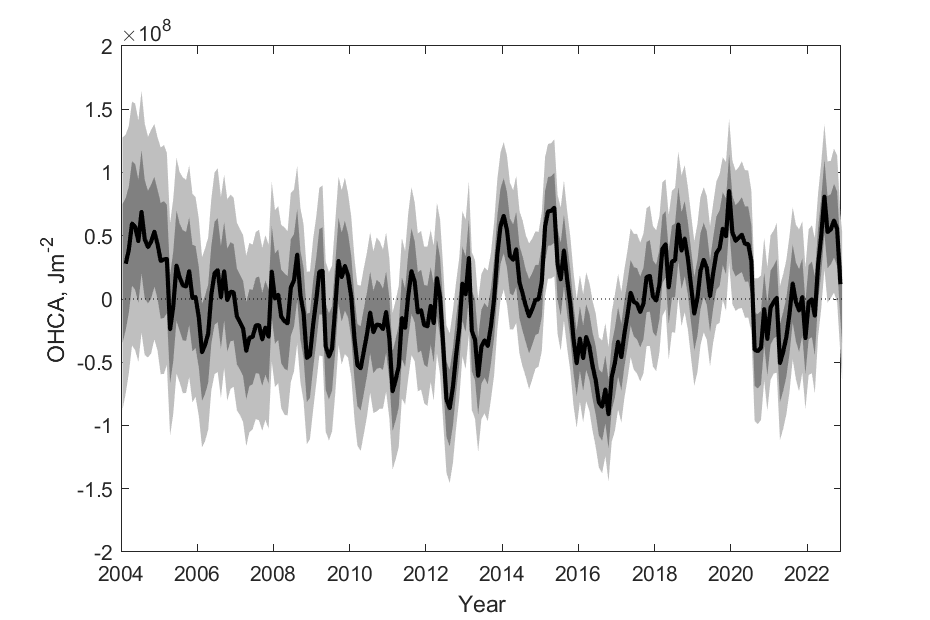}}
    \subfigure[12-month moving average]{
        \includegraphics[width=6cm,trim = 0.5cm 0cm 1cm 0.2cm,clip = true]{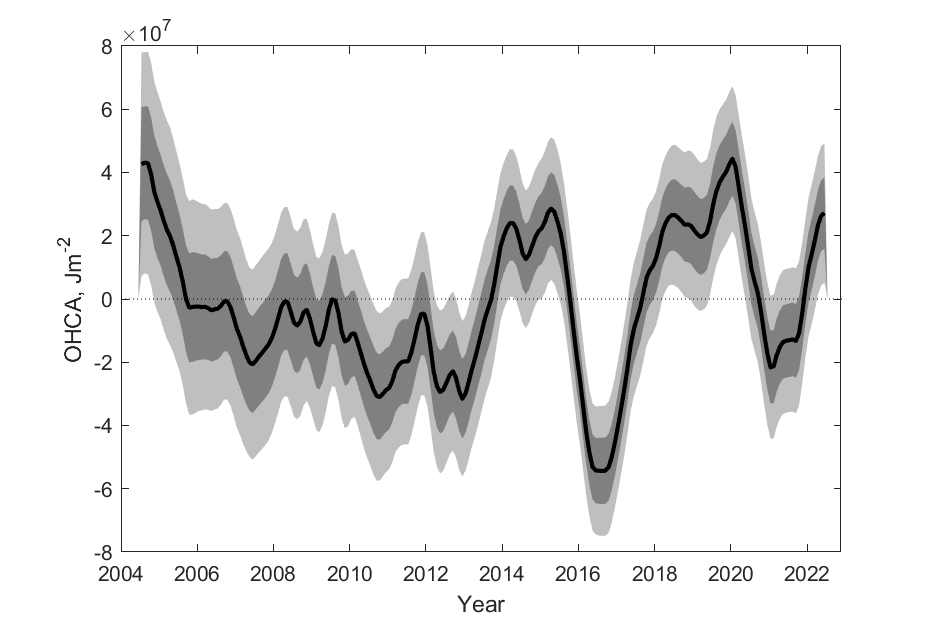}}\\
    \subfigure[24-month moving average]{
        \includegraphics[width=6cm,trim = 0.5cm 0cm 1cm 0.2cm,clip = true]{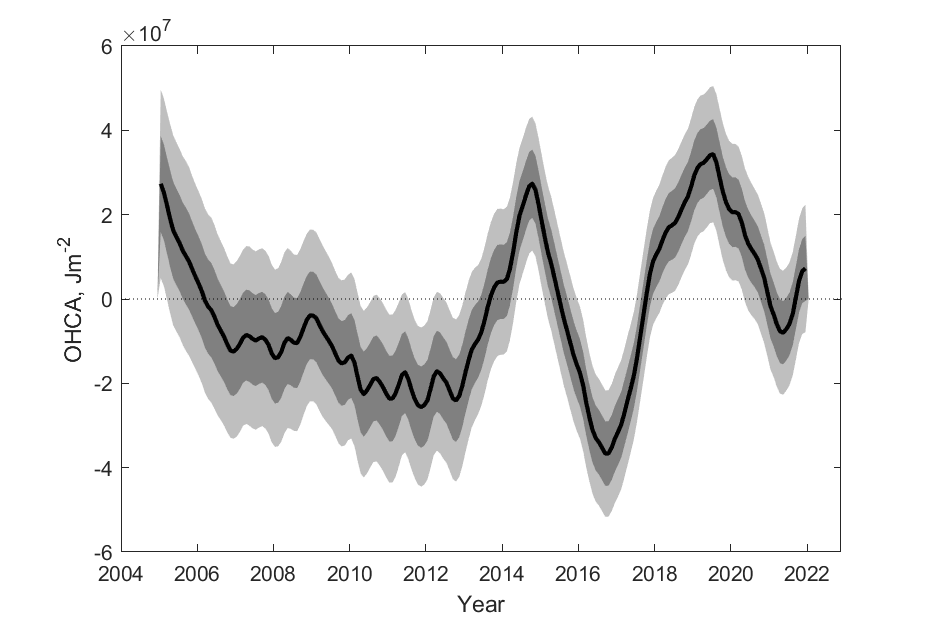}}
    \subfigure[36-month moving average]{
        \includegraphics[width=6cm,trim = 0.5cm 0cm 1cm 0.2cm,clip = true]{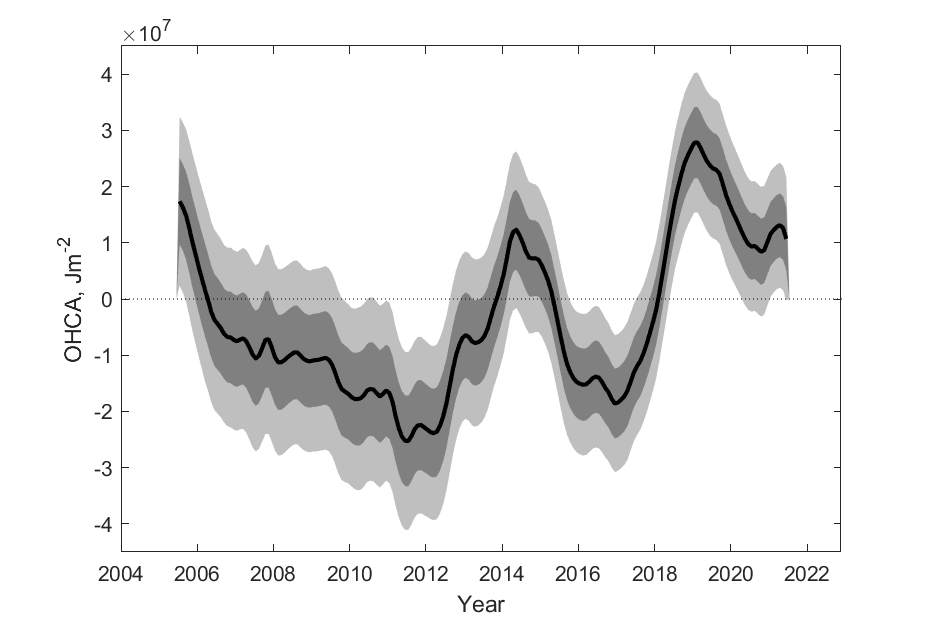}}\\
    \caption{Total global OHC (15--1850 dbar) anomaly time series with 68\% (dark gray) and 95\% (light gray) uncertainties obtained using the conditional simulation ensembles.}
    \label{fig:OHCA_sum}
\end{figure}

The time series of global OHC (deseasonalized, detrended) anomalies with uncertainties shows how uncertainties are overall larger in the earlier years (Figure~\ref{fig:OHCA_sum}; normalized to the area of the Argo-sampled global ocean), which is expected given the smaller number of floats (the uncertainty reduces in time as more profiles are available and more regions are observed). While most of the monthly anomalies are not significantly different from zero (i.e., the gray shading in Figure~\ref{fig:OHCA_sum}(a) often intersects zero), a clearer picture of the overall interannual variability emerges when we examine moving averages (i.e., when we create low-pass filtered time series; Figures~\ref{fig:OHCA_sum}(b)--(d)). In this case, we can confirm that, e.g., the negative anomaly that follows the 2016 El Ni\~no event (seen previously in, e.g., \cite{Hakuba2024}) is highly statistically significant. Like the OHC trend uncertainties, we can obtain the OHC anomaly moving average uncertainties by first computing the moving average time series for each conditional simulation ensemble member in each vertical section before summing, across the vertical sections, the monthly standard errors. OHC anomaly time series for the two vertical sections separately are given in Figures~\ref{fig:OHCA_upper_ocean} and \ref{fig:OHCA_midocean} in the appendix.

\subsection{Ocean heat uptake anomalies}

\begin{figure}[!h]
    \centering
    \subfigure[Monthly]{
        \includegraphics[width=6cm,trim = 0.5cm 0.5cm 1cm 0.5cm,clip = true]{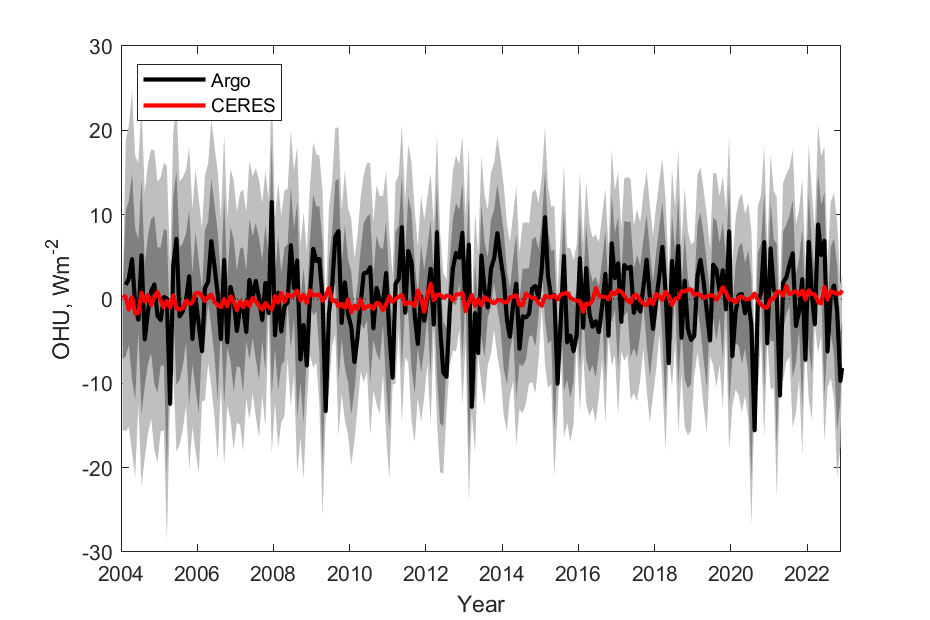}}
    \subfigure[12-month moving average]{
        \includegraphics[width=6cm,trim = 0.5cm 0.5cm 1cm 0.5cm,clip = true]{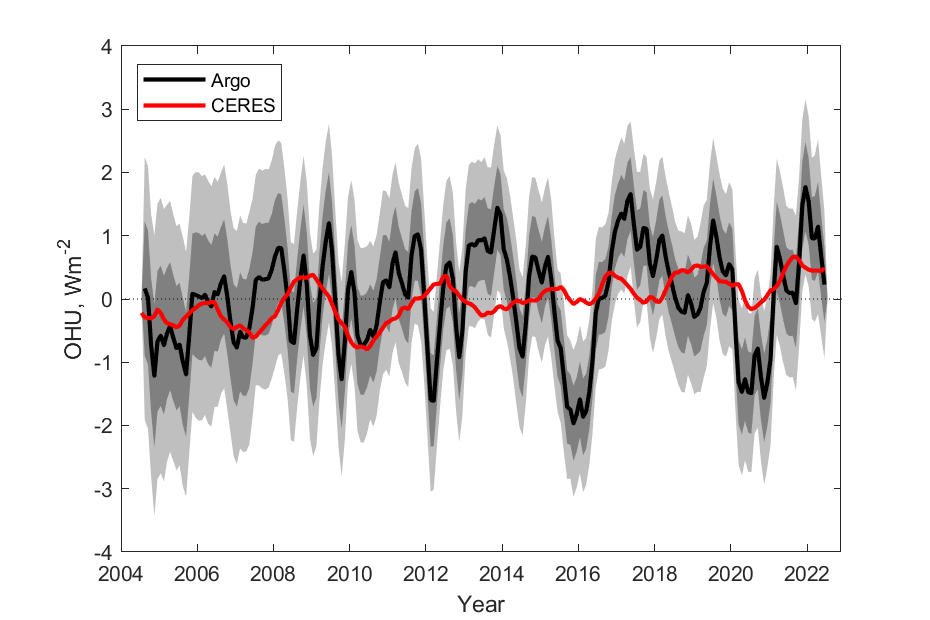}}\\
    \subfigure[24-month moving average]{
        \includegraphics[width=6cm,trim = 0.5cm 0.5cm 1cm 0.5cm,clip = true]{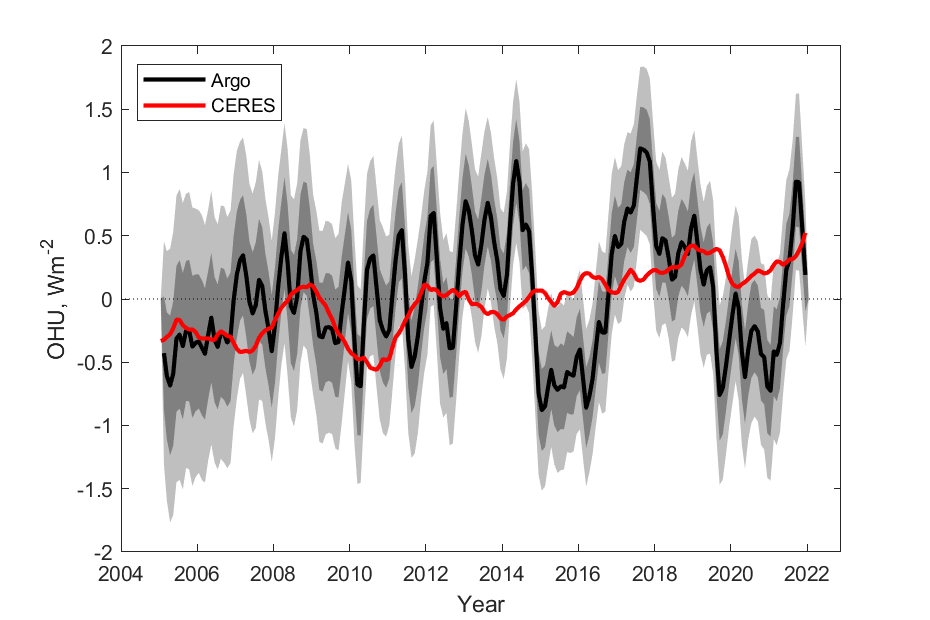}}
    \subfigure[36-month moving average]{
        \includegraphics[width=6cm,trim = 0.5cm 0.5cm 1cm 0.5cm,clip = true]{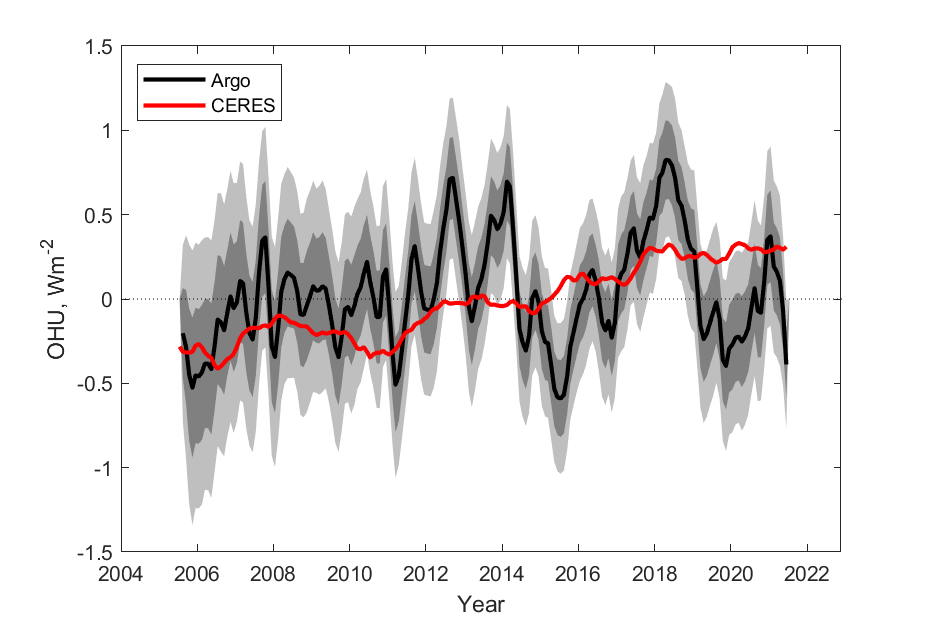}}\\
    \caption{Ocean heat uptake (15--1850 dbar) anomalies with 68\% (dark gray) and 95\% (light gray) uncertainties compared with CERES TOA radiative net flux anomalies.}\label{fig:ohu-vs-ceres}
\end{figure}

Another related quantity of interest is the ocean heat uptake (OHU), which is defined as the time derivative of OHC, $\text{OHU}(t) = \mathrm{d}/\mathrm{d}t \; \text{OHC}(t)$, and provides a measure of the rate of change in OHC. Since changes in OHC are ultimately driven by changes in top-of-atmosphere (TOA) radiative imbalance, OHU is often compared with TOA net flux observations (e.g., \cite{Johnson2016}). In Figure~\ref{fig:ohu-vs-ceres}, we plot our OHU anomaly estimates (converted to a flux relative to the area of the Earth and with the intercept and seasonal cycle removed) with uncertainties alongside CERES satellite TOA net flux anomalies from the EBAF-TOA 4.2 data product (\cite{Loeb2018}, \cite{Kato2018}). We approximate OHU using the first differences of the monthly OHC time series for 15--1850 dbar and obtain the uncertainties by taking the first differences of the OHC conditional simulation ensemble members for each vertical section and summing the monthly standard errors. While it is hard to discern a signal from the monthly estimates (Figure \ref{fig:ohu-vs-ceres}(a)) due to noise, we can see slight positive trends in both the OHU estimate and TOA net flux after applying a moving average low-pass filter (Figures~\ref{fig:ohu-vs-ceres}(b)--(d)). As described by other authors \citep{Storto_and_Yang2024}, the high-frequency variability in observed TOA fluxes differs from that in OHU-based estimates. As an example the  statistically significant drop in the Argo-resolved OHU in 2015--16 does not appear in the CERES observations,   as also shown in \cite{Trenberth_etal2022}. Differences between OHU and CERES observations start decreasing when a 36-month moving average is applied, consistent with the climate community’s increasing focus on comparing these complementary datasets at timescales longer than about three years \citep{Storto_etal2022,Marti_etal2022,meyssignac_etal2023}.

\subsection{Gridded OHC maps}

\begin{figure}[!h]
  \centering
  \noindent\includegraphics[width=10cm, trim=1.5cm 2.5cm 1cm 2cm, clip=true]{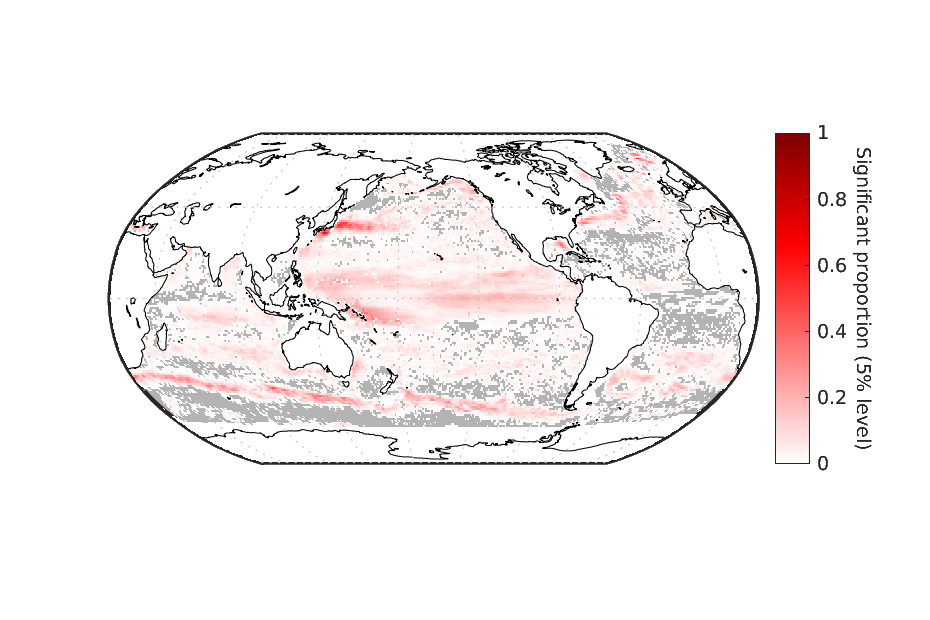}\\
  \caption{Proportion of statistically significant monthly OHC anomalies (15--1850 dbar; 5\% level).}\label{fig:ohc-sigprop}
\end{figure}

So far, we have only shown estimates and uncertainties for the globally integrated OHC and related quantities. However, our mapping pipeline and conditional simulation algorithm also make it possible to obtain maps and uncertainties at regional scales. This enables inferring more localized climatological signals and ocean dynamics. For example, in Figure~\ref{fig:ohc-sigprop}, we show regionally the fraction of statistically significant monthly OHC anomalies (15-1850 dbar) at the 5\% significance level between 2004--2022. We define the uncertainties for each grid point by computing the conservative standard error as in Eq.~\eqref{eq:UQ_UB} using the conditional simulation ensemble members. Examining the regional variation, we can see that regions near the Antarctic Circumpolar Current (ACC) and western boundary currents have the largest number of significant months (e.g., the Kuroshio Current region with more than 60\% significant months), while the equatorial region has a moderate number of significant months (around 30\% for the Pacific and less for the Atlantic and Indian Oceans).

\begin{figure}[!h]
  \centering
  \subfigure[Significant trends (5\% level)]{\includegraphics[height=5cm,trim = 1.5cm 2.1cm 0.8cm 2cm,clip = true]{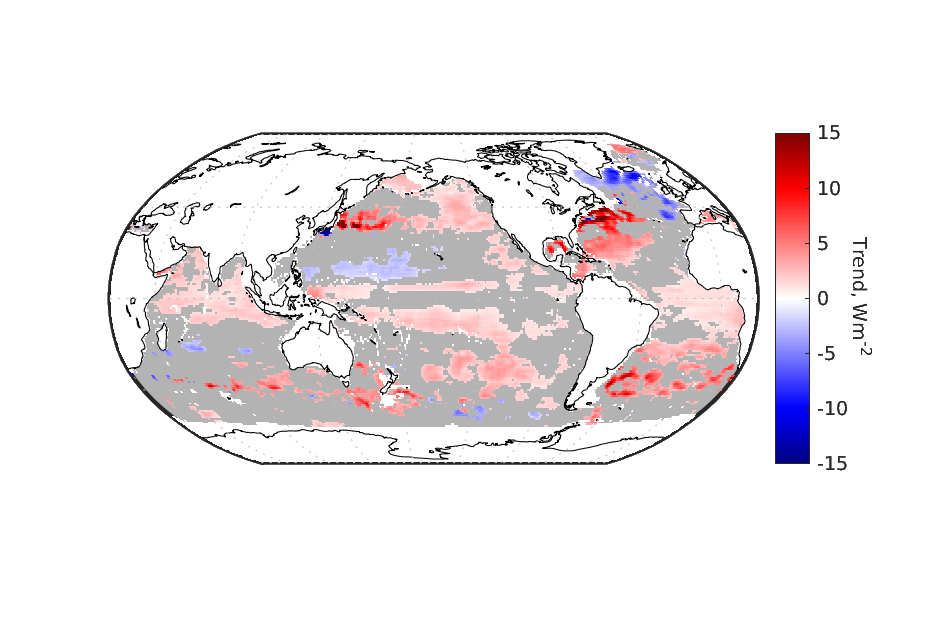} \label{fig:OHC_trend_map}}
  \subfigure[Zonal trends]{\includegraphics[height=5cm, trim = 0 0.5cm 0 0.5cm clip = true]{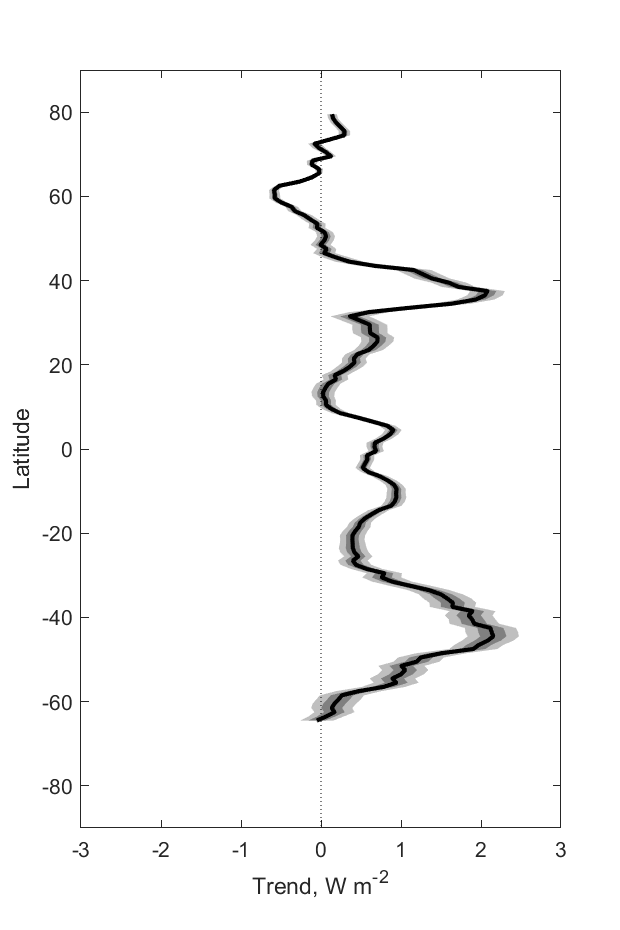} \label{fig:OHC_zonal_trends}}\\
  \caption{Regional OHC trends (15--1850 dbar). Figure~\subref{fig:OHC_trend_map}: Significant OHC linear trends at 5\% significance level. Figure~\subref{fig:OHC_zonal_trends}: Zonal OHC trends with 68\% (dark gray) and 95\% (light gray) uncertainties.}\label{fig:ohc-trend-pixelwise}
\end{figure}

Likewise, we plot a gridded map of statistically significant OHC linear trends (expressed as a heat flux) in Figure~\ref{fig:OHC_trend_map}. While the majority of the significant trends are warming trends---we estimate that the Kuroshio Current, Gulf Stream and ACC regions are warming particularly fast (dark red)---there are also sizable cooling trends off the coast of Japan and in the North Atlantic (dark blue). These patterns are broadly consistent with other recent OHC trend maps (e.g., \cite{Johnson2020,li_etal2023}).

Figure~\ref{fig:OHC_zonal_trends} shows the contribution to the global OHC trend by latitude with uncertainties (normalized to the Argo-sampled ocean area of each latitude band). Consistent with previous studies \citep[e.g.,][]{Trenberth2025}, most latitudes are warming on average, with the regions near $40^\circ$N and $40^\circ$S making the greatest contribution to the global trend. We also observe that the Southern Hemisphere trend appears to be more uncertain, which might be explained by the greater number of Argo profiles in the Northern Hemisphere. We note that providing a zonally averaged uncertainty requires accounting for the spatial dependence in the mapping uncertainties, which our conditional simulation approach allows us to do.

\subsection{Regional time series}

\begin{figure}[!h]
  \centering
  \subfigure[Ni\~{n}o~3.4]{
  \includegraphics[width=8cm,angle=0]{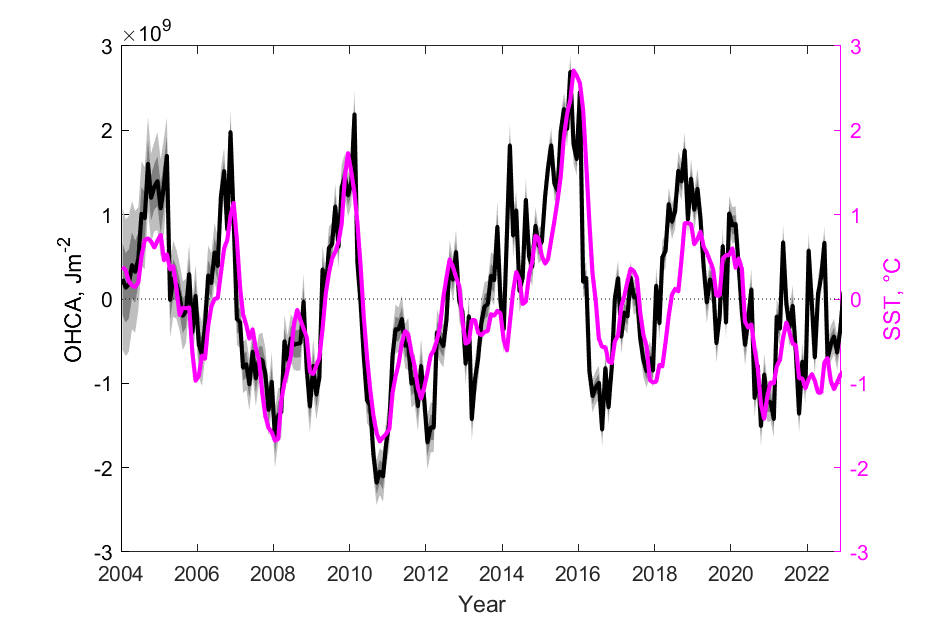}\label{fig:ohc-vs-oni}}
  \subfigure[Northeast Pacific]{
  \includegraphics[width=8cm,angle=0]{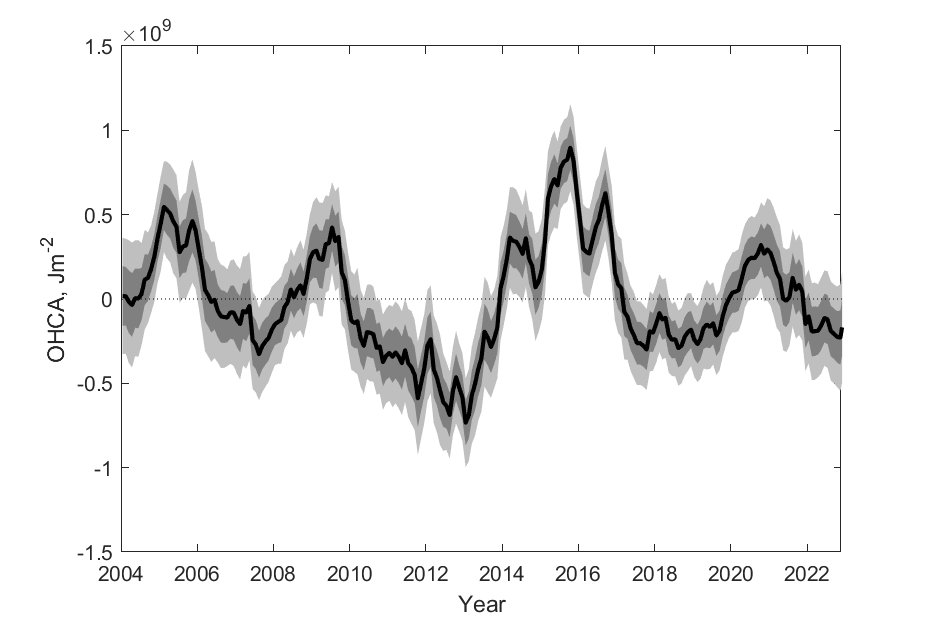}\label{fig:ohca-blob}}
  \caption{Regional monthly OHC (15--1850 dbar) anomaly time series (normalized to the area of each region) with 68\% (dark gray) and 95\% (light gray) uncertainties. Figure~\subref{fig:ohc-vs-oni}: OHC in the Ni\~{n}o~3.4 region (black) vs.~the monthly SST input to the ONI index (magenta). Figure~\subref{fig:ohca-blob}: OHC in the Northeast Pacific region ($40^\circ$N--$50^\circ$N, $135^\circ$W--$150^\circ$W).}
\end{figure}

In addition to examining regional patterns through gridded maps, we also plot OHC anomaly time series with uncertainties averaged over a few regions of interest. We first show the OHC anomaly time series (with trend and seasonal cycle removed) for the Ni\~{n}o~3.4 region ($5^\circ$N--$5^\circ$S, $120^\circ$W--$170^\circ$W) in Figure~\ref{fig:ohc-vs-oni}. The Oceanic Ni\~{n}o Index (ONI) is a measure used for classifying whether the El Ni\~{n}o Southern Oscillation (ENSO) phenomenon is present in a given month by examining whether the sea surface temperature (SST) exceeds a certain threshold in the Ni\~{n}o~3.4 region (\cite{ONI}). We overlay the monthly input to the ONI index (i.e., the monthly ERSST.v5 SST anomalies) in magenta in Figure~\ref{fig:ohc-vs-oni}. While OHC is a subsurface integral and SST is a surface measurement, the two time series match well, particularly between 2006--2013, reflecting the large contribution of the shallower layers to the OHC anomaly within 15--1850~dbar and giving us further confidence in our mapping pipeline. Overall, the relative size of the uncertainties is smaller than for the global OHC time series (Figure~\ref{fig:OHCA_sum}), which means that the OHC anomaly time series is resolved well in this particular region, consistent with the large number of Argo observations and long spatial decorrelation scales (Figures~\ref{fig:length_scales_upper_ocean} \& \ref{fig:length_scales_midocean}) in the region.

As another regional example, the OHC anomaly time series in Figure~\ref{fig:ohca-blob} for a region in the Northeast Pacific (specifically $40^\circ$N--$50^\circ$N, $135^\circ$W--$150^\circ$W) shows the signature of an intense marine heatwave event during 2013--2016  (\cite{Bond2015}; reflected here as the statistically significant positive anomalies during most of those years). This marine heatwave event contributed to drought conditions on the North American West Coast (e.g., \cite{Seager2015}) and had impacts on the biogeochemistry of the ocean due to both changes in temperature and ocean circulation (\cite{Mogen2022}). Notice that the time series in the Northeast Pacific has less high-frequency variation than the Ni\~{n}o 3.4 time series, which is due to the large difference in the estimated temporal decorrelation scales between the two regions (Figures~\ref{fig:length_scales_upper_ocean} \& \ref{fig:length_scales_midocean}).

\subsection{OHC and ENSO cross-correlation} \label{sec:OHC_ENSO_ccf}

Our final example application demonstrates that the conditional simulation ensembles can be used to obtain predictions and uncertainties for arbitrarily complex downstream quantities that depend on the OHC estimates. Specifically, we will consider the empirical cross-correlation between the global OHC anomalies (OHCA) and the SST input to the ONI index. The cross-correlation can be understood as a transformation $T$ of the OHCA time series, $\mathbf{OHCA} = (\text{OHCA}(t_1),\ldots,\text{OHCA}(t_n))$. So far, all the transformations we have considered have been linear, so the conditional expectation satisfies $\mathbb{E}(T(\mathbf{OHCA}) | \mathrm{data}) = T(\mathbb{E}(\mathbf{OHCA} | \mathrm{data}))$. In other words, we can simply transform the GP prediction and this gives us the best prediction of the transformed quantity. However, when $T$ is nonlinear, as is the case for cross-correlation, we need to keep track of the entire predictive distribution $p(T(\mathbf{OHCA}) | \mathrm{data})$ and use its expectation and quantiles for inference. This is easy to do using the conditional simulations. As for the other examples before, we apply $T$ on each ensemble member: this gives us a sample from $p(T(\mathbf{OHCA}) | \mathrm{data})$, which we can use to estimate the derived quantity of interest (as the mean across ensembles) and related uncertainties (based on quantiles across ensemble members).

\begin{figure}[t]
  \centering
  \noindent\includegraphics[width=10cm, trim = 0.5cm 0 1cm 0.5cm, clip=true]{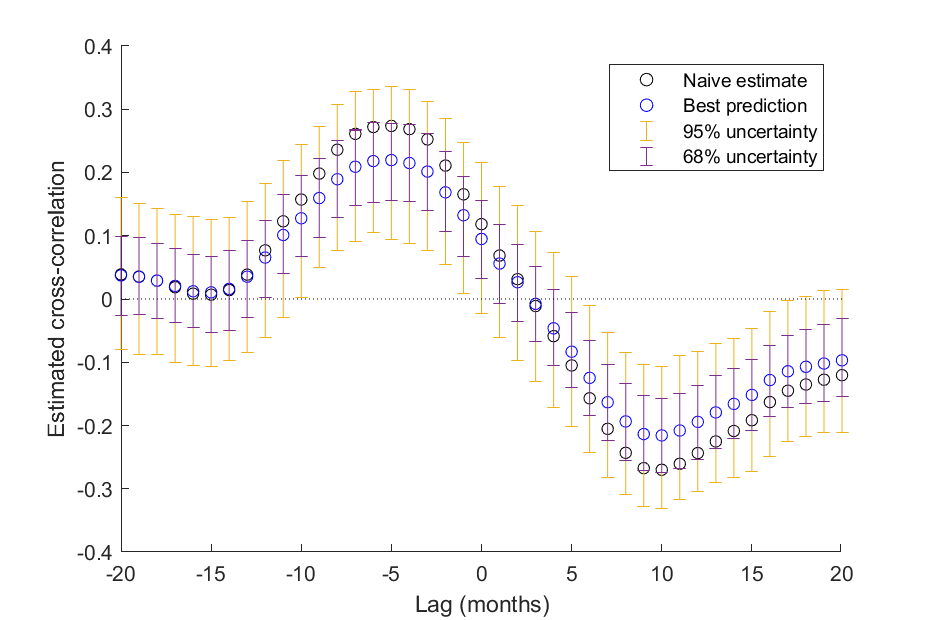}\\
  \caption{Cross-correlation between global upper ocean (15--975 dbar) monthly OHC and Ni\~{n}o 3.4 region SST with 68\% and 95\% uncertainties. A naive cross-correlation estimate that ignores nonlinearity is also shown for reference.}\label{fig:ohc-vs-oni-ccf}
\end{figure}

In Figure~\ref{fig:ohc-vs-oni-ccf}, we show the cross-correlation between the global upper ocean (15--975 dbar) monthly OHC anomalies and the monthly input SST to the ONI index with 68\% and 95\% uncertainty intervals. The estimated cross-correlation is obtained as the mean of the cross-correlation predictive distribution and the uncertainties are defined using the appropriate quantiles of the predictive distribution. We also show the naive estimate $T(\mathbb{E}(\mathbf{OHCA} | \mathrm{data}))$ to illustrate that it is markedly different from the best prediction $\mathbb{E}(T(\mathbf{OHCA}) | \mathrm{data})$. The strongest cross-correlation magnitude is found around lags -5 and +10 months, which are both statistically significant. This implies that, with moderate positive correlation, global OHC anomalies, as observed by Argo, lead ENSO by approximately 5 months. This observation is consistent with the accumulation of heat in the Western Pacific in advance of El Ni\~{n}o events (\cite{McPhaden2012}, \cite{Cheng_etal2019}).

Since in this work we do not model the dependence between the vertical sections, we only report here the cross-correlation and its uncertainties for the upper ocean section. Unlike the previous quantities, the total cross-correlation cannot be expressed as a sum of the two sections, so one cannot directly obtain its best prediction or use the conservative upper bound in Eq.~\eqref{eq:UQ_UB} for uncertainty quantification. A future improvement will involve modeling the vertical dependence to produce estimates and uncertainties for the cross-correlation between the full 15--1850~dbar OHC anomalies and ENSO.

\section{Discussion and conclusions}
\label{sec:discussion}

We introduced a new mapping and uncertainty quantification framework for estimating ocean heat content from Argo observations. Based on the mapping foundation in \cite{Kuusela2018}, we used fully data-driven decorrelation scales and variances in a locally stationary spatio-temporal Gaussian process model to map OHC. We developed a principled and flexible uncertainty quantification method through local conditional simulation ensembles that incorporates dependence in space and time into the uncertainties for not only OHC but also related downstream quantities. We used cross-validation to validate the modeling choices and uncertainties. We implemented this mapping pipeline in an open-source, modular codebase that enables full reproducibility of the results. We used the pipeline to produce OHC maps for 2004--2022 and demonstrated the resulting estimates and uncertainties for the global OHC trend and anomalies, OHU, regional OHC estimates and cross-correlation between OHC anomalies and ENSO. A data product using this pipeline \citep[LocalGP]{Giglio_etal2026_Zenodo} has already contributed to several climate assessments, oceanographic studies, and OHC intercomparison efforts and will enable new regional analyses and principled uncertainty quantification in future studies of OHC and EEI.

There are several potential avenues for future improvements to this mapping framework:
\begin{itemize}
    \item In this work, we used plug-in estimates of the mean field and the GP parameters. Both of these estimates are considered fixed when quantifying the uncertainty of the mapped fields so the final uncertainties do not account for the uncertainty of these estimates. One future direction would be to explore ways of propagating these sources of uncertainty into the final uncertainties.
    \item Throughout this work, we reported conservative estimates of the uncertainties for quantities involving the sum of the vertical sections. In ongoing follow-up work, we seek to improve the total OHC uncertainties by modeling the dependence between the vertical sections using a bivariate locally stationary Gaussian process, which can also improve the mapped anomalies since the sparsely sampled midocean can borrow strength from the more densely sampled upper ocean.
    \item We saw in Section~\ref{sec:validation}\ref{sec:val_UQ} that in our current model there is some amount of unmodeled conditional dependence at longer time scales. A potential direction for future research could be to change how space and time interact in the covariance function through more general space-time covariance classes \citep{Porcu2021}, including asymmetric covariances \citep{ma2025asymmetric} and neural network covariances \citep{Wilson2016}.
    \item \cite{Park2023} introduced a bias correction for the mean field and a way to jointly estimate the mean field and covariance parameters within the locally stationary modeling framework. These are both improvements that could be incorporated into future versions of the OHC mapping pipeline.
    \item In this work, we divided the ocean into two vertical sections. While this is sufficient to demonstrate our framework for the vertically integrated OHC, it only allows to use profiles that fully cover at least one of the layers considered (e.g., profiles shallower than 975~dbar are not included, which limits our ability to use in-situ observations beyond Argo). Also, understanding the vertical distribution of heat better would require finer vertical resolution in the mapping by either using more vertical layers or switching to full 4D maps by adding the vertical dimension to the covariance model. In addition to the resulting modeling challenges, a major challenge with this extension would be the computational cost associated with producing the maps. Recent advances in amortized neural inference for spatial models \citep{Gerber2021, Lenzi2023, SainsburyDale2024, SainsburyDale2025, Walchessen2024, Walchessen2025} are likely to provide a path forward in this setting.
    \item \cite{Lyman2023} showed that remotely sensed sea surface height (SSH) and sea surface temperature (SST) contain useful information for mapping the subsurface ocean heat content. Their mapping approach is based on a random forest regression method. In future work, it would be interesting to consider how to incorporate SSH/SST observations into a spatio-temporal Gaussian process model, which might provide benefits over random forests in terms of model interpretability and uncertainty quantification. Here again, a major challenge with this extension would be the computational cost, as satellite data are available at a much higher spatio-temporal sampling resolution than in-situ data.
\end{itemize}

In addition to improvements to the mapping methodology, another important direction for future work is to build on previous OHC intercomparison efforts, such as \cite{von_Schuckmann2023, Hakuba2024}, to better understand how the different modeling and data analysis choices contribute to the differences across  OHC products. 
Our validation studies in Section~\ref{sec:validation} provide indications that could explain some of the differences. Specifically, consistent with \cite{Cheng2015}, our results indicate that the handling of the climatological trend in the mean field can have a significant impact on the estimated OHC trend. Additionally, we showed that the details of the spatio-temporal modeling can significantly impact the estimated temporal and spatial variation. While new assessments of mapping methods are ongoing \citep[][leveraging synthetic profiles from models and the comparison of  full field reconstructions with the model truth]{ME4OH}, a major open direction is a comprehensive assessment of OHC uncertainties across products, which is challenging as mapping uncertainty estimates are often not available in existing products.

The mapping and uncertainty quantification pipeline presented here is not limited to OHC and can be easily deployed to map other oceanographic variables based on Argo data and other in-situ oceanographic observations. The software framework we developed is specifically designed to be easily adaptable to mapping other oceanographic fields with uncertainties, which is a direction that we are currently actively pursuing.

\acknowledgments TS and MK were supported by NOAA grant NA21OAR4310258. DG was supported by NOAA grant NA21OAR4310261. We would like to thank members of the CMU STAMPS Research Center and the SAMSI Statistical Oceanography Working Group, as well as participants of the GEWEX Earth Energy Imbalance Assessment Workshops, for helpful comments and feedback on this work. This work used Bridges-2 at Pittsburgh Supercomputing Center through allocation MTH230014 from the Advanced Cyberinfrastructure Coordination Ecosystem: Services \& Support (ACCESS) program, which is supported by National Science Foundation grants \#2138259, \#2138286, \#2138307, \#2137603, and \#2138296.


%
%
\datastatement

Argo data were collected and made freely available by the international Argo project and the national programs that contribute to it. The Argo Program is part of the Global Ocean Observing System.
The codebase for the mapping pipeline and case studies presented in this paper is available at \url{https://github.com/ttsukianto/LocalGP_OHC}. The preprocessed data files, parameter estimates, OHC maps and conditional simulation ensembles produced as part of this work are available at \url{https://doi.org/10.5281/zenodo.18273718}.


%

\clearpage

\appendix





%
\subsection{Estimated covariance parameters} \label{sec:covParamPlots}

\begin{figure}[!h]
    \centering
    \subfigure[Latitude]{
        \includegraphics[width=7.5cm,trim = 1cm 3.5cm 1cm 3cm,clip = true]{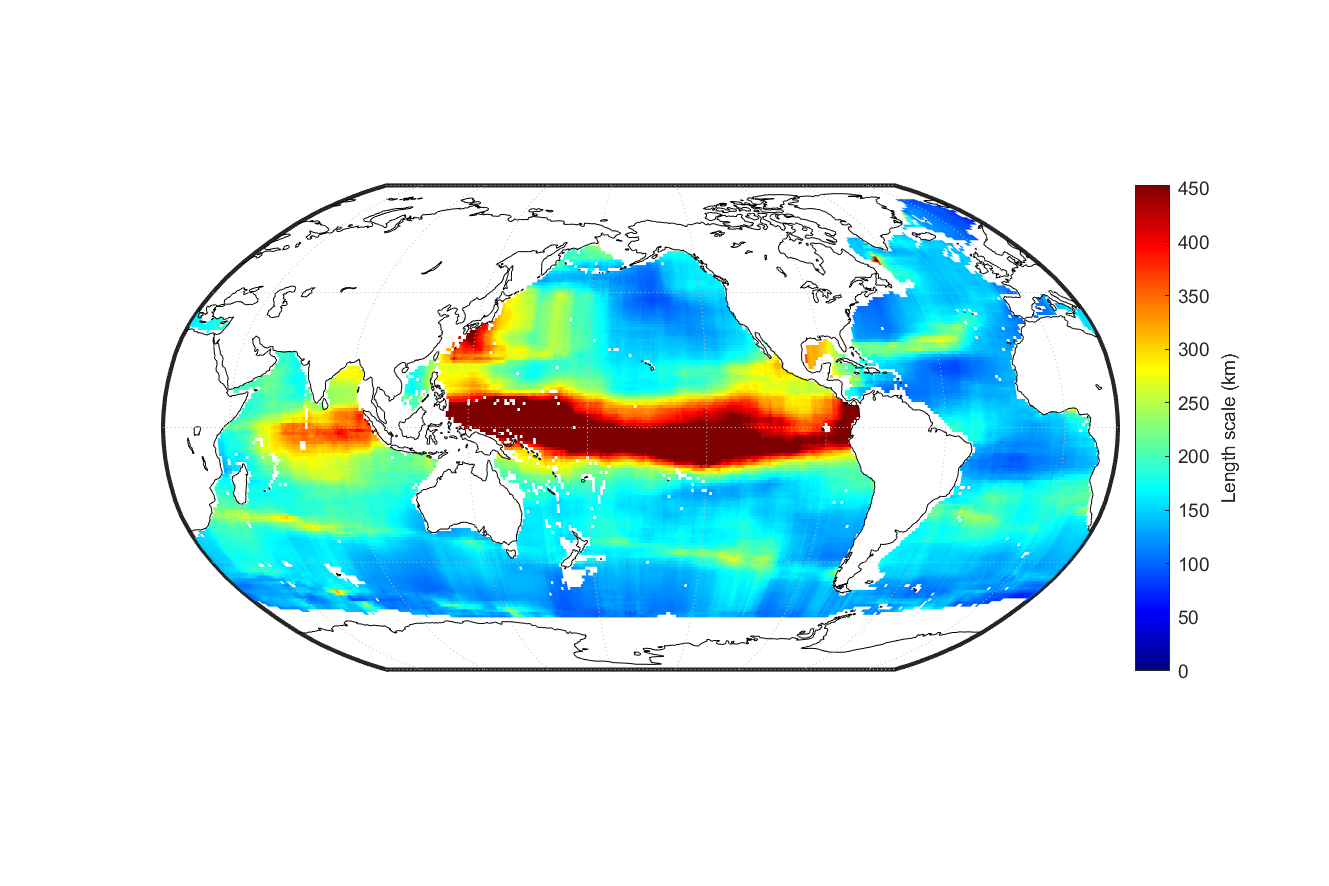}}
    \subfigure[Longitude]{
        \includegraphics[width=7.5cm,trim = 1cm 3.5cm 1cm 3cm,clip = true]{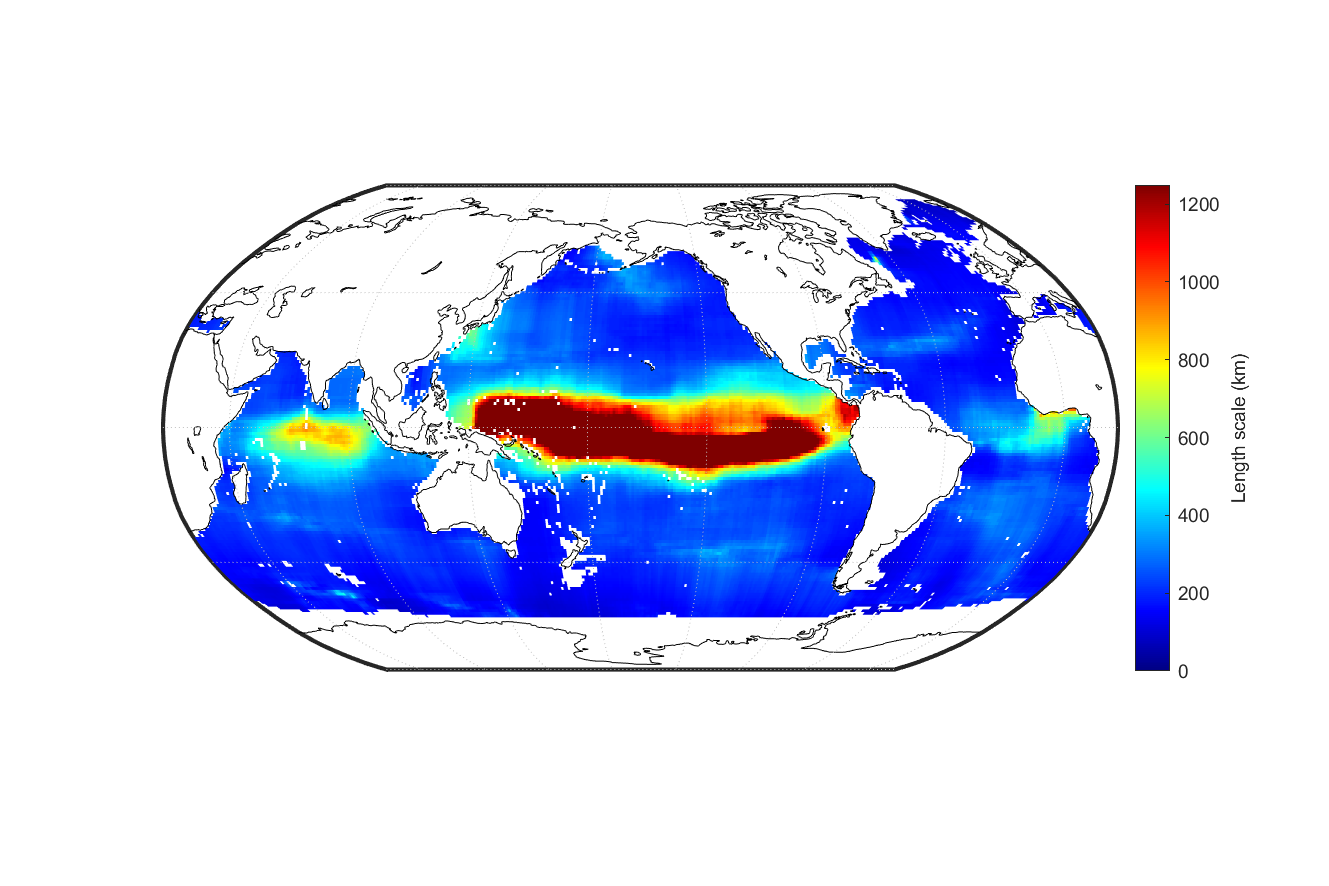}}
    \subfigure[Longitude vs.~latitude]{
        \includegraphics[width=7.5cm,trim = 1cm 3.5cm 1cm 3cm,clip = true]{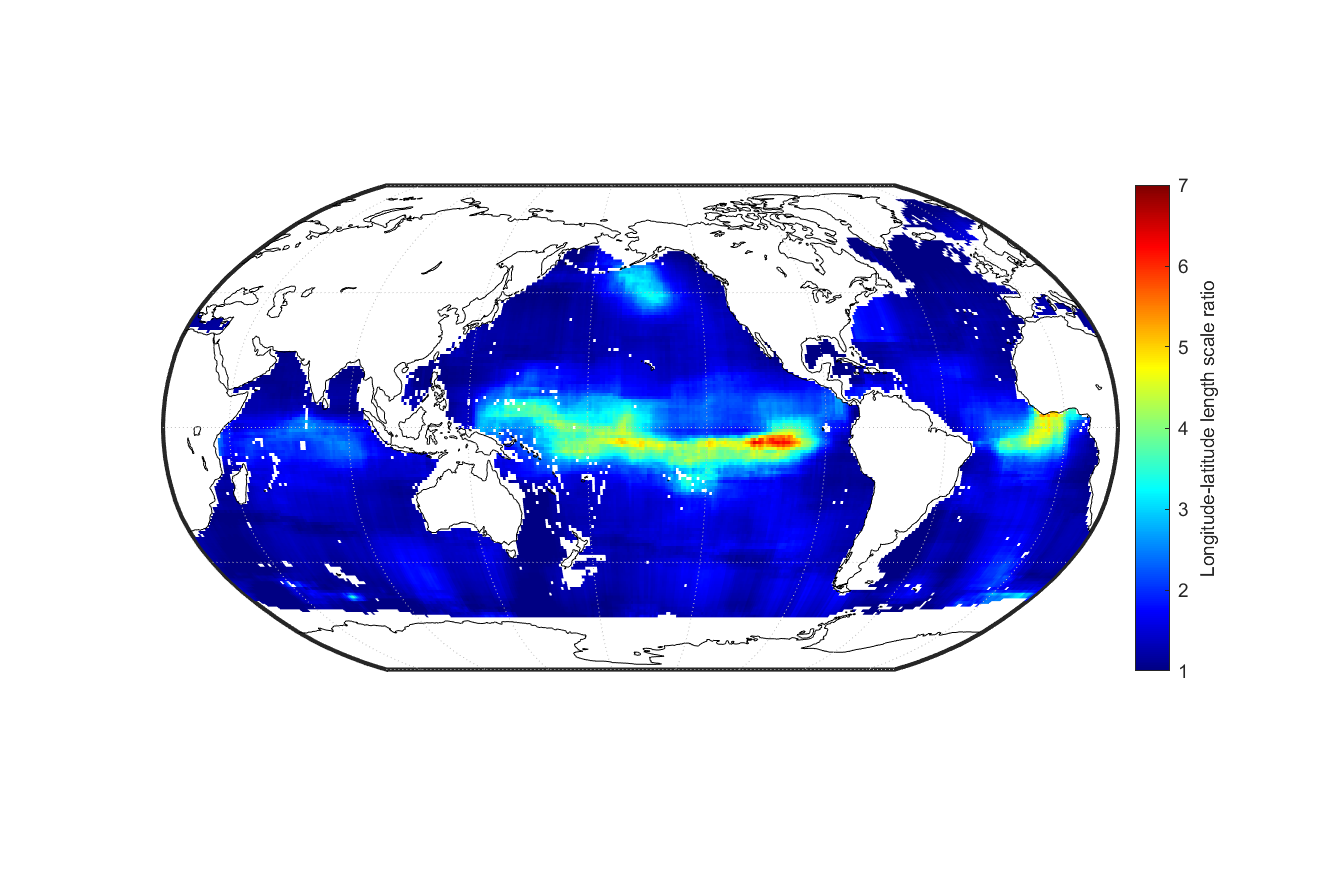}}
    \subfigure[Time]{
        \includegraphics[width=7.5cm,trim = 1cm 3.5cm 1cm 3cm,clip = true]{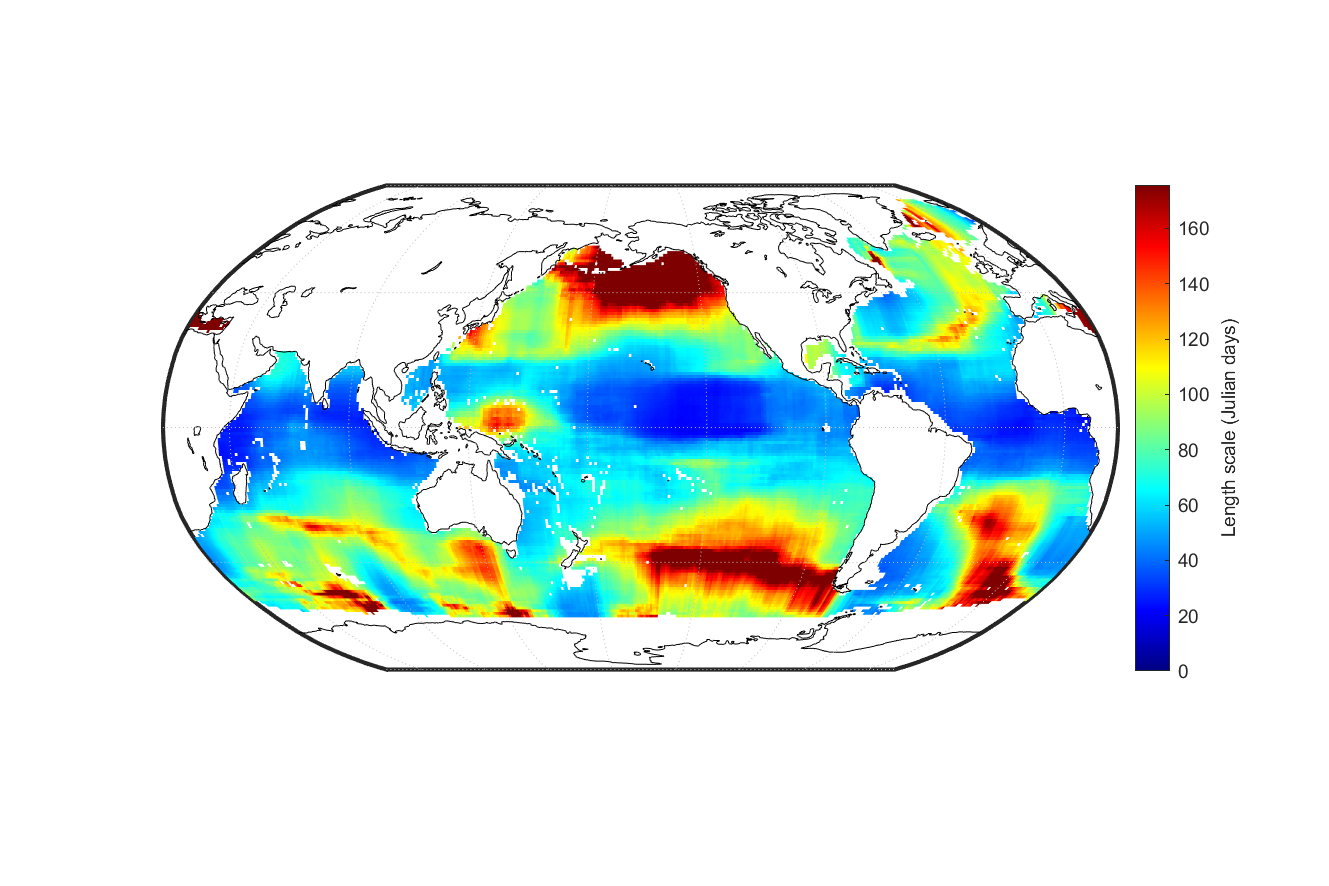}}
    \caption{Space-time model length scale parameter estimates for the upper ocean section.}
    \label{fig:length_scales_upper_ocean}
\end{figure}

\vspace{1cm}

\begin{figure}[!h]
    \centering
    \subfigure[Latitude]{
        \includegraphics[width=7.5cm,trim = 1cm 3.5cm 1cm 3cm,clip = true]{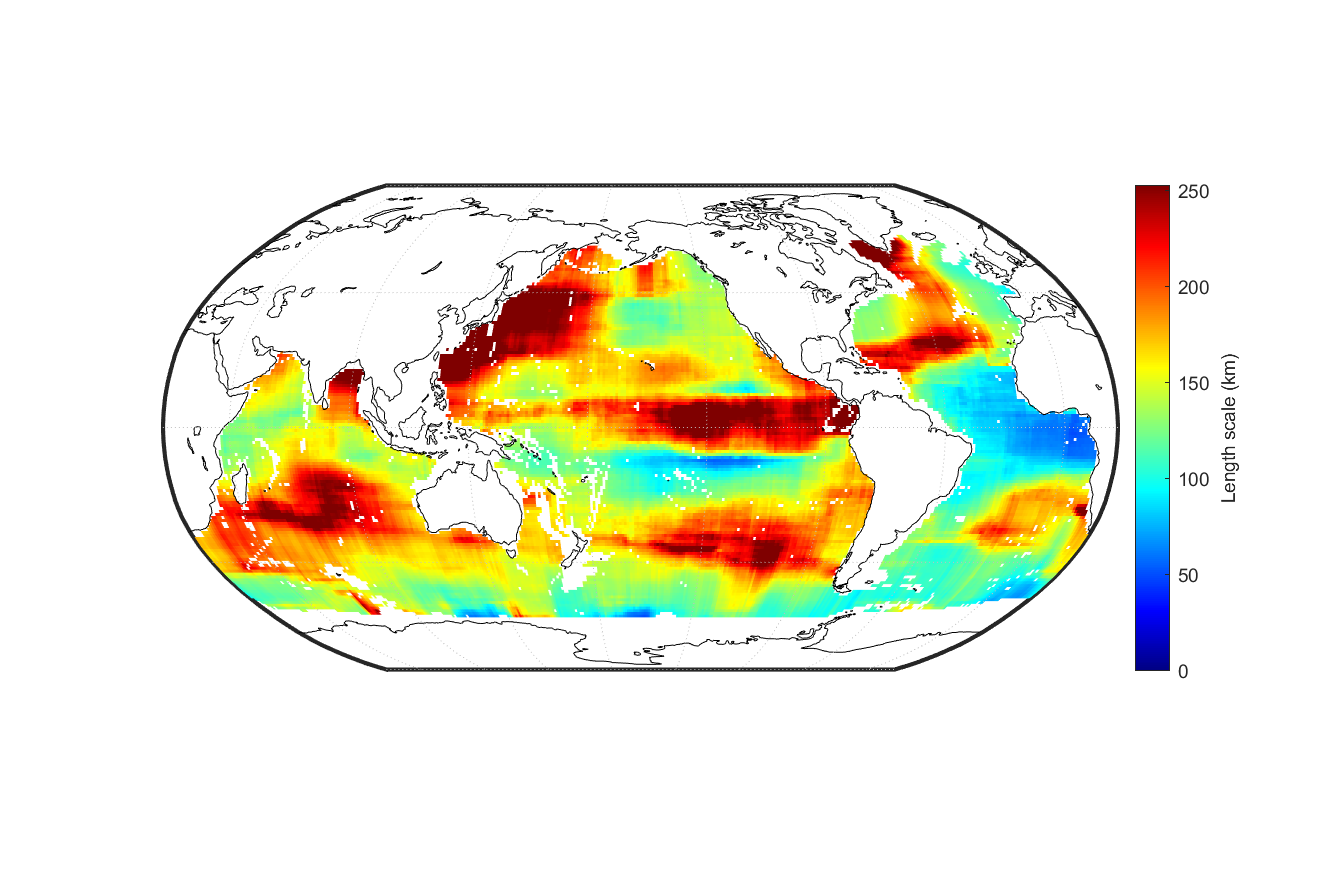}}
    \subfigure[Longitude]{
        \includegraphics[width=7.5cm,trim = 1cm 3.5cm 1cm 3cm,clip = true]{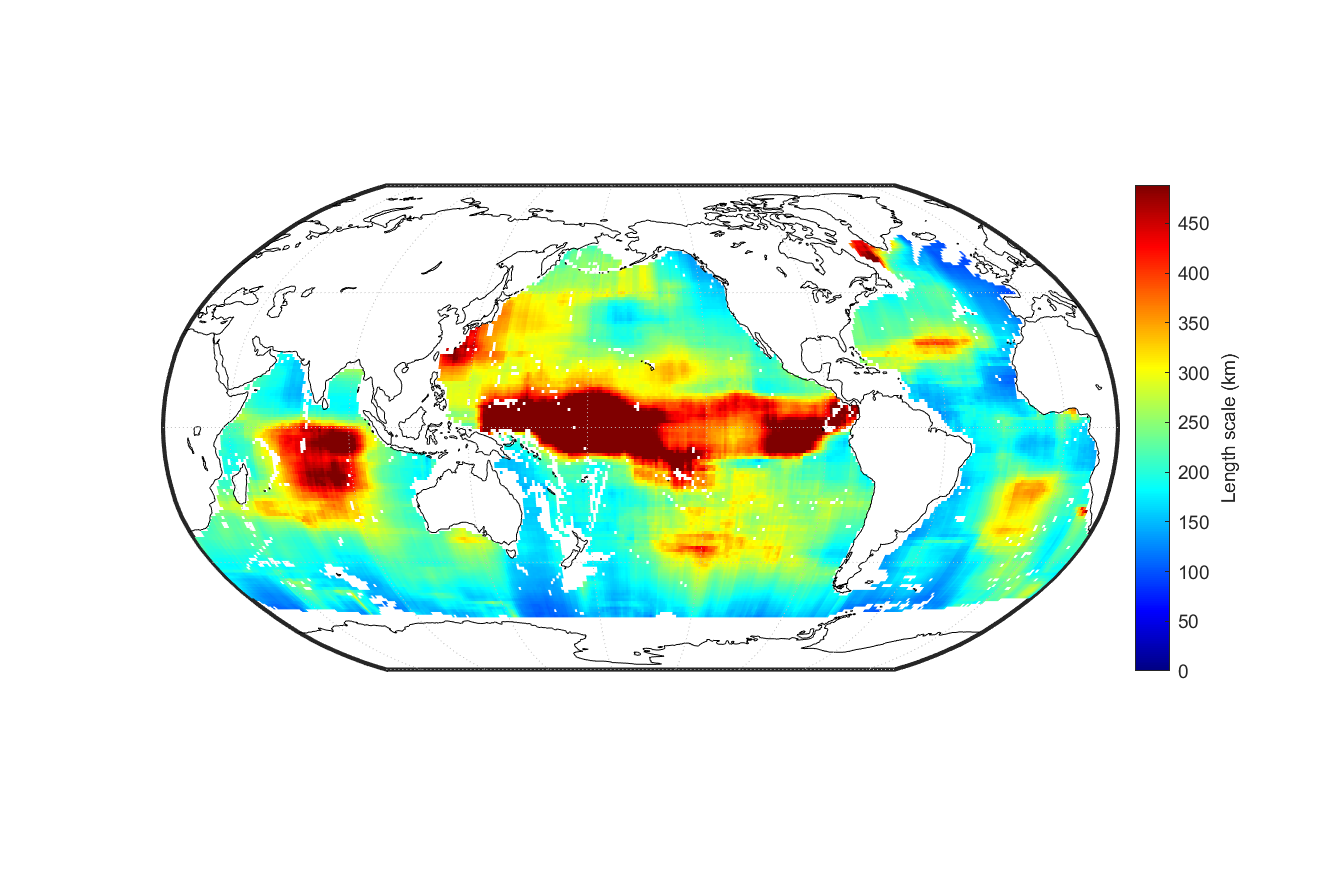}}
    \subfigure[Longitude vs.~latitude]{
        \includegraphics[width=7.5cm,trim = 1cm 3.5cm 1cm 3cm,clip = true]{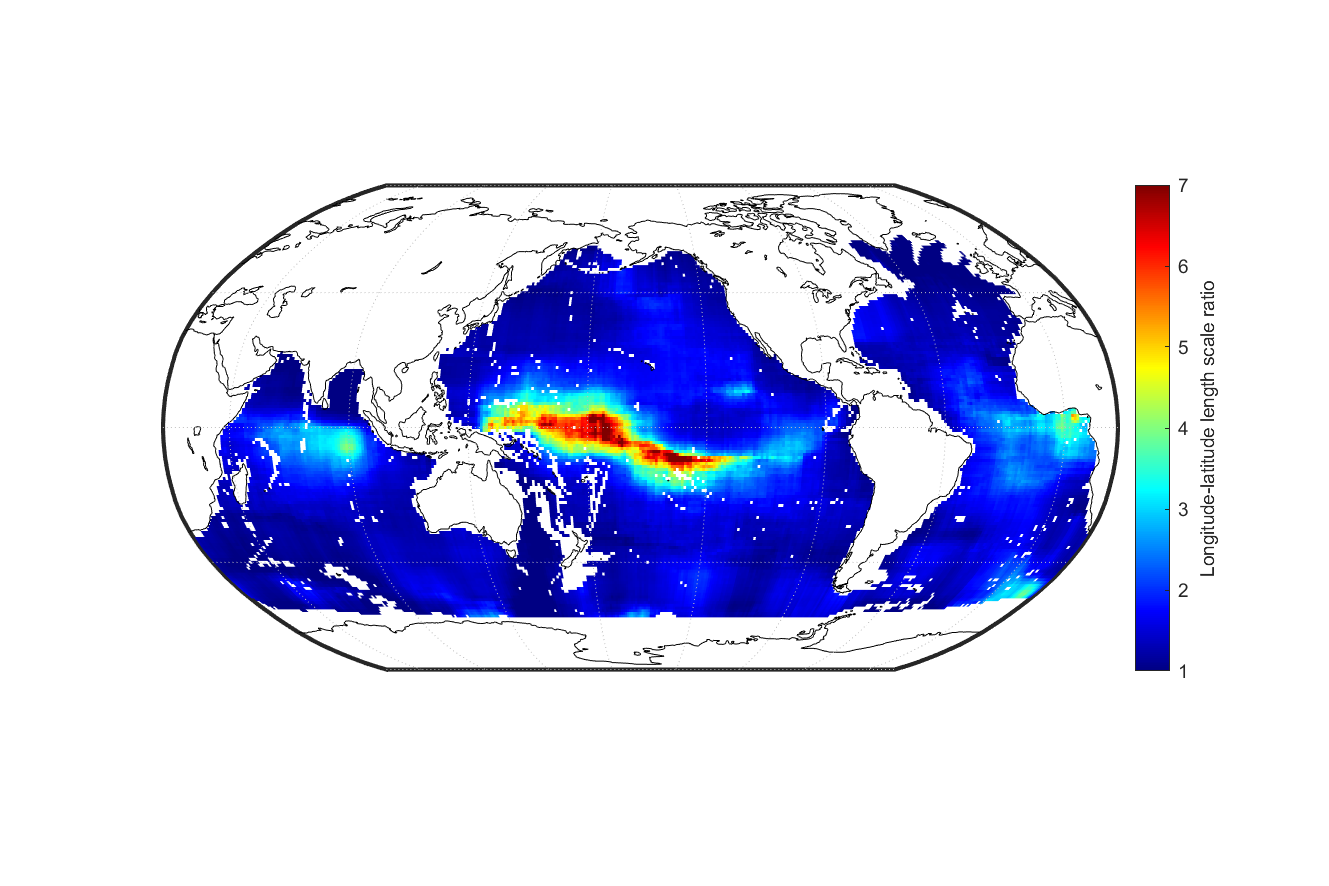}}
    \subfigure[Time]{
        \includegraphics[width=7.5cm,trim = 1cm 3.5cm 1cm 3cm,clip = true]{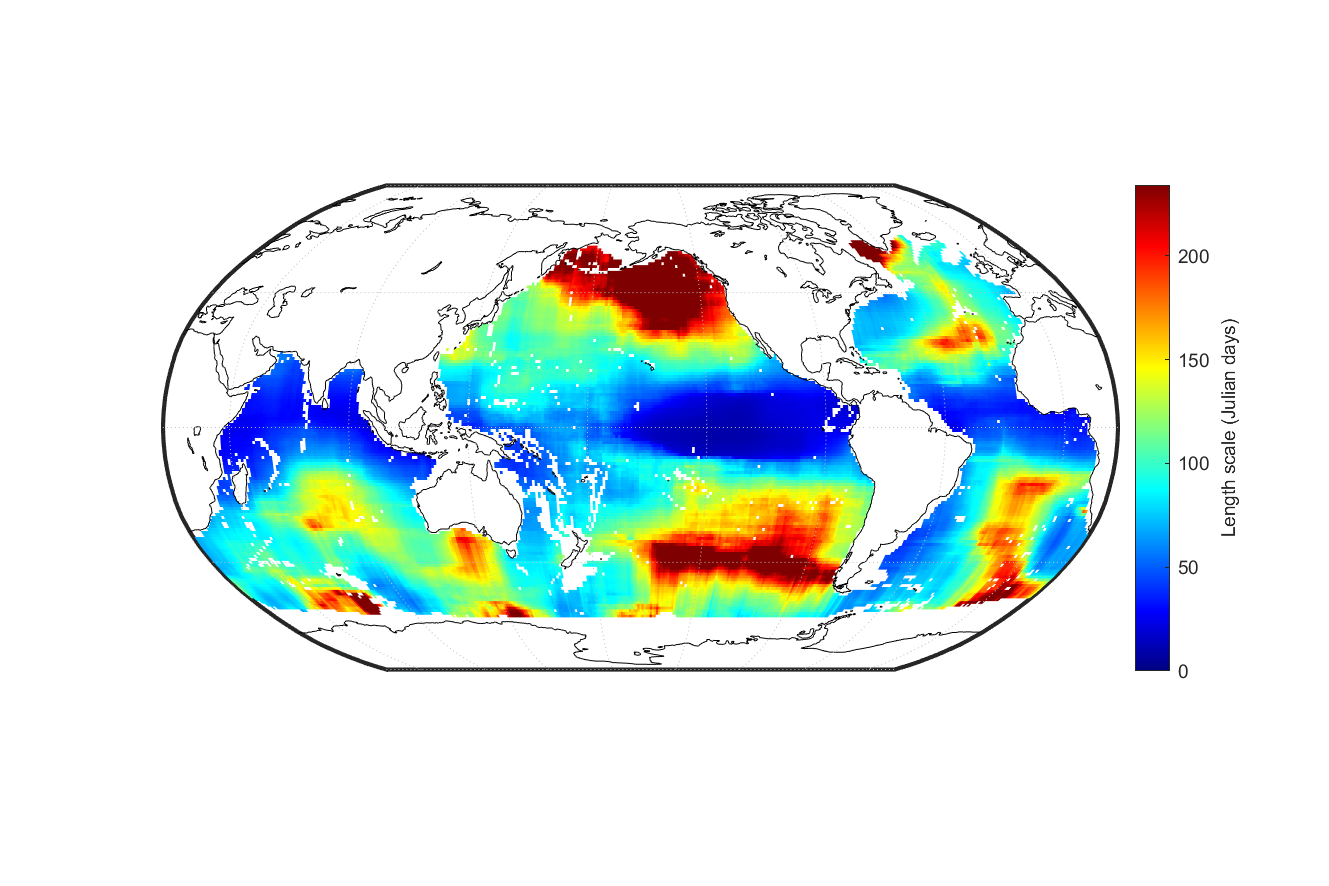}}
    \caption{Space-time model length scale parameter estimates for the midocean section.}
    \label{fig:length_scales_midocean}
\end{figure}

\begin{figure}[!h]
    \centering
    \subfigure[Log field variance]{
        \includegraphics[width=7.5cm,trim = 1cm 3.5cm 1cm 3cm,clip = true]{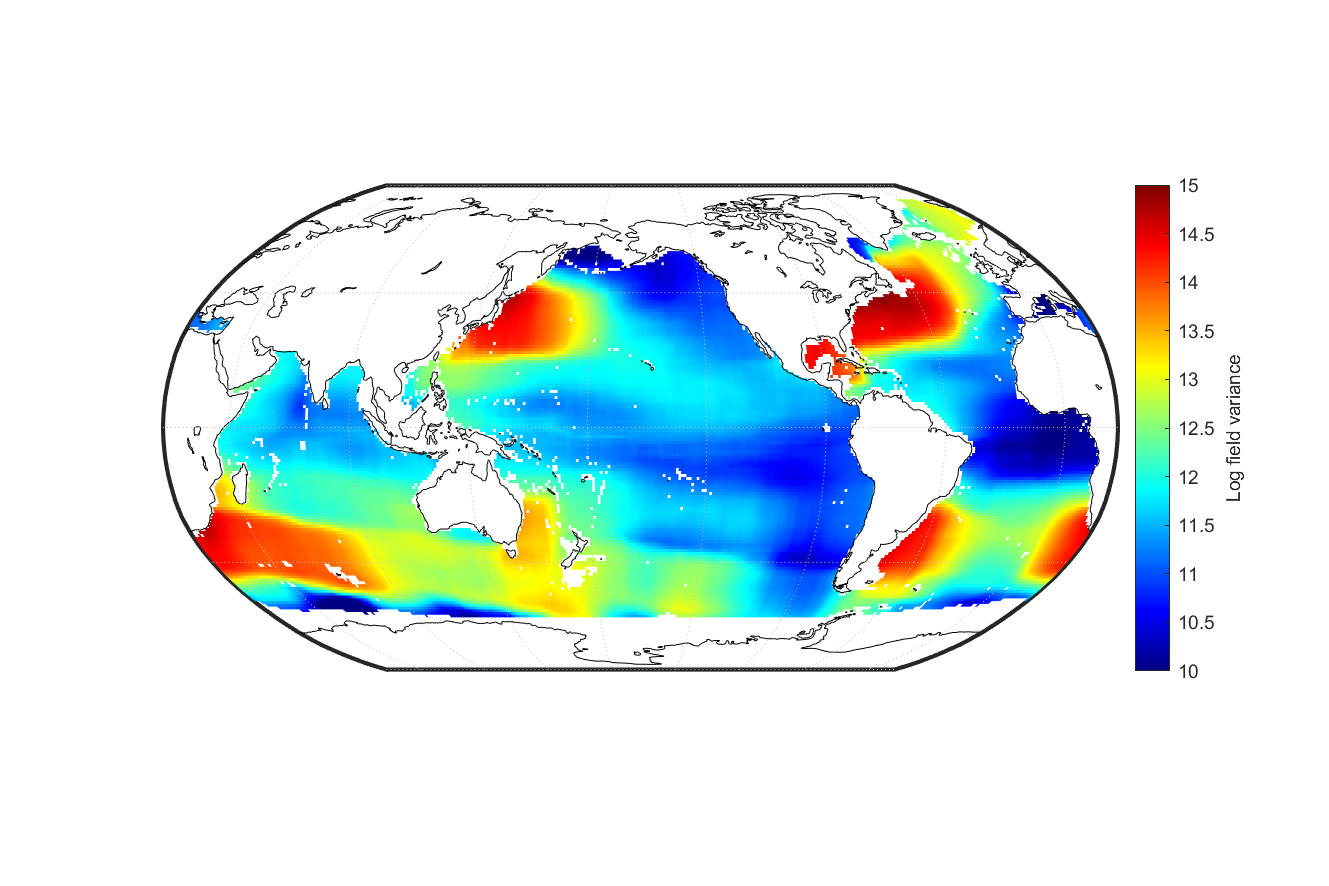}}
    \subfigure[Log nugget variance]{
        \includegraphics[width=7.5cm,trim = 1cm 3.5cm 1cm 3cm,clip = true]{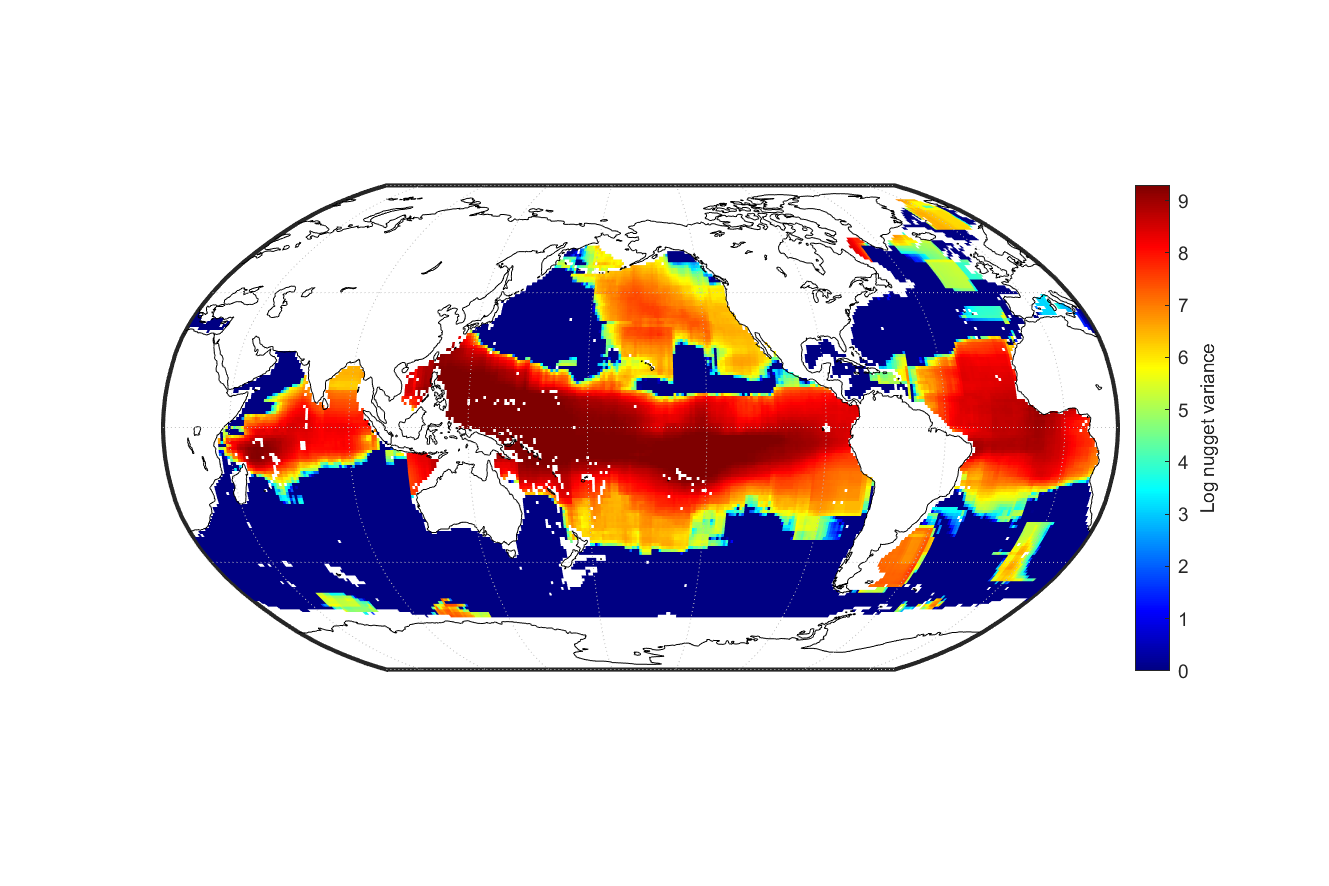}}\\
    \subfigure[Log total variance]{
        \includegraphics[width=7.5cm,trim = 1cm 3.5cm 1cm 3cm,clip = true]{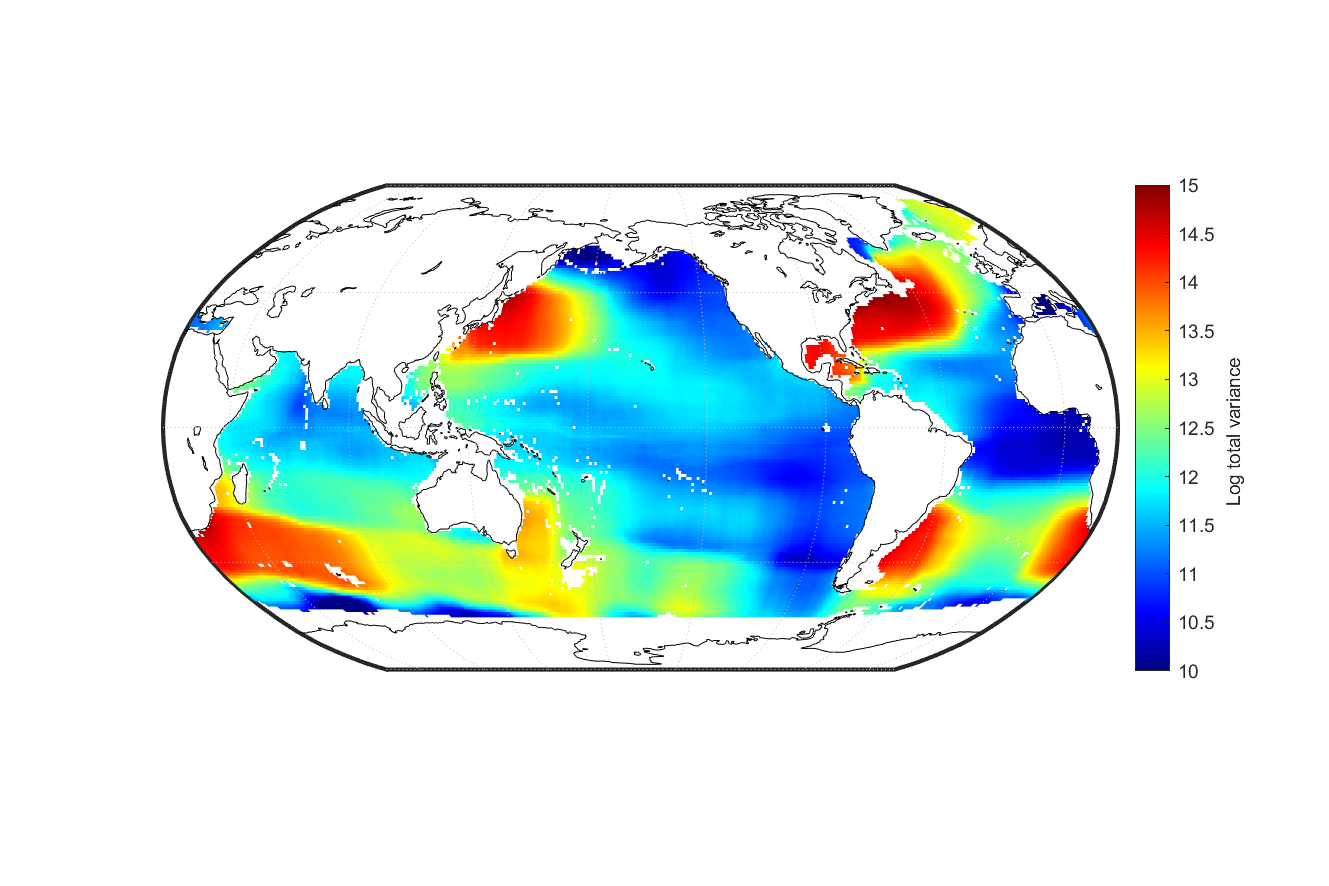}}
    \subfigure[Nugget variance vs.~total variance]{
        \includegraphics[width=7.5cm,trim = 1cm 3.5cm 1cm 3cm,clip = true]{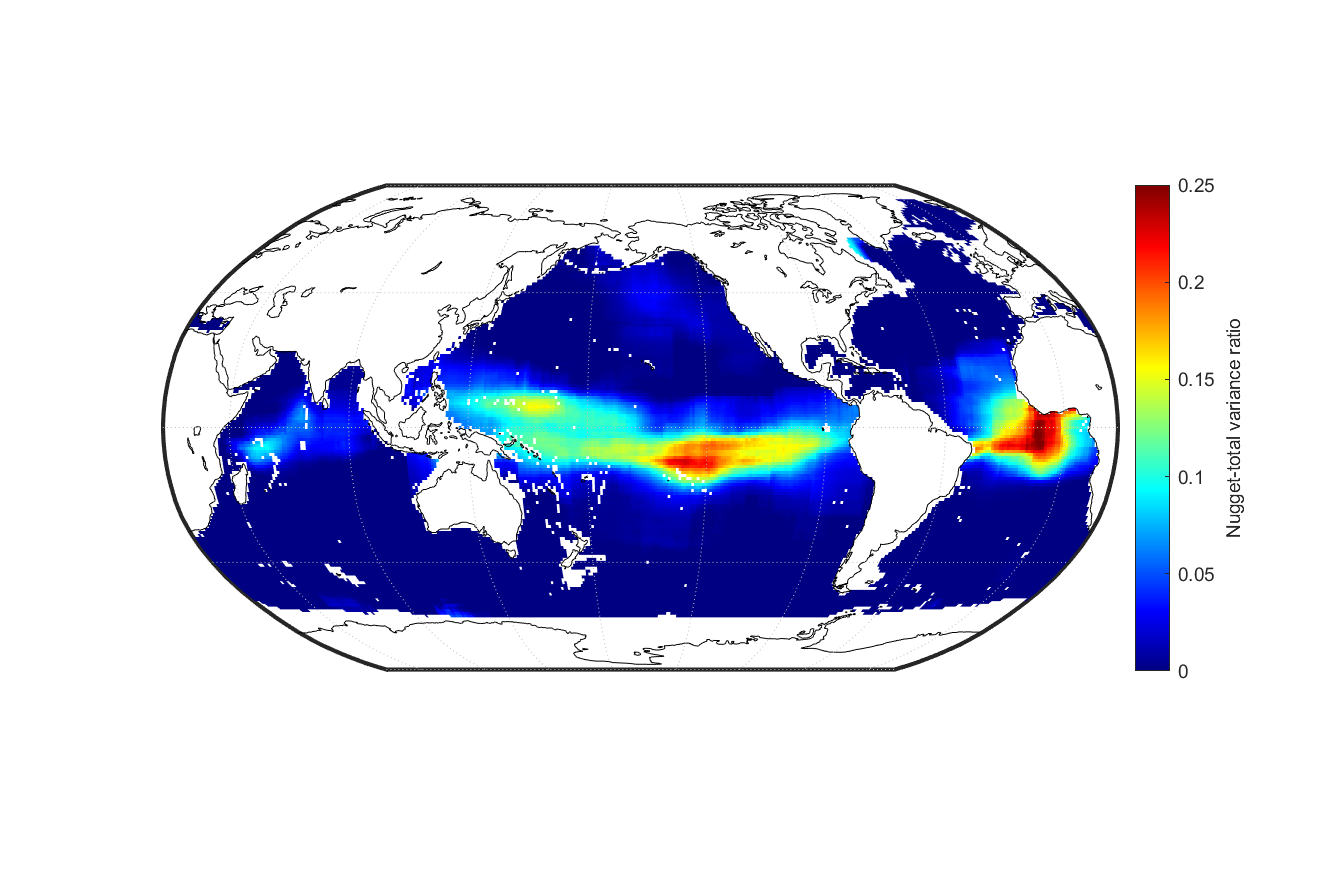}}\\
    \caption{Space-time model field and nugget variance estimates for the upper ocean section.}
\end{figure}

\begin{figure}[!h]
    \centering
    \subfigure[Log field variance]{
        \includegraphics[width=7.5cm,trim = 1cm 3.5cm 1cm 3cm,clip = true]{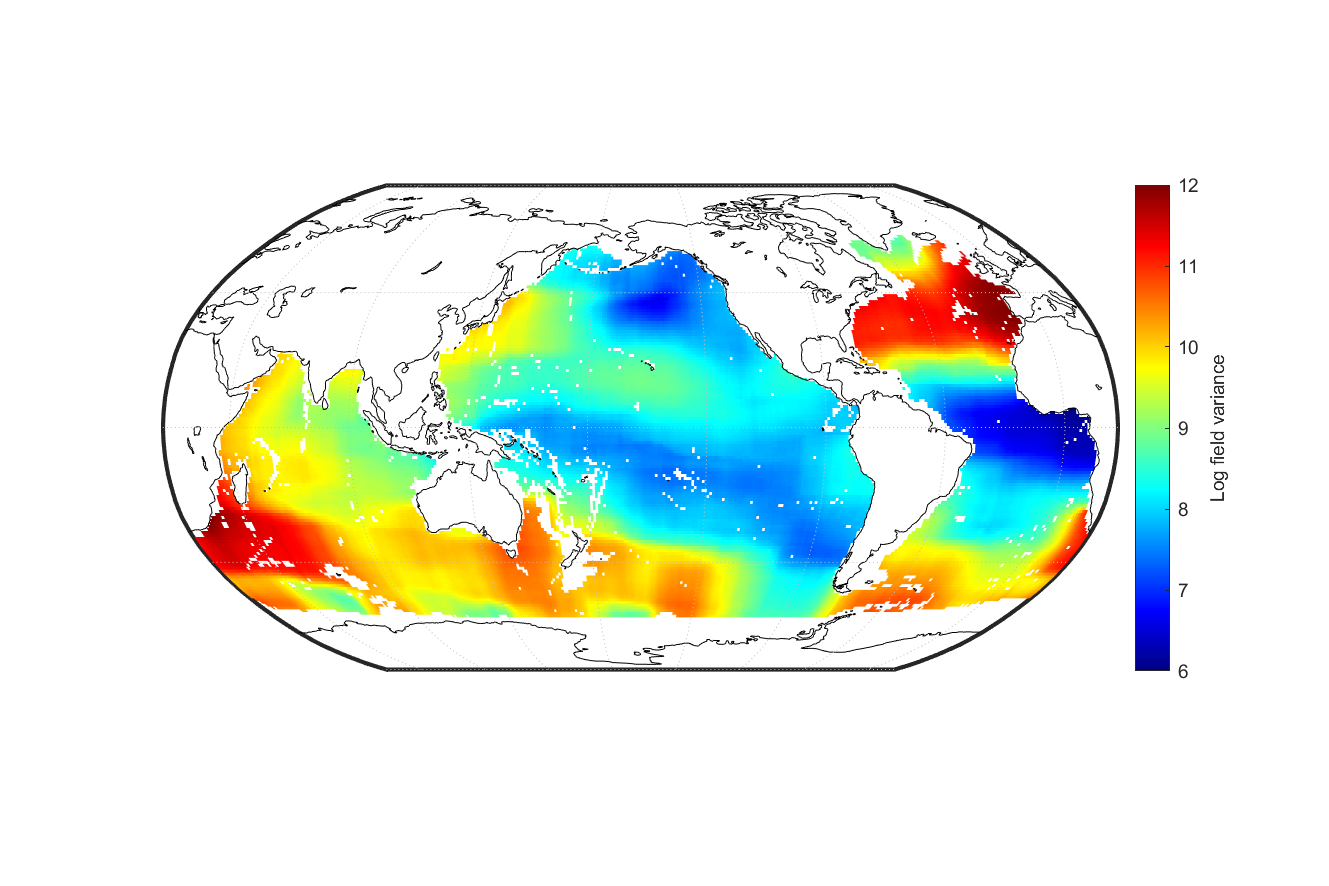}}
    \subfigure[Log nugget variance]{
        \includegraphics[width=7.5cm,trim = 1cm 3.5cm 1cm 3cm,clip = true]{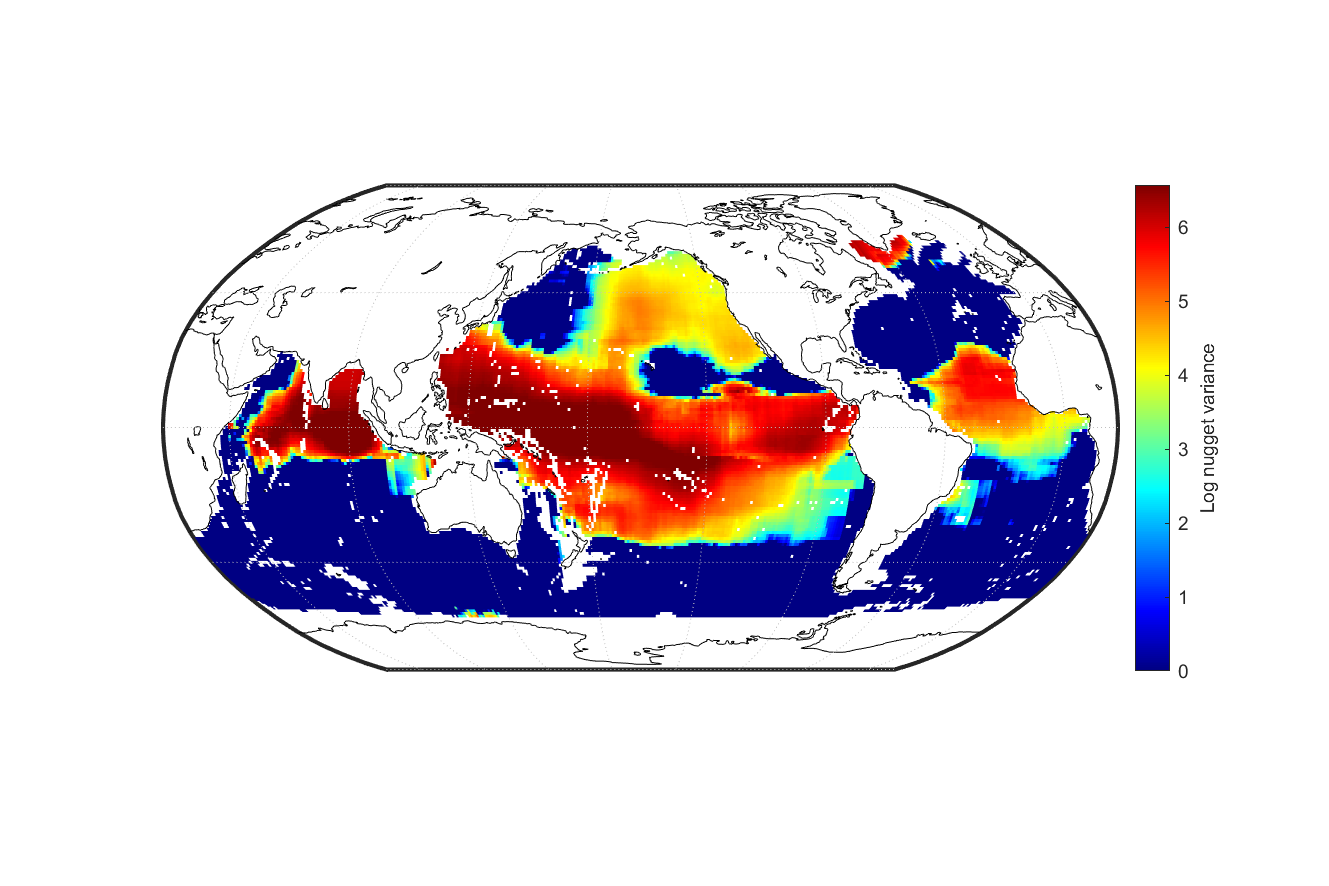}}\\
    \subfigure[Log total variance]{
        \includegraphics[width=7.5cm,trim = 1cm 3.5cm 1cm 3cm,clip = true]{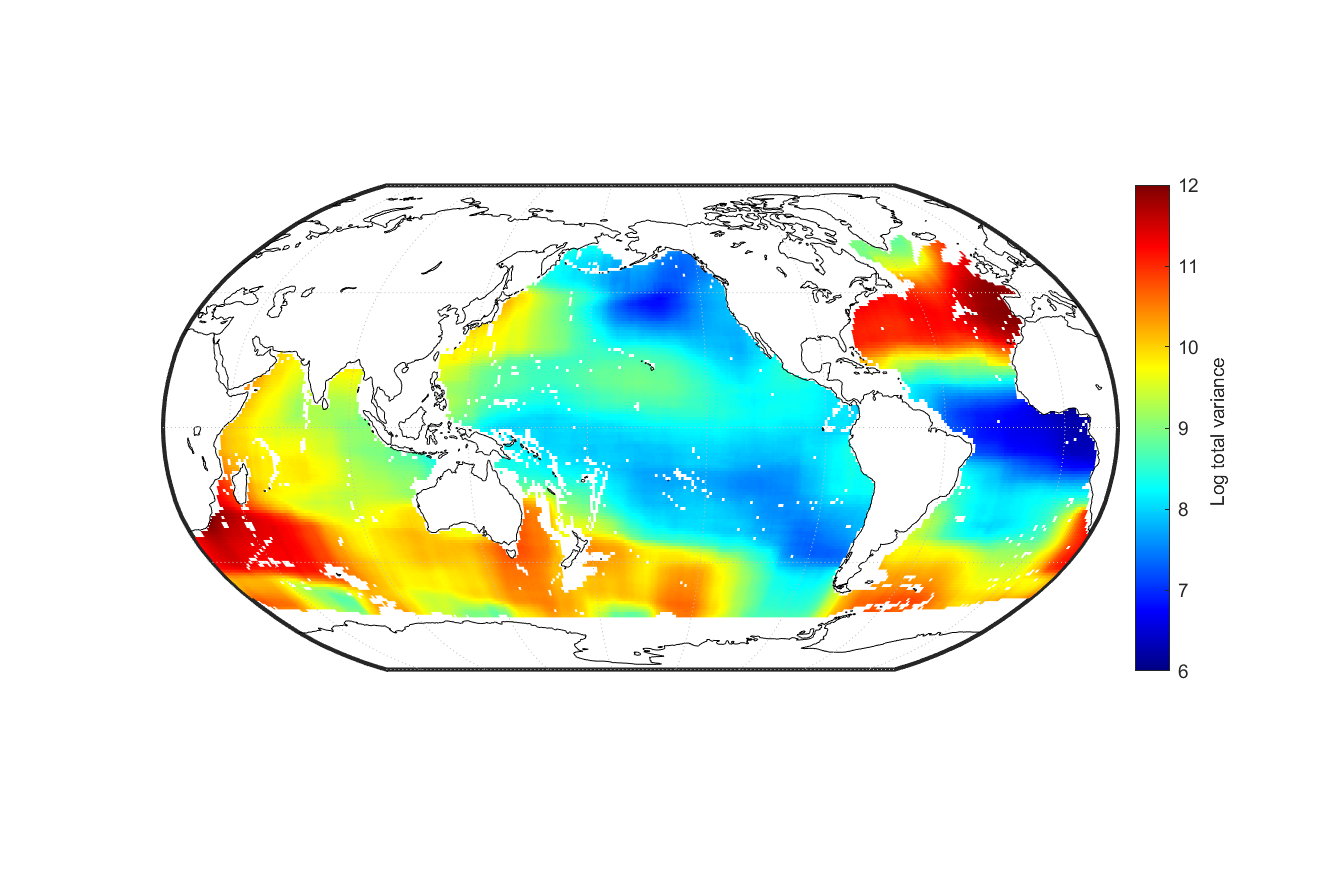}}
    \subfigure[Nugget variance vs.~total variance]{
        \includegraphics[width=7.5cm,trim = 1cm 3.5cm 1cm 3cm,clip = true]{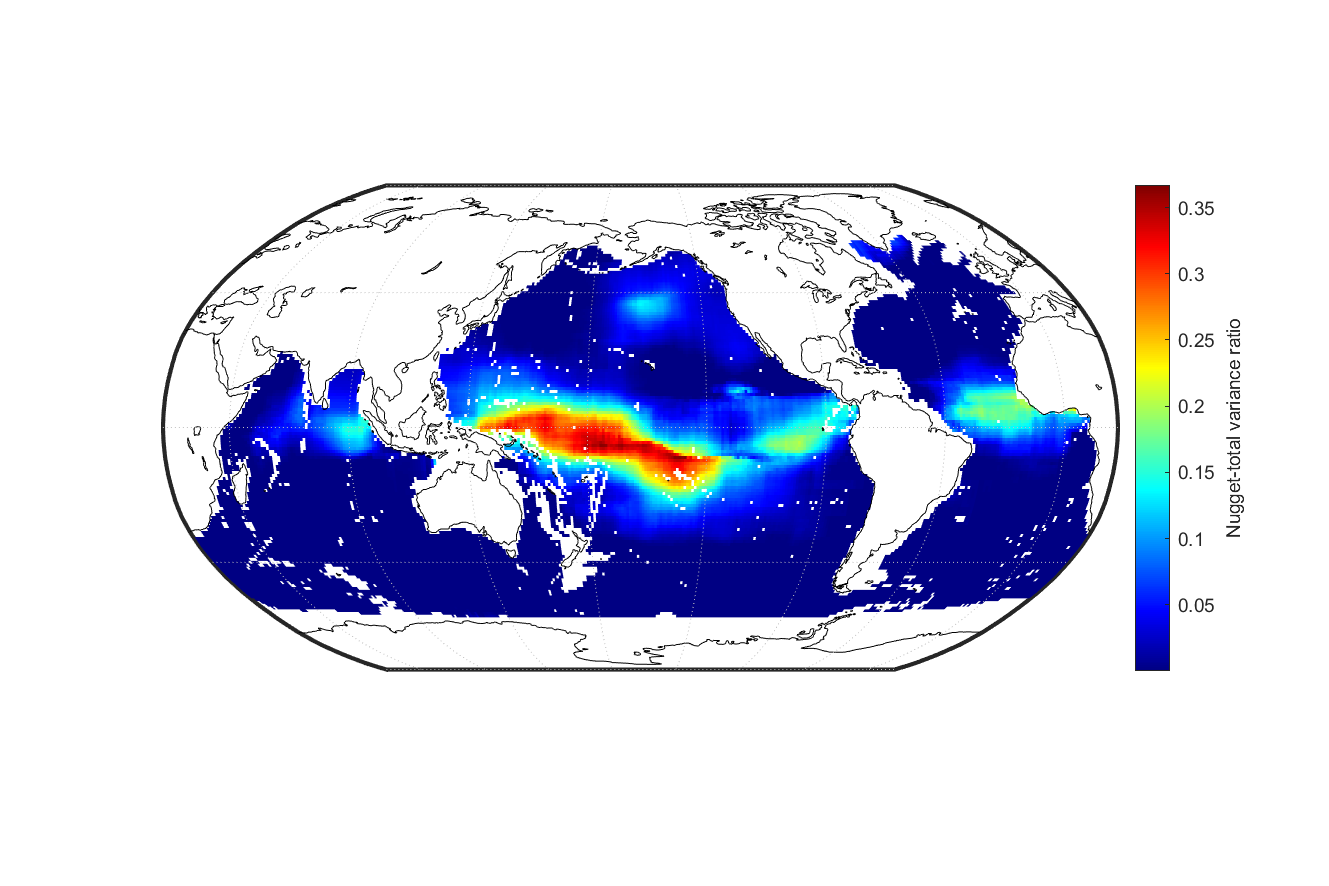}}\\
    \caption{Space-time model field and nugget variance estimates for the midocean section.}
\end{figure}

\clearpage

\subsection{Mapped anomalies and conditional simulations} \label{sec:condSimExamples}

\begin{figure}[!h]
    \centering
    \subfigure[Mapped anomalies]{
        \includegraphics[width=7.5cm,trim = 1cm 3.5cm 1cm 2.5cm,clip = true]{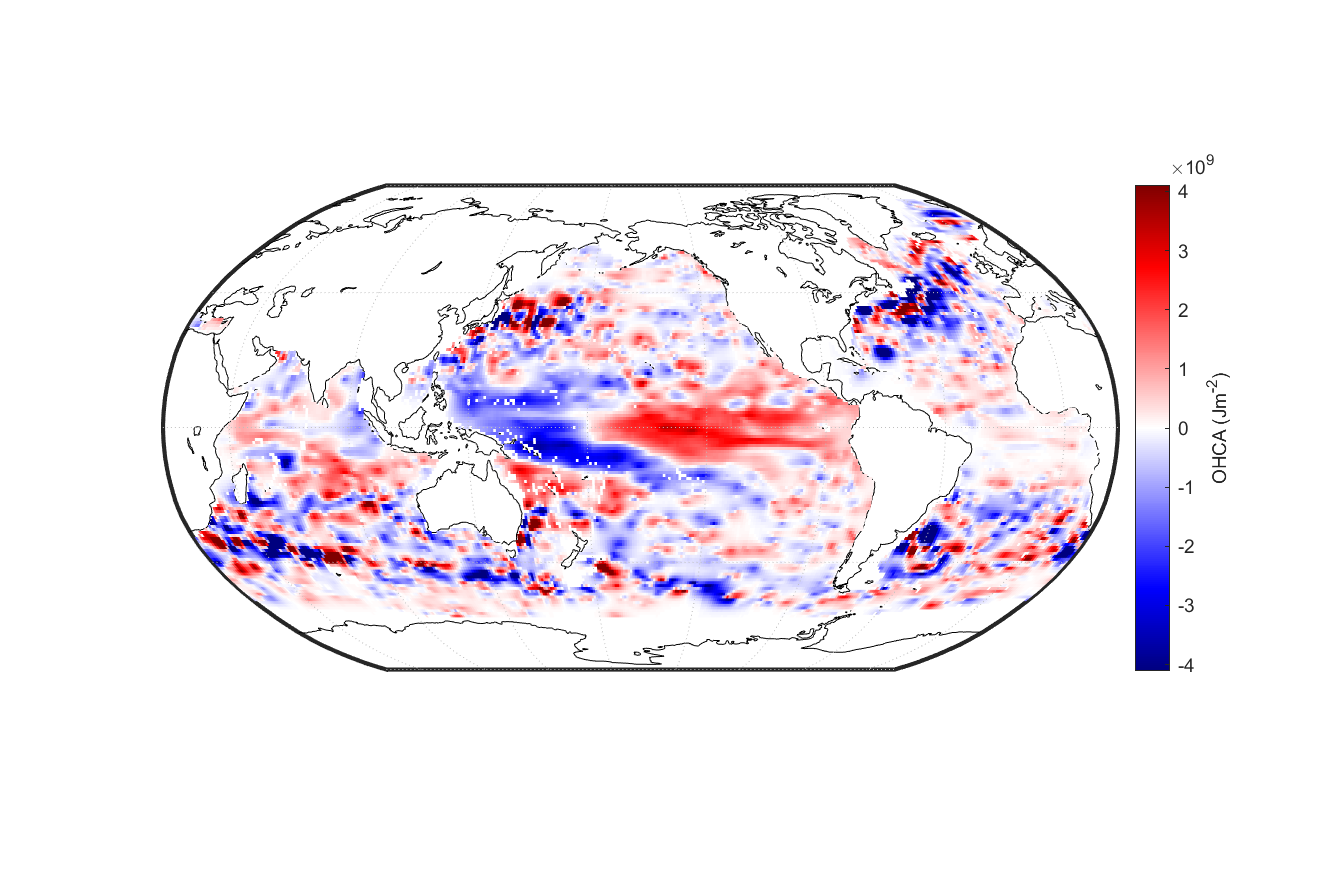}}
    \subfigure[Conditional simulation realization \#1]{
        \includegraphics[width=7.5cm,trim = 1cm 3.5cm 1cm 2.5cm,clip = true]{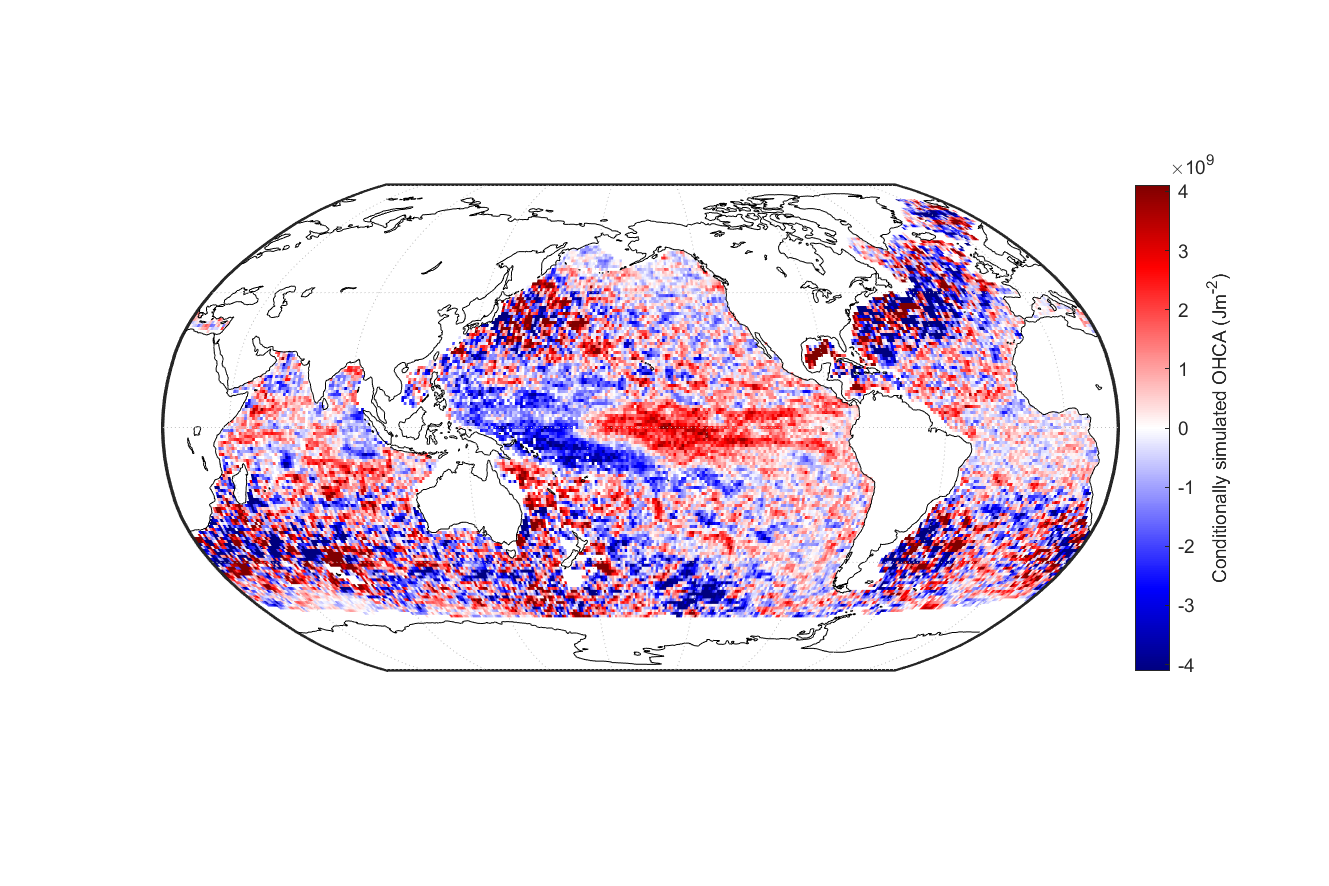}}\\
    \subfigure[Conditional simulation realization \#2]{
        \includegraphics[width=7.5cm,trim = 1cm 3.5cm 1cm 2.5cm,clip = true]{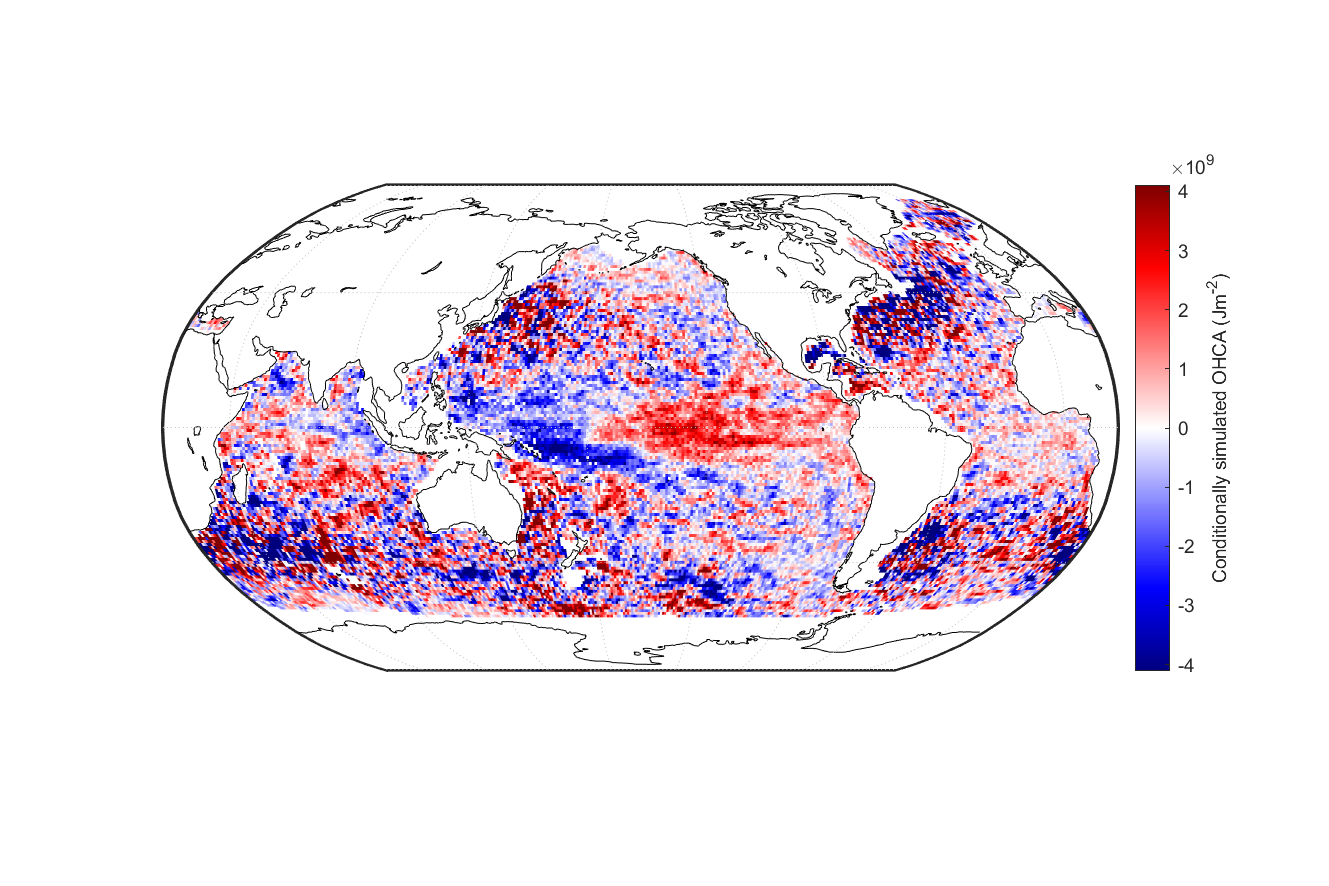}}
    \subfigure[Conditional simulation realization \#3]{
        \includegraphics[width=7.5cm,trim = 1cm 3.5cm 1cm 2.5cm,clip = true]{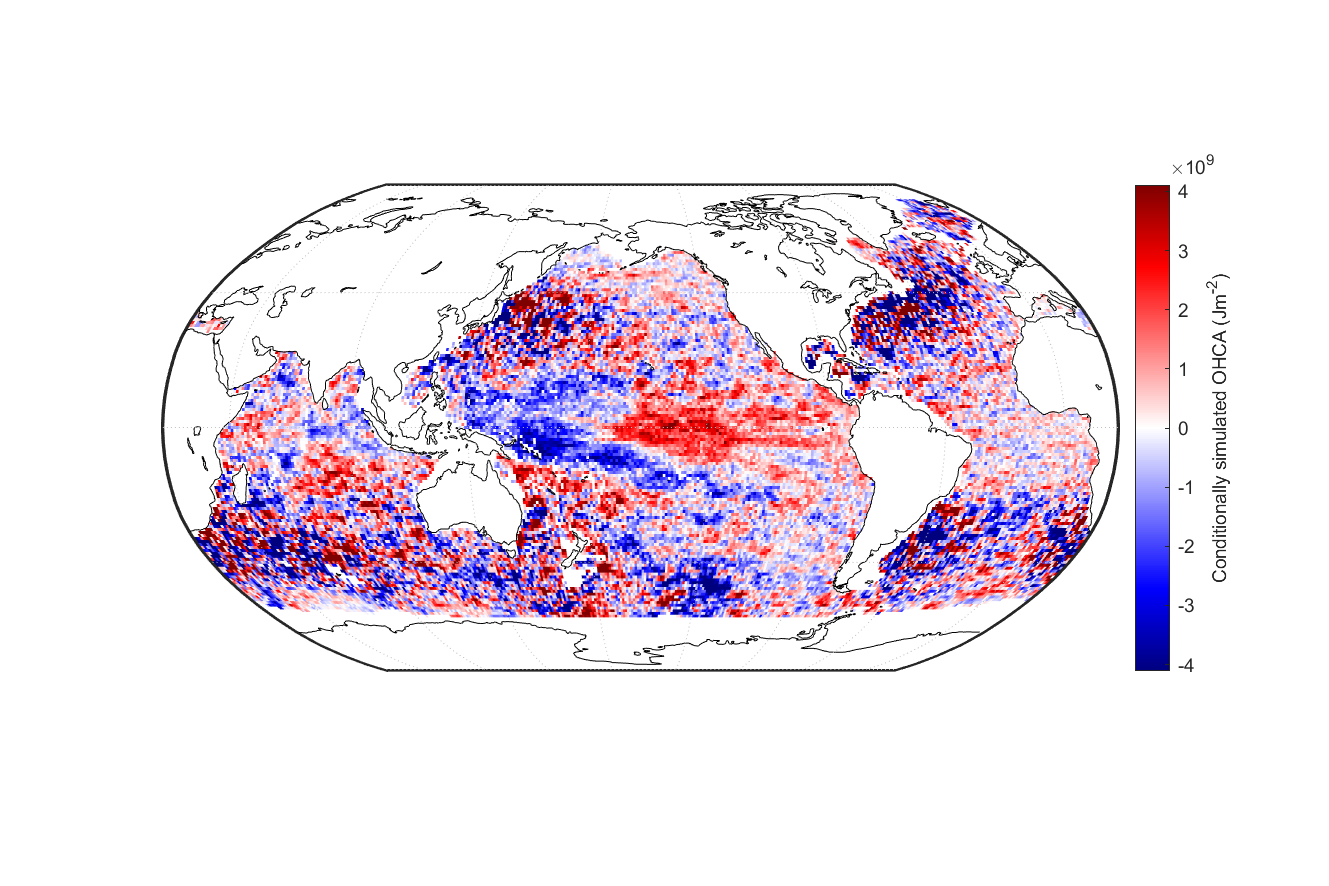}}\\
    \caption{Mapped upper ocean OHC anomaly and example conditional simulation realizations for February 2010.}
\end{figure}

\vspace{1cm}

\begin{figure}[!h]
    \centering
    \subfigure[Mapped anomalies]{
        \includegraphics[width=7.5cm,trim = 1cm 3.5cm 1cm 2.5cm,clip = true]{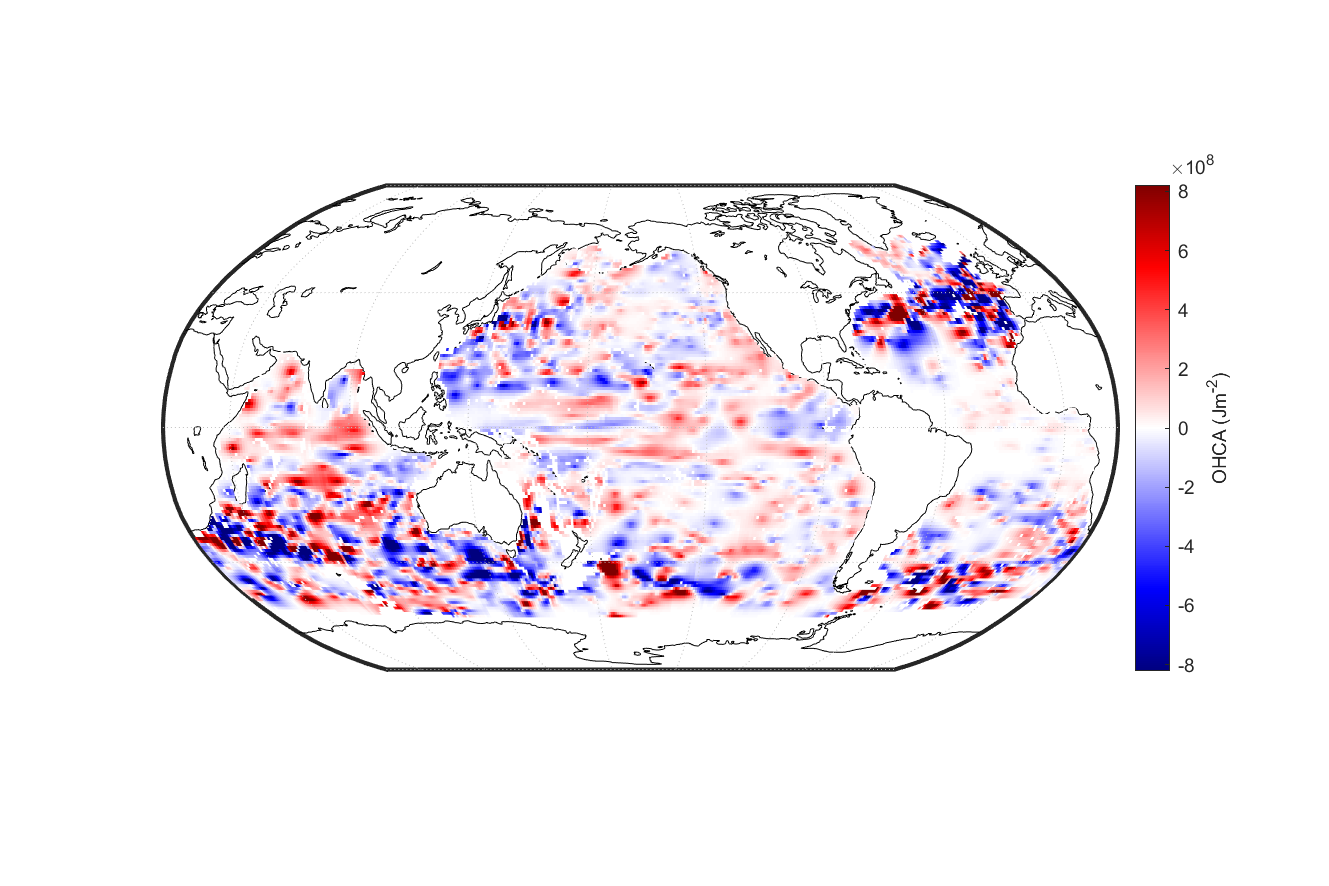}}
    \subfigure[Conditional simulation realization \#1]{
        \includegraphics[width=7.5cm,trim = 1cm 3.5cm 1cm 2.5cm,clip = true]{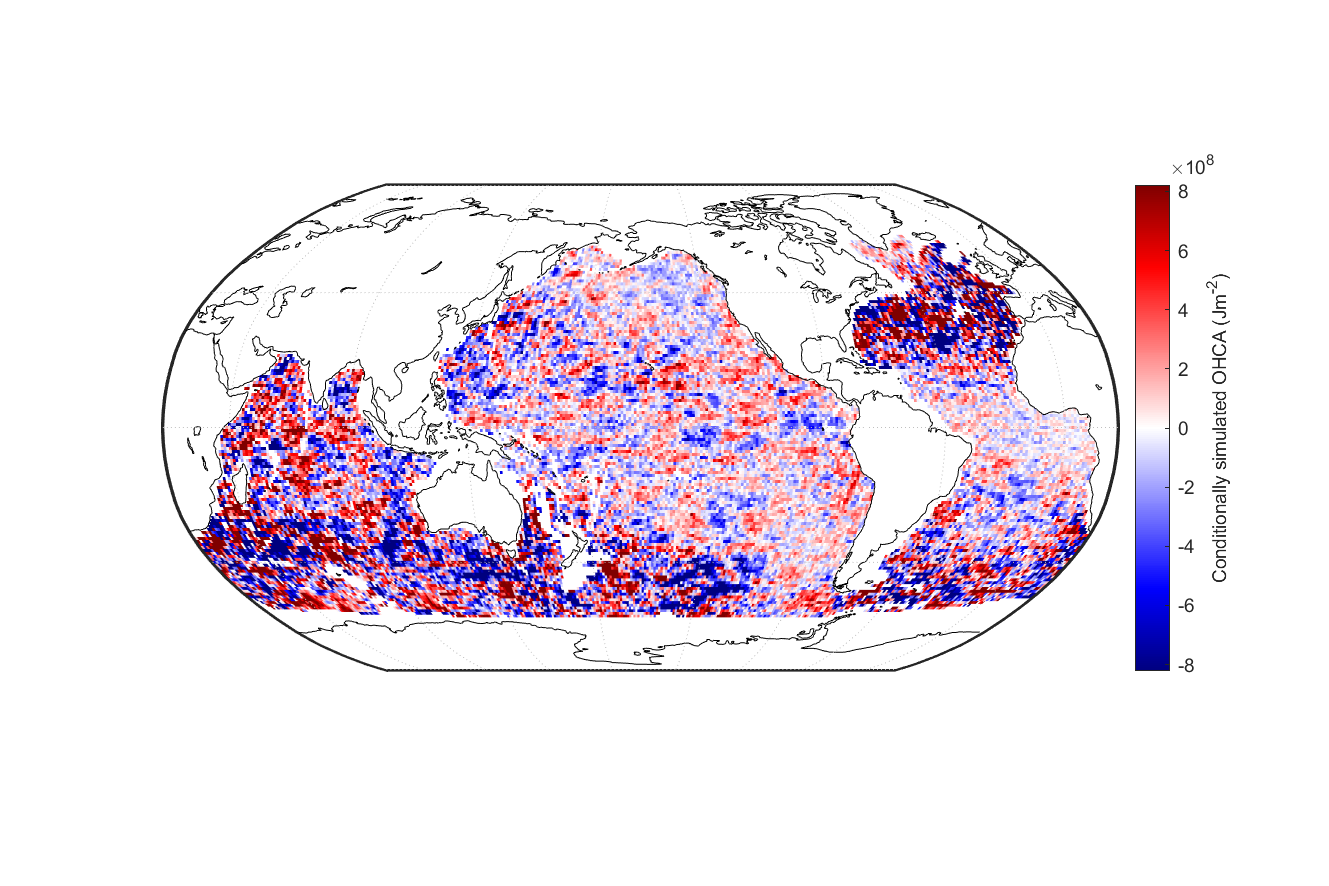}}\\
    \subfigure[Conditional simulation realization \#2]{
        \includegraphics[width=7.5cm,trim = 1cm 3.5cm 1cm 2.5cm,clip = true]{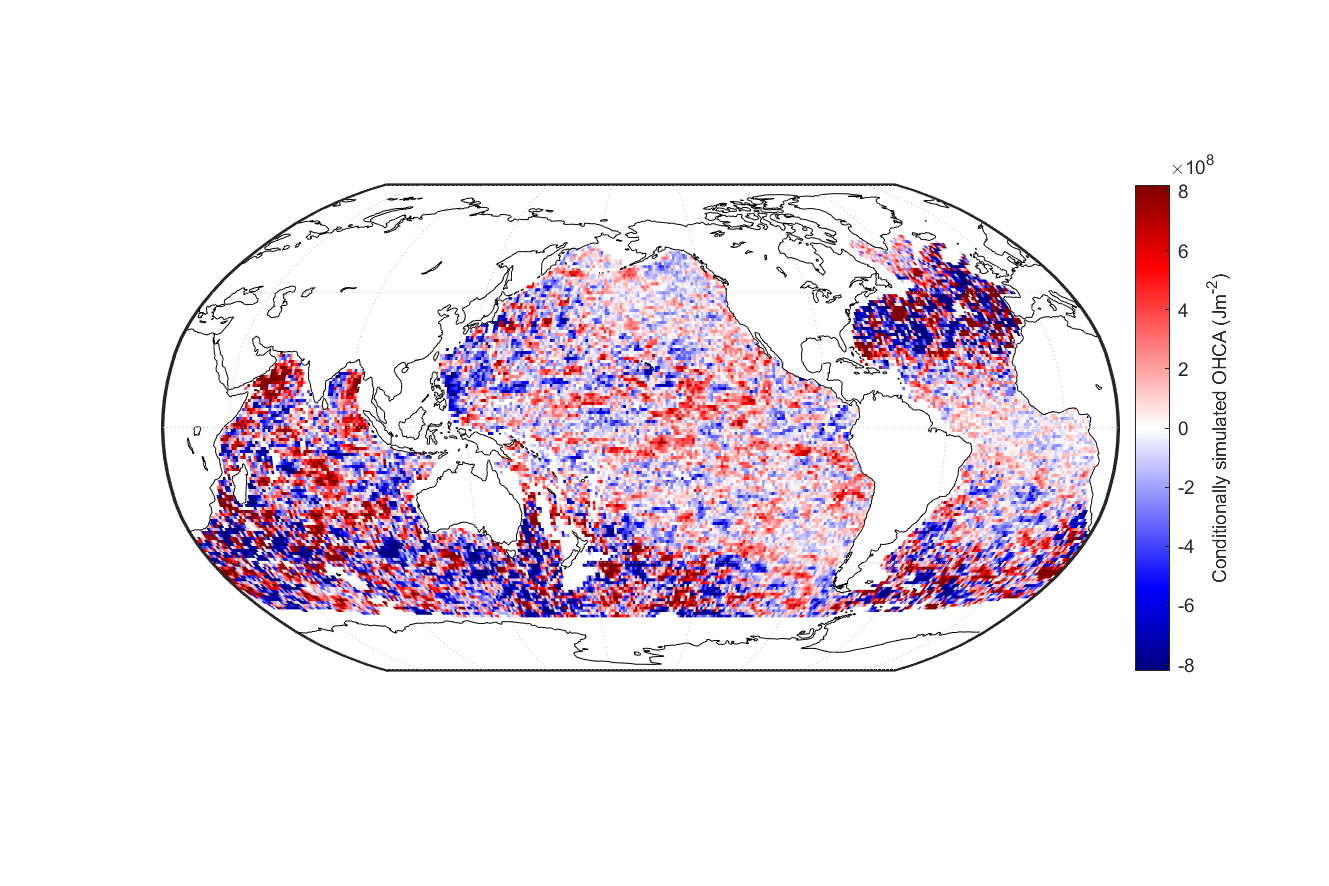}}
    \subfigure[Conditional simulation realization \#3]{
        \includegraphics[width=7.5cm,trim = 1cm 3.5cm 1cm 2.5cm,clip = true]{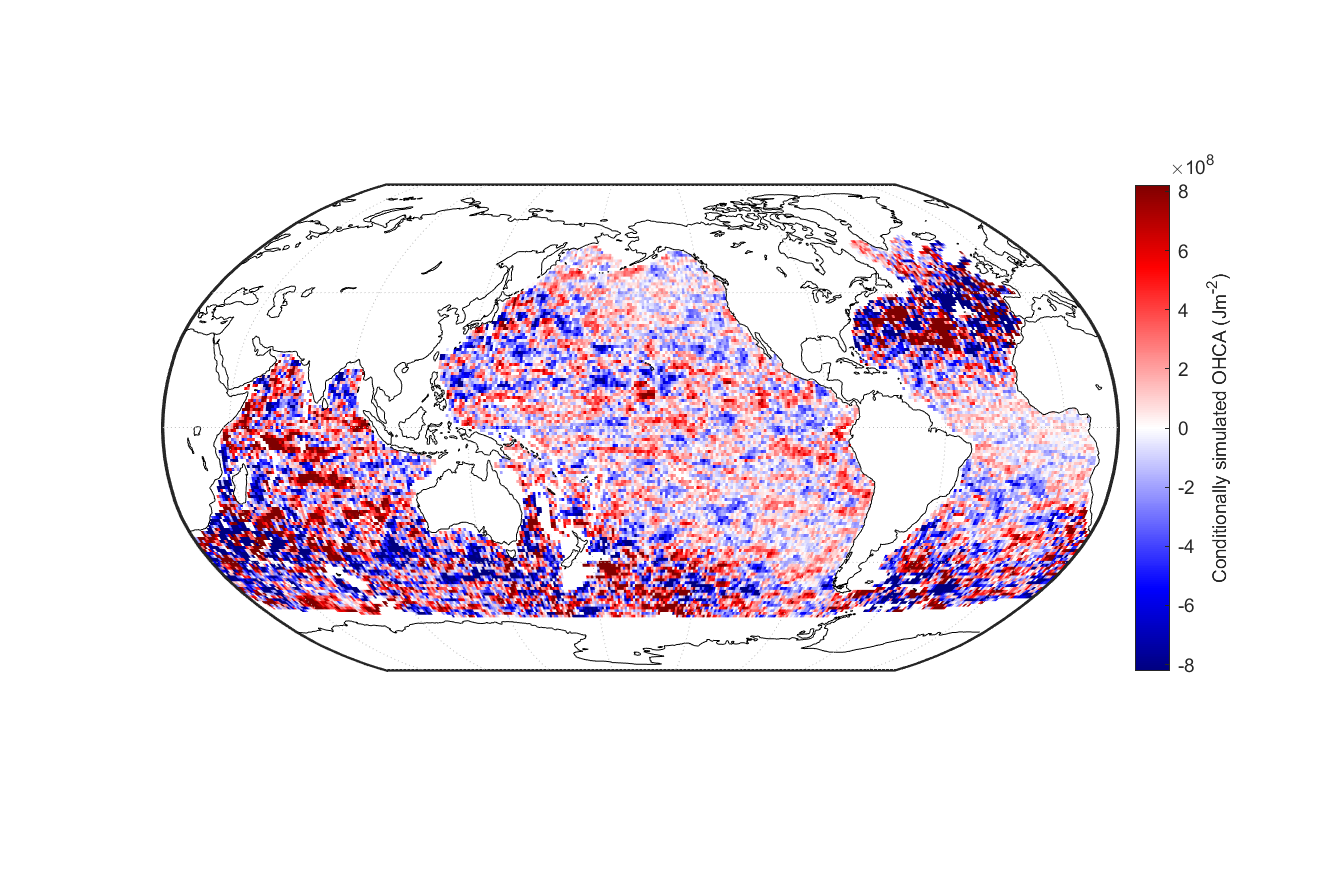}}\\
    \caption{Mapped midocean OHC anomaly and example conditional simulation realizations for February 2010.}
\end{figure}

\clearpage

\subsection{Additional validation results}

\begin{figure}[!h]
    \centering
    \setcounter{subfigure}{0}
    \subfigure[Space-only covariance, LOOO]{
        \includegraphics[width=6cm,trim = 0.5cm 0cm 1cm 0cm,clip = true]{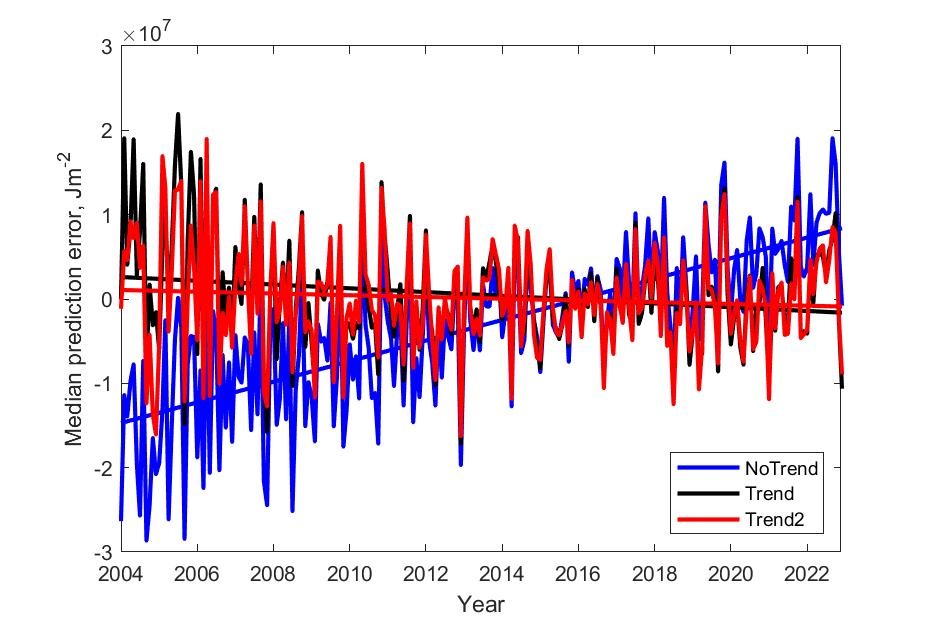}}
    \subfigure[Space-time covariance, LOOO]{
        \includegraphics[width=6cm,trim = 0.5cm 0cm 1cm 0cm,clip = true]{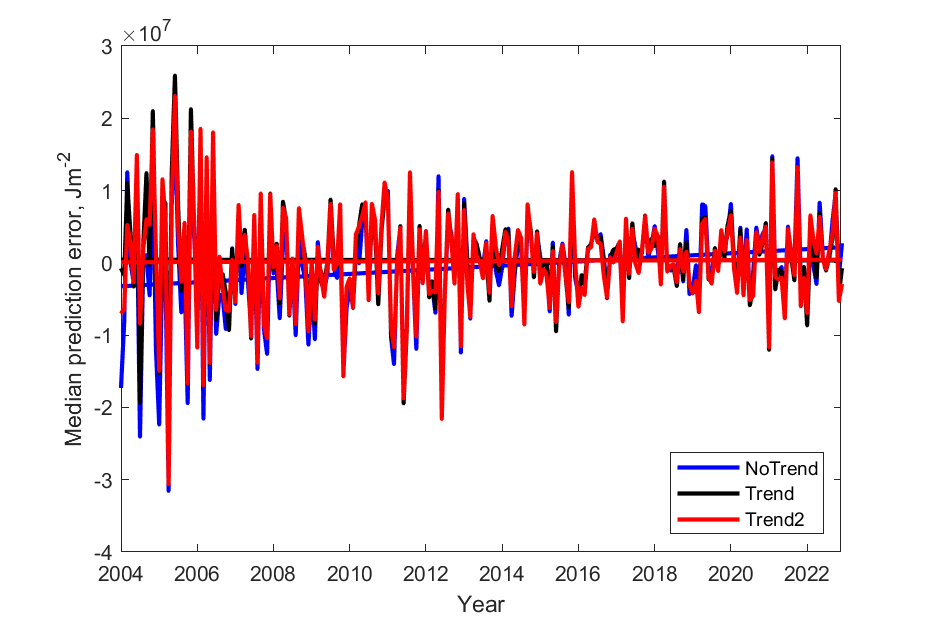}}\\
    \subfigure[Space-only covariance, LOFO]{
        \includegraphics[width=6cm,trim = 0.5cm 0cm 1cm 0cm,clip = true]{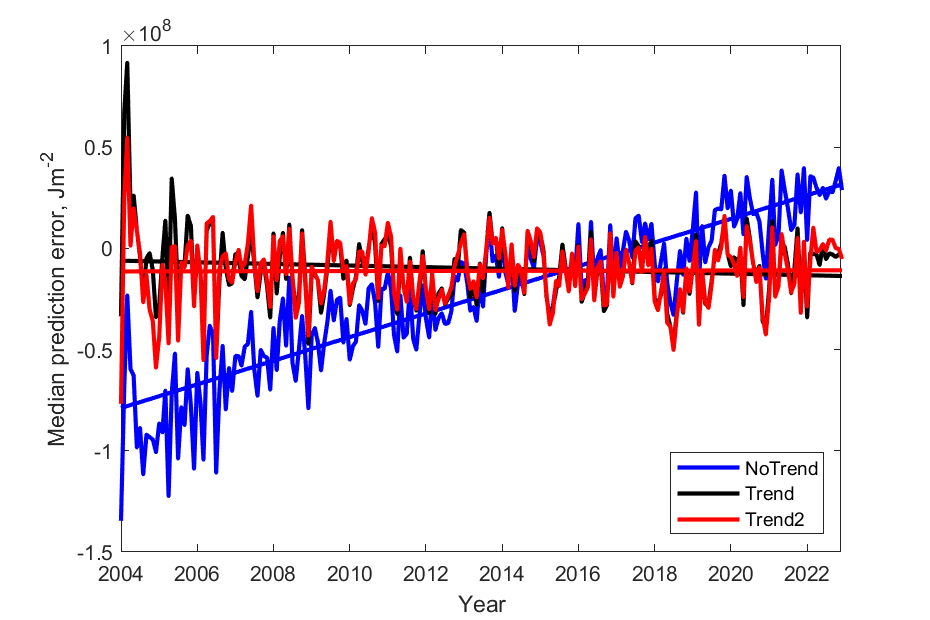}}
    \subfigure[Space-time covariance, LOFO]{
        \includegraphics[width=6cm,trim = 0.5cm 0cm 1cm 0cm,clip = true]{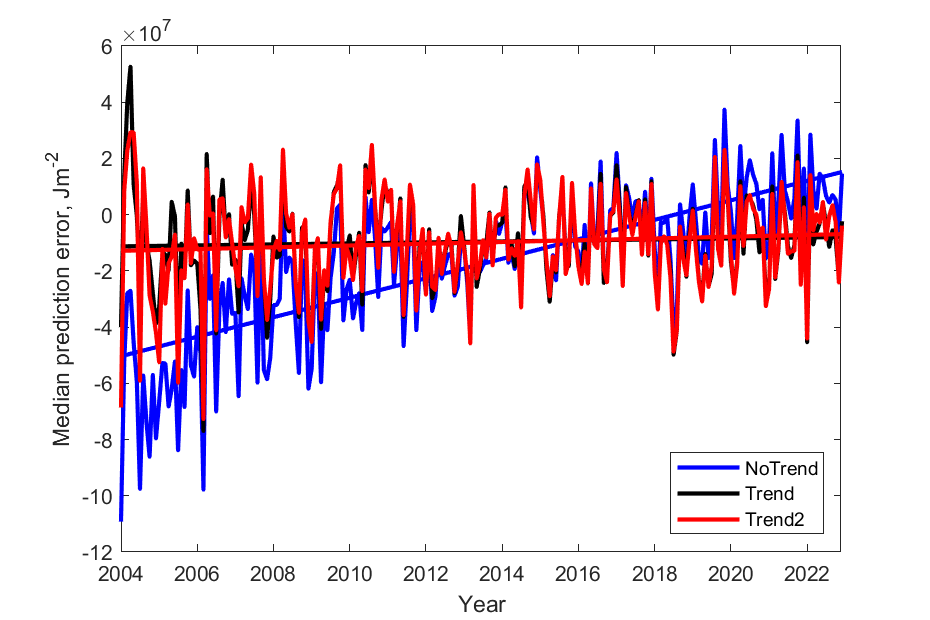}}\\
    \caption{Cross-validated monthly median prediction errors (upper ocean) for different modeling choices.}
    \label{fig:OHC_val_trend_CV_upper_ocean}
\end{figure}

\begin{figure}[!h]
    \centering
    \subfigure[Midocean, LOOO]{
        \includegraphics[width=8cm,trim = 0.5cm 0cm 1cm 0cm,clip = true]{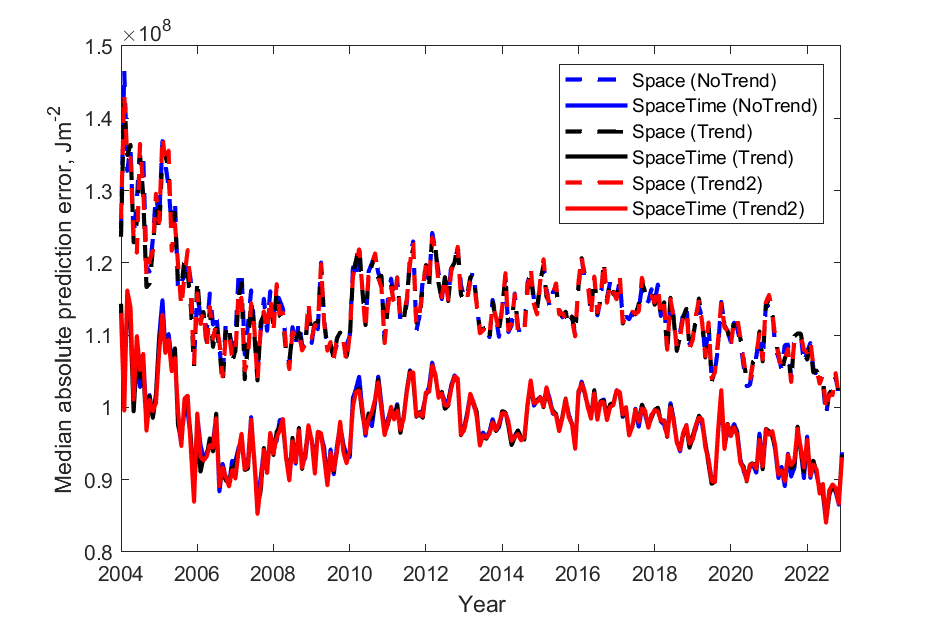}}
    \subfigure[Midocean, LOFO]{
        \includegraphics[width=8cm,trim = 0.5cm 0cm 1cm 0cm,clip = true]{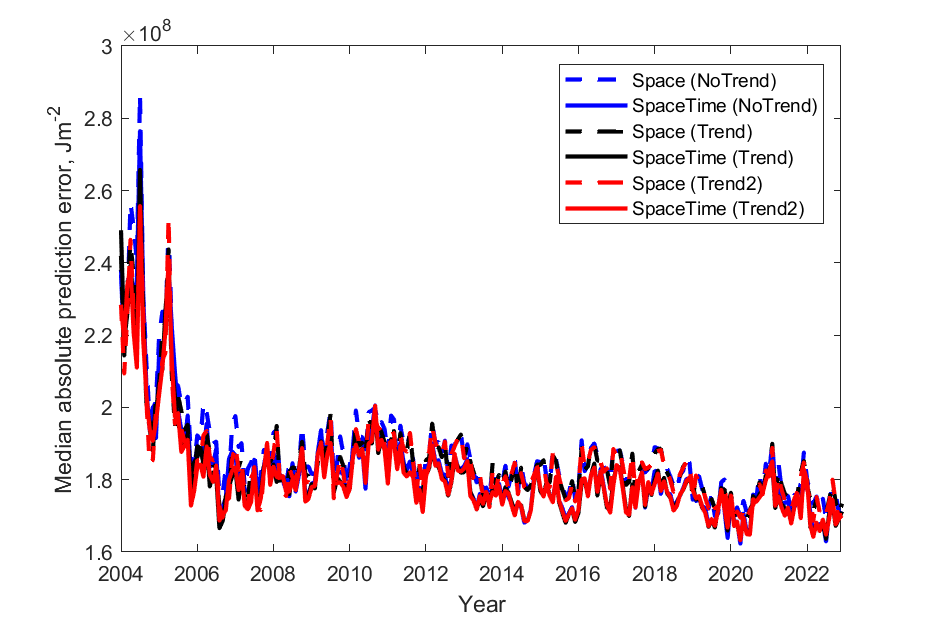}}
    \caption{Cross-validated monthly median absolute prediction errors (midocean) for different modeling choices.}
    \label{fig:OHC_val_cov_CV_midocean}
\end{figure}

\begin{figure}[!h]
    \centering
    \subfigure[Upper ocean]{
        \includegraphics[width=8cm,trim = 0.5cm 0cm 1cm 0.5cm,clip = true]{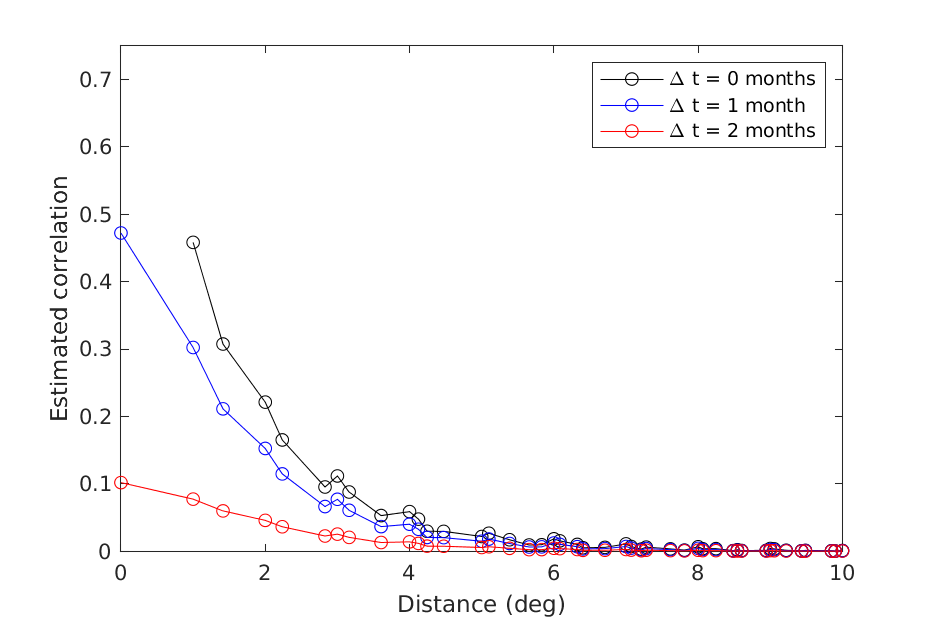}}
    \subfigure[Midocean]{
        \includegraphics[width=8cm,trim = 0.5cm 0cm 1cm 0.5cm,clip = true]{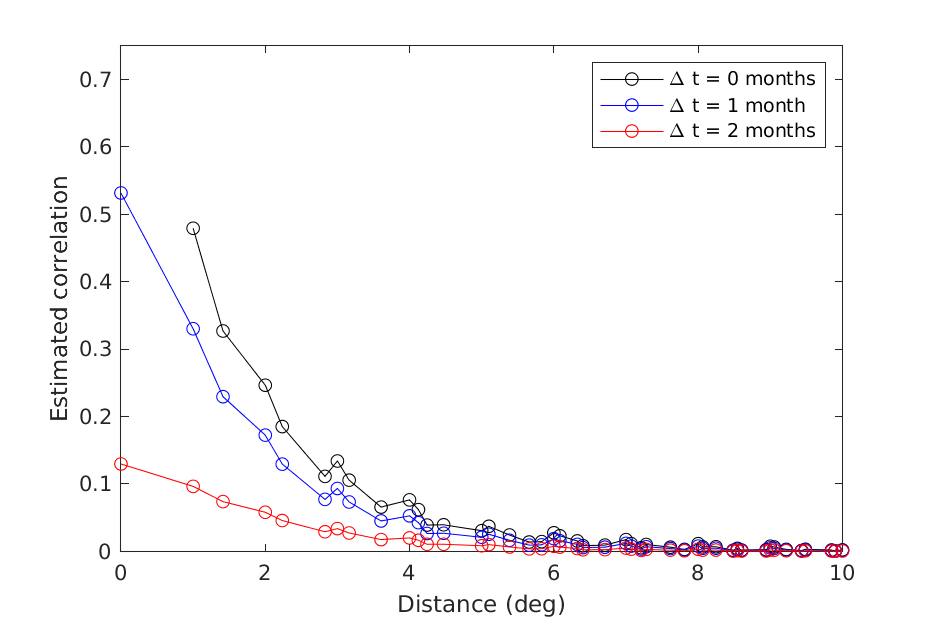}}
    \caption{Conditional simulation aggregate empirical spatio-temporal correlation functions in 2005.}
    \label{fig:OHC_val_cond_sim}
\end{figure}

\clearpage

\subsection{Global ocean heat content trend}
\begin{figure}[!h]
    \centering
    \noindent\includegraphics[width=10cm,angle=0]{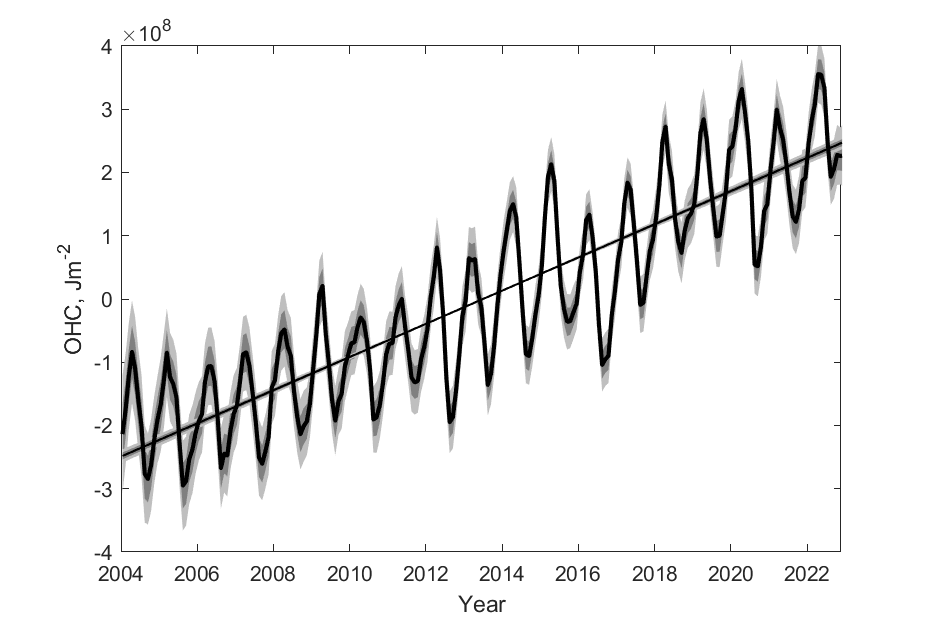}\\
    \caption{Global upper ocean OHC (15--975 dbar) time series with fitted trend. The shaded regions show 68\% (dark gray) and 95\% (light gray) uncertainties obtained using the conditional simulation ensemble. The fitted trend and its 68\% uncertainty are $0.843 \pm 0.018$ W/m$^2$.}
    \label{fig:OHC_trend_upper_ocean}
\end{figure}

\begin{figure}[!h]
    \centering
    \noindent\includegraphics[width=10cm,angle=0]{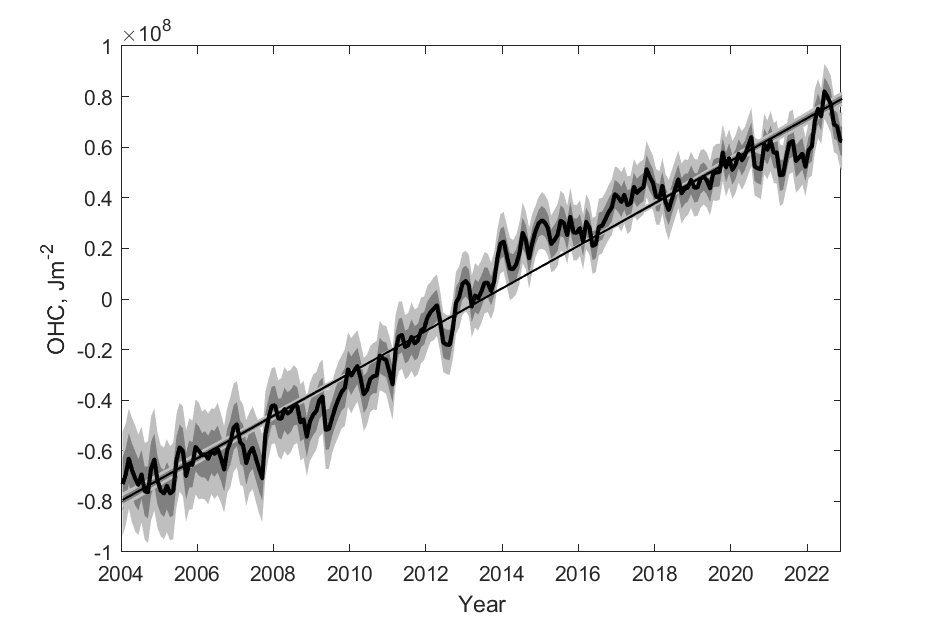}\\
    \caption{Global midocean OHC (975--1850 dbar) time series with fitted trend. The shaded regions show 68\% (dark gray) and 95\% (light gray) uncertainties obtained using the conditional simulation ensemble. The fitted trend and its 68\% uncertainty are $0.270 \pm 0.005$ W/m$^2$.}
    \label{fig:OHC_trend_midocean}
\end{figure}

\clearpage

\subsection{Global ocean heat content anomalies}

\begin{figure}[!h]
    \centering
    \subfigure[Monthly]{
        \includegraphics[width=5cm,trim = 0.5cm 0cm 1cm 0.2cm,clip = true]{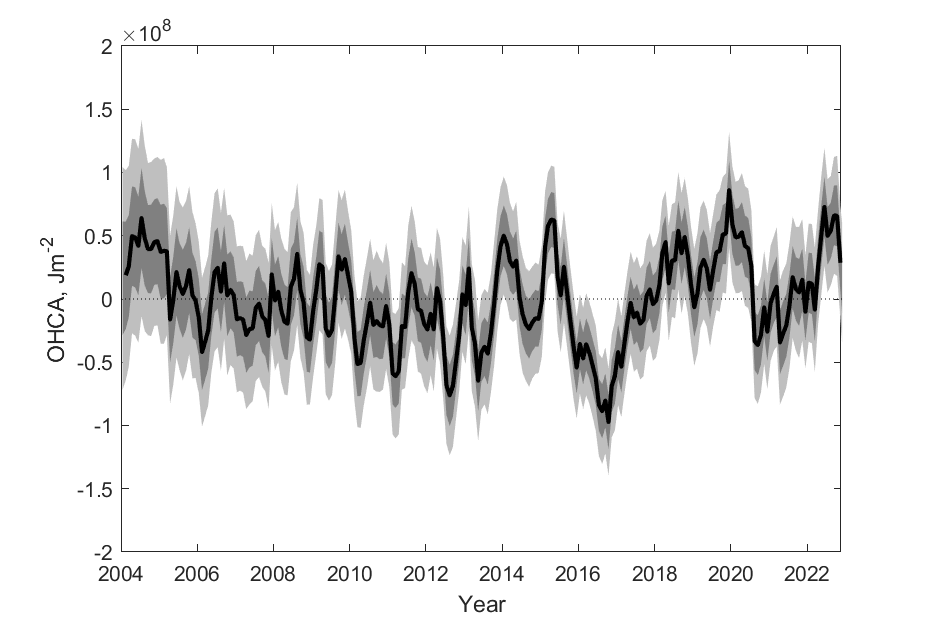}}
    \subfigure[12-month moving average]{
        \includegraphics[width=5cm,trim = 0.5cm 0cm 1cm 0.2cm,clip = true]{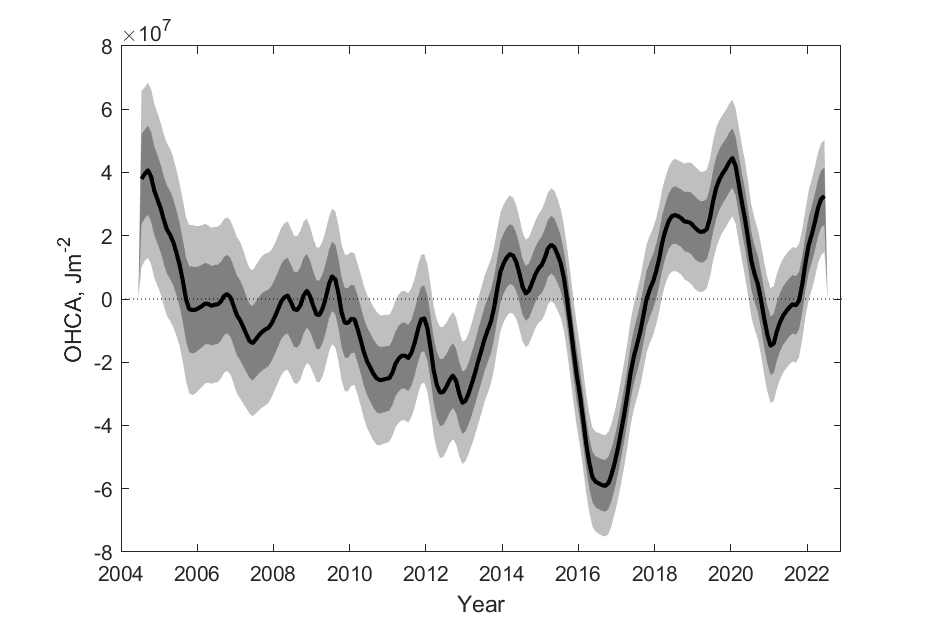}}\\
    \subfigure[24-month moving average]{
        \includegraphics[width=5cm,trim = 0.5cm 0cm 1cm 0.2cm,clip = true]{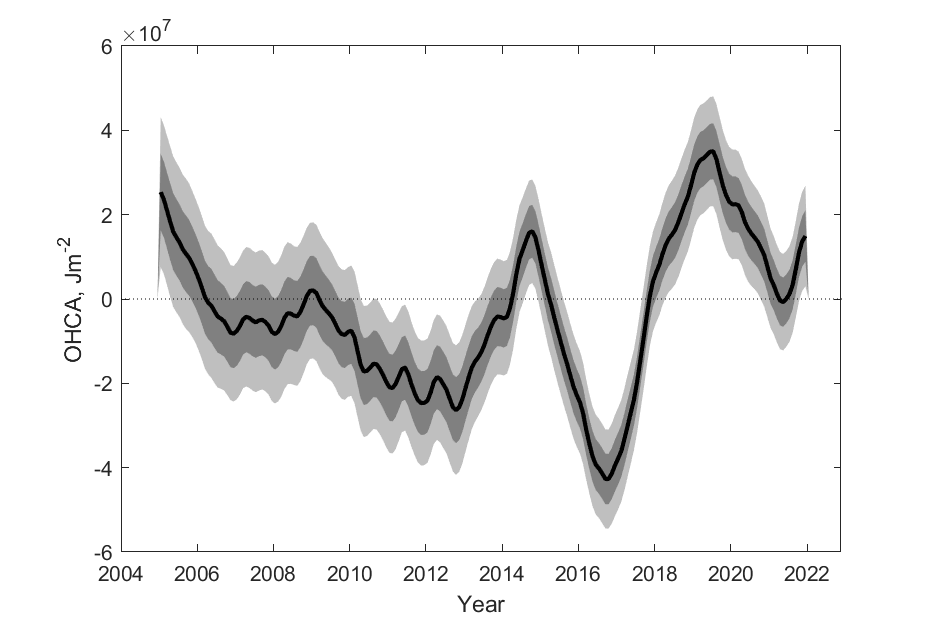}}
    \subfigure[36-month moving average]{
        \includegraphics[width=5cm,trim = 0.5cm 0cm 1cm 0.2cm,clip = true]{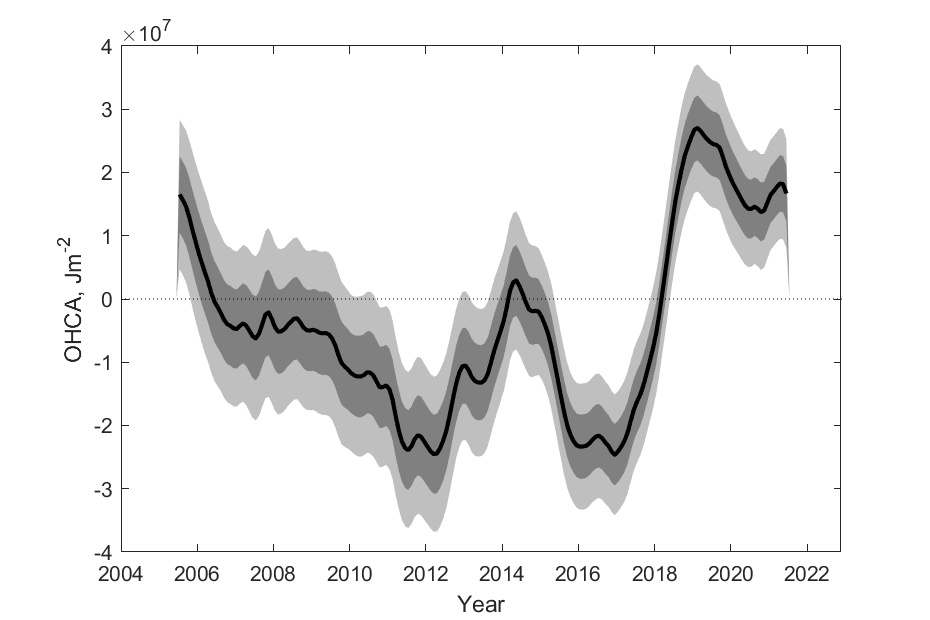}}\\
        \caption{Global upper ocean OHC (15--975 dbar) anomaly time series with 68\% (dark gray) and 95\% (light gray) uncertainties obtained using the conditional simulation ensemble.}
    \label{fig:OHCA_upper_ocean}
\end{figure}

\begin{figure}[!h]
    \centering
    \subfigure[Monthly]{
        \includegraphics[width=5cm,trim = 0.5cm 0cm 1cm 0.2cm,clip = true]{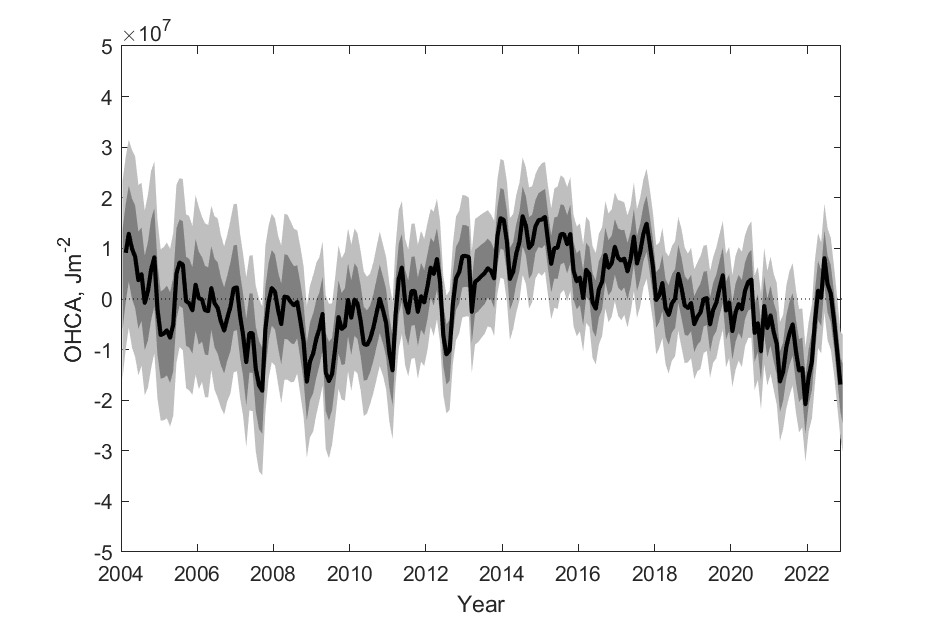}}
    \subfigure[12-month moving average]{
        \includegraphics[width=5cm,trim = 0.5cm 0cm 1cm 0.2cm,clip = true]{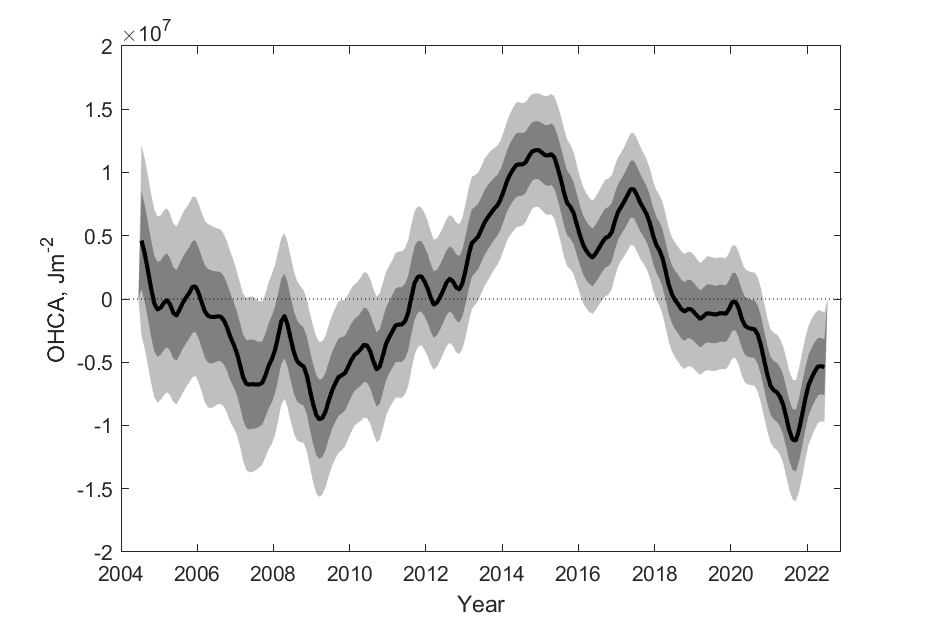}}\\
    \subfigure[24-month moving average]{
        \includegraphics[width=5cm,trim = 0.5cm 0cm 1cm 0.2cm,clip = true]{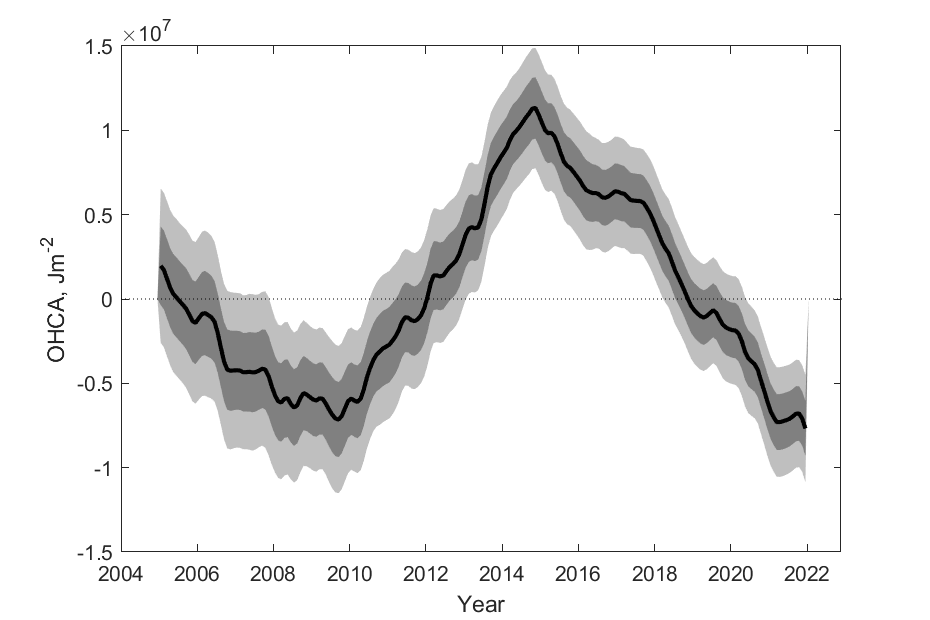}}
    \subfigure[36-month moving average]{
        \includegraphics[width=5cm,trim = 0.5cm 0cm 1cm 0.2cm,clip = true]{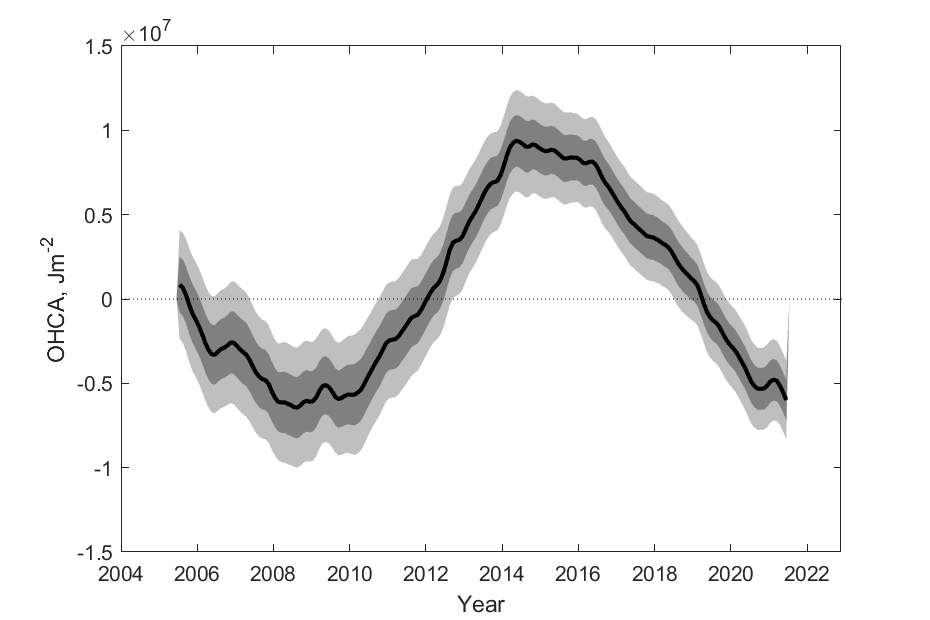}}\\
        \caption{Global midocean OHC (975--1850 dbar) anomaly time series with 68\% (dark gray) and 95\% (light gray) uncertainties obtained using the conditional simulation ensemble.}
    \label{fig:OHCA_midocean}
\end{figure}



\bibliographystyle{ametsocV6}
\bibliography{references}

\end{document}